\documentclass[10pt]{article}

\usepackage{amssymb}
\usepackage{amsthm}
\usepackage{amsmath}

\usepackage{graphicx} 
\usepackage{psfrag}
\usepackage{subfigure}

\newcommand{\proofend}{\hfill\rule{0.2cm}{0.2cm}}
\newcommand{\sfrac}[2]{{\textstyle{\frac{#1}{#2}}}}
\theoremstyle{plain}
\newtheorem{theorem}{Theorem}[section]

\newtheorem*{definition}{Definition}

\theoremstyle{remark}

\newtheorem*{example}{Example}
\newtheorem*{remark}{Remark}
\newtheorem{assumption}{Assumption}

\begin{document}

%%%%%%%%%%%%%%%%%%%%%%%%%%%%%%%%%%%%%%%%%%%%%%%%%%%%%%%%%%%%%%%%%%%
\title{\sc Newtonian stellar models}
%%%%%%%%%%%%%%%%%%%%%%%%%%%%%%%%%%%%%%%%%%%%%%%%%%%%%%%%%%%%%%%%%%%
\author{\sc 
J.\ Mark Heinzle$^{1}$\thanks{Electronic address: {\tt
Mark.Heinzle@ap.univie.ac.at}}\ ,\
\ and
Claes Uggla$^{2}$\thanks{Electronic address:
{\tt Claes.Uggla@kau.se}}\\
$^{1}${\small\em Institute for Theoretical Physics, University of Vienna}\\
{\small\em A-1090 Vienna, Austria}\\
$^{2}${\small\em Department of Physics, University of Karlstad}\\
{\small\em S-651 88 Karlstad, Sweden}}

%%%%%%%%%%%%%%%%%%%%%%%%%%%%%%%%%%%%%%%%%%%%%%%%%%%%%%%%%%%%%%%%%%%
\date{\normalsize{November 22, 2002}}
%%%%%%%%%%%%%%%%%%%%%%%%%%%%%%%%%%%%%%%%%%%%%%%%%%%%%%%%%%%%%%%%%%%
\maketitle
%%%%%%%%%%%%%%%%%%%%%%%%%%%%%%%%%%%%%%%%%%%%%%%%%%%%%%%%%%%%%%%%%%%
%%%%%%%%%%%%%%%%%%%%%%%%%%%%%%%%%%%%%%%%%%%%%%%%%%%%%%%%%%%%%%%%%%%
\begin{abstract}
%%%%%%%%%%%%%%%%%%%%%%%%%%%%%%%%%%%%%%%%%%%%%%%%%%%%%%%%%%%%%%%%%%%
We investigate static spherically symmetric perfect fluid models 
in Newtonian gravity for baro\-tropic 
equations of state that are asymptotically polytropic at low and high pressures. 
This is done by casting the equations into a 
3-dimensional regular dynamical system with bounded dependent 
variables. The low and high central 
pressure limits correspond to two 2-dimensional boundary subsets, 
described by homology invariant equations for exact polytropes. 
Thus the formulation naturally places work about polytropes 
in a more general context. The introduced framework yields a 
visual aid for obtaining qualitative information about the 
solution space and is also suitable for numerical investigations. 
Last, but not least, it makes a host of mathematical tools 
from dynamical systems theory available. This allows 
us to prove a number of theorems about 
the relationship between the equation of state and properties 
concerning total masses and radii.

%%%%%%%%%%%%%%%%%%%%%%%%%%%%%%%%%%%%%%%%%%%%%%%%%%%%%%%%%%%%%%%%%%%
\end{abstract}
%%%%%%%%%%%%%%%%%%%%%%%%%%%%%%%%%%%%%%%%%%%%%%%%%%%%%%%%%%%%%%%%%%%
\bigskip
\centerline{\noindent PACS numbers: 
04.40.--b, % self-gravitating systems; continuous media
97.10.--q, % stellar properties
02.90.+p.} % mathematical methods in physics

\vspace{10pt}

\begin{center}
\begin{minipage}{0.8\textwidth}
\begin{tabbing}
\> asymptotically polytropic equations of state.
\end{tabbing}
\end{minipage}
\end{center}

\vfill
\newpage
%%%%%%%%%%%%%%%%%%%%%%%%%%%%%%%%%%%%%%%%%%%%%%%%%%%%%%%%%%%%%%%%%%%

%%%%%%%%%%%%%%%%%%%%%%%%%%%%%%%%%%%%%%%%%%%%%%%%%%%%%%%%%%%%%%%%%%%%%%%%
%%%%%%%%%%%%%%%%%%%%%%%%%%%%%%%%%%%%%%%%%%%%%%%%%%%%%%%%%%%%%%%%%%%%%%%%
%%%%%%%%%%%%%%%%%%%%%%%%%%%%%%%%%%%%%%%%%%%%%%%%%%%%%%%%%%%%%%%%%%%%%%%%
%%%%%%%%%%%%%%%%%%%%%%%%%%%%%%%%%%%%%%%%%%%%%%%%%%%%%%%%%%%%%%%%%%%%%%%%
\section{Introduction}
\label{introduction}
%%%%%%%%%%%%%%%%%%%%%%%%%%%%%%%%%%%%%%%%%%%%%%%%%%%%%%%%%%%%%%%%%%%%%%%%

Spherically symmetric static Newtonian perfect fluid models are the
starting point for many discussions about stellar structure and
evolution. Among the least complicated models are those described by
barotropic equations of state $p=p(\rho)$, where $p$ is the pressure
and $\rho$ the mass density. Within this class of models, polytropic
equations of state, $p = K\rho^{1 + 1/n}$ ($K$, $n$ non-negative constants), 
have been studied thoroughly, and discussions about polytropes can be found
in many textbooks on astrophysics (e.g., 
Chandrasekhar~\cite{Chandrasekhar:1939}, 
Kippenhahn and Weigert~\cite{book:KippWeig1994},
Shapiro and Teukolsky~\cite{Shapiro/Teukolsky:1983}).
The reason for this is at least
two-fold: Firstly, they are of physical interest, since they often
constitute good approximations for certain regions of various types of
stars. Secondly, the polytropic equations of state give rise to
symmetries, which makes the equations more tractable. 
In particular, one can
introduce so-called homology invariants and thereby reduce the
hydrodynamical problem to two coupled autonomous first order ordinary
differential equations (see e.g., 
Kippenhahn and Weigert~\cite{book:KippWeig1994}, and references therein).

More realistic stellar models must necessarily go beyond the
pure polytropes.
The next level of complexity is to consider
more general classes of barotropic
equations of state. In astrophysics numerous equations of state 
describing different kinds of matter, e.g., a stellar plasma or a 
degenerate ideal Fermi gas,
have been considered and some of the 
corresponding stellar models have been studied in detail, although mostly
numerically. It is worth noticing that most of the equations of state
that are of astrophysical interest behave like polytropes for low and 
for high pressures, i.e., they belong to the class of asymptotically 
polytropic equations of state.
However, a unified analytical treatment of this important class,
comparable to that of the polytropes, does not exist so far in
the literature, and thus there is a gap that needs to be closed.

In this paper we consider static spherically symmetric Newtonian
perfect fluid models with barotropic equations of state that are
asymptotically polytropic for low and high pressures. This is done by
introducing new variables so that a regular set of three autonomous
ordinary differential equations with bounded dependent variables is
obtained. Two of the variables are asymptotically homology
invariant. Moreover, the variables are chosen so that the equation of
state is encoded in a single function appearing at only one place in
the equation system. For polytropes this function reduces to the
polytropic index $n$. Thus the formulation not only naturally
incorporates our knowledge about polytropes but also allows us to
deduce features valid for the general class of asymptotically
polytropic equations of state. The main goals of the paper are to
provide:

\begin{itemize} 
\item a framework yielding a visual representation of 
the solution space, which allows us to obtain qualitative
information about solutions,
\item a framework suitable for numerical investigations, 
\item a formulation that makes a host of mathematical tools 
from dynamical systems theory available, i.e., a framework that makes it 
possible to prove theorems about e.g., the relationship between 
the equation of state and mass- and radius-properties. 
\end{itemize}

The outline of the paper is the following:  in
Section~\ref{equationsofstate} we define and discuss the family of
asymptotically polytropic equations of state. In
Section~\ref{dynamicalsystemsformulation} we introduce new bounded
variables, and derive the 3-dimensional system of autonomous
equations. In Section~\ref{dynamicalsystemsanalysis} we discuss the
state space associated with our formulation and do a local dynamical
systems analysis that yields asymptotic expressions for e.g.,
the mass and gravitational potential for the various solutions. In
addition we give a theorem describing the global dynamics of the state
space. This theorem (and most of the following theorems) 
is proved later in Appendix~\ref{A}, to make the
paper more accessible to the general reader. In
Section~\ref{examples} we give a number of examples that illustrate the
usefulness and pedagogical value of our approach. For instance, it is
shown that the problem corresponding to Chandrasekhar's equation of
state for white dwarfs has a particularly simple and elegant
description. We also give a variety of mass-radius diagrams showing how
extremely complicated mass-radius diagrams can become (at least for
unstable models). In Section~\ref{qualitativeresults} we state a
number of theorems. Firstly, we give criteria for the equation of
state that ensure that the corresponding models have finite radii and
masses. Secondly, we state theorems about how mass-radius
relationships depend on the equation of state. Most of the proofs are
given in Appendix~\ref{A}. In Section~\ref{concludingremarks} we conclude 
with a discussion about possible
generalizations. Appendix~\ref{A} contains a brief introduction to
some relevant background material from dynamical systems theory, 
and most of the proofs of the theorems. Finally, in Appendix~\ref{B}
we discuss some miscellaneous results.

%%%%%%%%%%%%%%%%%%%%%%%%%%%%%%%%%%%%%%%%%%%%%%%%%%%%%%%%%%%%%%%%%%%%%%%%
%%%%%%%%%%%%%%%%%%%%%%%%%%%%%%%%%%%%%%%%%%%%%%%%%%%%%%%%%%%%%%%%%%%%%%%%
%%%%%%%%%%%%%%%%%%%%%%%%%%%%%%%%%%%%%%%%%%%%%%%%%%%%%%%%%%%%%%%%%%%%%%%%
%%%%%%%%%%%%%%%%%%%%%%%%%%%%%%%%%%%%%%%%%%%%%%%%%%%%%%%%%%%%%%%%%%%%%%%%
\section{Equations of state}
\label{equationsofstate}
%%%%%%%%%%%%%%%%%%%%%%%%%%%%%%%%%%%%%%%%%%%%%%%%%%%%%%%%%%%%%%%%%%%%%%%%

Consider a perfect fluid with barotropic equation of state $\rho=\rho(p)$.
Define
\begin{equation}\label{definitionofeta}
\eta = \int_0^p dp^\prime {\rho}^{-1}(p^\prime) \quad.
\end{equation}
 
\begin{assumption}\label{etaass}
We assume that $\rho$ is defined for $p\geq 0$ and
positive for $p > 0$ and also that $\eta$ exists when $p>0$.  
\end{assumption}

For static perfect fluids the 
quantity $\eta$ is directly related to the Newtonian potential $v$.
Integrating the fluid equation $dp + \rho dv = 0$ (Euler's equation) 
we obtain $\eta= v_S - v(p)$, where $v_S = v|_{p=0}$ is
the so-called surface potential. 

From the above assumption it follows that $\eta$ is a monotone function of
$p$ and $\eta|_{p=0} = 0$. Consequently, both $p$ and $\rho$ can be viewed as
functions of $\eta$ and the equation of state is implicitly given by $\rho(\eta)$ 
and $p(\eta)$, where $p = \int_0^\eta \rho(\eta) d\eta$. Moreover, 
monotonicity of $\rho(p)$ implies monotonicity
of $\rho(\eta)$ and conversely.

\begin{example} 
The polytropic equations of state are given by 
$p =\frac{1}{n+1}\,\rho_-^{-1/n} \rho^{(n+1)/n}$, where $\rho_-$ is a
positive constant and the so-called polytropic index $n$ is a
non-negative constant. In terms of $\eta$ this equation of state is
given implicitly by  $\rho = \rho_-\eta^n\,  ,\, p =
\frac{1}{n+1}\,\rho_- \eta^{n+1}$.
\end{example}

We now introduce the {\it index-function} $n(\eta)$. Let $\rho(\eta)$ be
continuous and sufficiently smooth.%
	\footnote{For example, ${\mathcal C}^2$
		is sufficient. We will not discuss to what extent this restriction can be softened,
		since we will describe below how to handle less restrictive situations like
		phase transitions.}
Now define 
\begin{equation}\label{ndef} 
n(\eta) = \frac{\eta}{\rho}\:\frac{d\rho}{d\eta} = \eta\,\frac{d\rho}{dp}\quad.
\end{equation}

\begin{assumption}\label{n0n1exists}
We assume that there exists some $a_0>0$ such that 
$n(\eta)-n_0 = O(\eta^{a_0})$ ($\eta\rightarrow 0$) and some $a_1>0$ such that
$n(\eta)-n_1 = O(\eta^{-a_1})$ ($\eta\rightarrow \infty$).%
	\footnote{In particular $a_0$ (and $a_1$) may be less than $1$.
		Hence, the class of asymptotically polytropic equations
		of state comprises the occasionally
		considered ``essentially polytropic'' or
		``quasi-polytropic''  equations of state \cite{Makino:1998}.}
We also assume that 
$n(\eta)$ is ${\mathcal C}^1$ on $(0,\infty)$ and bounded on $[0,\infty]$.
\end{assumption}

%The above assumption implies that we restrict our attention to 
%{\it asymptotically polytropic} equations of state, i.e., we consider 
%equations of state that for low pressures 
%(i.e., for $p \rightarrow 0\,;\, \eta\rightarrow 0$) 
%take the form 
%$\rho(\eta) = \rho_-\, \eta^{n_0} ( 1 + O(\eta^{a_0}) )\, ,\,
%p(\eta) = \frac{1}{n_0+1}\,\rho_- \eta^{n_0+1}( 1 + O(\eta^{a_0}) )$, 
%or equivalently
%$\rho(p) \propto p^{n_0/(n_0+1)}( 1 + O(p^{a_0/(n_0+1)}) )$,  
%and that for high pressures
%($p \rightarrow \infty\,;\, \eta\rightarrow \infty$) 
%take the form
%$\rho(\eta) = \rho_+\, \eta^{n_1} ( 1 + O(\eta^{-a_1}) )\, ,\,
%p(\eta) = \frac{1}{n_1+1}\,\rho_+ \eta^{n_1+1}( 1 + O(\eta^{-a_1}) )$, 
%or equivalently 
%$\rho(p) \propto p^{n_1/(n_1+1)} ( 1 + O(p^{-a_1/(n_1+1)}) )$.%

The above assumption implies that we restrict our attention to 
{\it asymptotically polytropic} equations of state, i.e., we consider 
equations of state that for low pressures 
(i.e., for $p \rightarrow 0\,;\, \eta\rightarrow 0$) 
take the form 
$\rho(\eta) = \rho_-\, \eta^{n_0} ( 1 + O(\eta^{a_0}) )\, ,\,
p(\eta) = \frac{1}{n_0+1}\,\rho_- \eta^{n_0+1}( 1 + O(\eta^{a_0}) )$, 
and that for high pressures
(i.e., for $p \rightarrow \infty\,;\, \eta\rightarrow \infty$) 
take the form
$\rho(\eta) = \rho_+\, \eta^{n_1} ( 1 + O(\eta^{-a_1}) )\, ,\,
p(\eta) = \frac{1}{n_1+1}\,\rho_+ \eta^{n_1+1}( 1 + O(\eta^{-a_1}) )$. This corresponds to
$\rho(p) \propto p^{n_0/(n_0+1)}( 1 + O(p^{a_0/(n_0+1)}) )$ ($p\rightarrow 0$) and  
$\rho(p) \propto p^{n_1/(n_1+1)} ( 1 + O(p^{-a_1/(n_1+1)}) )$ ($p\rightarrow \infty$).

It is possible to soften the above assumptions, but then it is
preferable to incorporate such generalizations into the formalism
itself. This leads to more cumbersome formulations than the one
introduced below, and, to show the value of our type of approach, we 
want to keep things relatively simple in this paper. Note that linear
equations of state $p= const\:\rho$ yield ``$n = \infty$'' because
$\eta$ does not exist. Equations of state that are asymptotically 
linear are therefore examples that need slightly different treatment
than those presently considered. 
In Appendix~\ref{B} we indicate how to treat equations of state 
with asymptotic behavior $\rho \propto p^{\nu}$ ($\nu\geq 0$;
where in particular $\nu$ might take the value $1$).

Given $n(\eta)$, an associated equation of state can be written as
$\rho(\eta) = \rho_- c(\eta) \eta^{n(\eta)}$, where the function
$c(\eta)=\eta^{n_0-n(\eta)} \exp(\int_0^\eta \eta^{-1} (n(\eta)-n_0) d\eta)$ 
($c(0)=1$) and $\rho_- =\mbox{constant}$. It follows that
$\frac{d\rho}{d\eta} = \rho_- c(\eta) n(\eta) \eta^{n(\eta)-1}$. (The
above description is adapted to the low pressure limit, however, if
one is so inclined, one can equally well adapt to the high pressure
limit, and introduce an associated constant $\rho_+$).
It follows that a given $n(\eta)$ gives rise to a 1-parameter set of
equations of state proportional to each other, parametrized by
$\rho_-$.

Note that $n(\eta)\equiv n = \mbox{const}$ for the polytropes and that
$n(\eta)$ can be interpreted as a local polytropic index. Note also that
monotonicity of $\rho(p)$ is equivalent to non-negativity of $n(\eta)$.  
Even though it is not necessary, we will for simplicity assume a non-negative
$n(\eta)$ throughout the paper, apart from Appendix~\ref{B}.

%%%%%%%%%%%%%%%%%%%%%%%%%%%%%%%%%%%%%%%%%%%%%%%%%%%%%%%%%%%%%%%%%%%%%%%%
%%%%%%%%%%%%%%%%%%%%%%%%%%%%%%%%%%%%%%%%%%%%%%%%%%%%%%%%%%%%%%%%%%%%%%%%
%%%%%%%%%%%%%%%%%%%%%%%%%%%%%%%%%%%%%%%%%%%%%%%%%%%%%%%%%%%%%%%%%%%%%%%%
%%%%%%%%%%%%%%%%%%%%%%%%%%%%%%%%%%%%%%%%%%%%%%%%%%%%%%%%%%%%%%%%%%%%%%%%
\section{Dynamical systems formulation}
\label{dynamicalsystemsformulation}
%%%%%%%%%%%%%%%%%%%%%%%%%%%%%%%%%%%%%%%%%%%%%%%%%%%%%%%%%%%%%%%%%%%%%%%%

In Newtonian gravity the spherically symmetric equilibrium of a 
self-gravitating perfect fluid is governed by
\begin{equation}\label{mpeq}
\frac{dm}{dr} = 4\pi r^2 \rho ,\quad \frac{dp}{dr} = -\frac{m\rho}{r^2} ,
\end{equation}
where $m(r)$ is the mass inside $r$; 
the gravitational constant has been set to $1$. 
Equation~(\ref{mpeq}) is to be understood in connection with 
an equation of state $\rho(p)$. We will focus on asymptotically polytropic 
equations of state.

Note incidentally that static stellar models are necessarily spherically symmetric, 
see \cite{Lindblom:1993} for an overview.
Existence and uniqueness of solutions of~(\ref{mpeq}) has been proved
for rather general equations of state comprising 
the equations of state considered in the present context, 
see e.g.,~\cite{Schaudt:2000}.

The above equations~(\ref{mpeq}) are singular when $r=0$, which causes some problems, 
but even worse is that $\rho(p)$ is not ${\mathcal C}^1$ at $p=0$ for 
asymptotically polytropic equations of state.
In this section, we will introduce new bounded dimensionless
variables that lead to a completely regular autonomous system of 
equations. We will do this in two steps since the intermediate step 
is useful in itself. 

To obtain an autonomous system we "elevate" $r$ to an dependent variable and 
introduce $\xi$, defined by $\xi=\ln r$, as a new independent variable.
We also make a variable transformation from $(m,p,r)$ to
the following three variables,
\begin{equation}\label{uqom}
u = \frac{4\pi r^3 \rho}{m}\ , \quad
q = \frac{m}{r \eta}\ , \quad
\omega = \eta^a \ ,
\end{equation}
where $a$ is some positive constant (the choice depending on the equation
of state, see discussion below). Note that $u= d\ln m /d\ln r$ and 
$q = - d\ln\eta/d \ln r$.
This yields the following system:
\begin{equation}\label{uqomeq}
\frac{du}{d\xi} = u\, (3 - u - n(\omega)\, q) \, ,\quad
\frac{dq}{d\xi} = q\,(-1 + u + q) \, ,\quad
\frac{d\omega}{d\xi} = -a\,\omega\, q \ .
\end{equation}
Here $n(\omega)$ is the index-function introduced in (\ref{ndef}), i.e.,
$n(\omega) = \frac{\eta}{\rho}\:\frac{d\rho}{d\eta}\:|_{\eta=\omega^{1/a}}\:$.

It is worth noting that for an (exact) polytrope 
$p =  \frac{1}{n+1}\,\rho_-^{-1/n}\,\rho^{1+1/n}$ ($n$ a
constant), the variable $q$ equals $q=\frac{1}{n+1} \,\frac{m\rho}{r p}$, and
hence $u$ and $q$ are so-called homology invariants (see, e.g., 
Kippenhahn and Weigert~\cite{book:KippWeig1994}, p.200). 
In this case the equation for $\omega$ decouples and
leaves a 2-dimensional coupled system for $u$ and $q$.

\begin{remark}
If $(m(r),p(r), \rho(r))$ is a solution of~(\ref{mpeq}), then so is 
$(r_0 m(r/r_0),r_0^{-2} p(r/r_0),r_0^{-2} \rho(r/r_0))$.
However, the latter is associated with an equation of state
that differs by a factor of $r_0^{-2}$ from that of
the first one, i.e., it is associated with $(r_0^{-2} \rho(\eta), r_0^{-2} p(\eta))$.
In the $(u,q,\omega;\xi)$ formulation, this feature corresponds to translatory
invariance of solutions, i.e., the second solution is given by
$(u(\xi-\xi_0),q(\xi-\xi_0),\omega(\xi-\xi_0))$, where
$r_0=e^{\xi_0}$, since $\xi=\ln r$. 
Accordingly the system~(\ref{uqomeq})
describes an equivalence class of proportional equations of state
$\{\rho(\eta) = \rho_-\, c(\eta)\, \eta^{n(\eta)}\:, \, p=\int_0^\eta
\rho(\eta)d\eta\:|\:\rho_- >0\}$, parametrized by $\rho_-$.
It is even possible to incorporate an additional degree of freedom
into the dynamical system~(\ref{uqomeq}) by choosing the
$\omega$-variable in a more general way, 
so that the system describes a two-parameter class of equations of state.

However, to simplify matters, we are 
going to take the viewpoint that we always consider a
given equation of state $\rho(\eta)$ (i.e., we fix $\rho_-$), and thus
that an orbit $\{(u(\xi),q(\xi),\omega(\xi))| \xi\in \mathbb{R}\}$ is
associated with a single  perfect fluid solution $(m(r), p(r),
\rho(r))$ of~(\ref{mpeq}).
\end{remark}

We now proceed by defining new bounded variables. By our assumptions it follows
that $\omega$ is positive and that $u, q$ are either both positive as well or
both negative depending on whether $m$ is positive or negative. We now consider
the positive mass case, while the negative mass case is treated
briefly in Appendix~\ref{B}. 
The new bounded variables $U,Q,\Omega$ ($U,Q,\Omega >0$) are defined by:

\begin{equation}\label{boundedvar}
U = \frac{u}{1+u}\ , \quad
Q = \frac{q}{1+q}\ , \quad
\Omega = \frac{\omega}{1+\omega} \ .
\end{equation}

Introducing a new independent variable $\lambda$ according to 
$d\lambda/d\xi = (1 - U)^{-1}(1 - Q)^{-1}$ yields the following
equations:

\begin{subequations}\label{UQOmega}
\begin{align}
\label{Ueq}
  \frac{dU}{d\lambda} &=  
  U(1-U)[(1-Q)(3-4U) - n(\Omega)\,Q(1-U)] \\
\label{Qeq}
  \frac{dQ}{d\lambda} &= 
  Q(1-Q)[(2U-1)(1-Q) + Q(1-U)] \\
\label{Omegaeq}
  \frac{d\Omega}{d\lambda} &= -a\Omega(1-\Omega)Q(1-U)\, . 
\end{align}
\end{subequations}

Here, $n(\Omega)$ is again the index-function,
$n(\Omega) = n(\eta(\Omega))= \frac{\eta}{\rho}\:\frac{d\rho}{d\eta}\:|_{\eta(\Omega)} =
[\Omega/(1-\Omega)]^{1/a} \:\frac{d\rho}{dp}$.
We now include the boundaries in our analysis so that
our state space consists of the unit cube:
$[0,1]^3$.

To be able to discuss this dynamical system (e.g., to do a fixed point
analysis) it is required that $n(\Omega)$ is 
${\mathcal C}^1$-differentiable on $[0,1]$.
This can be achieved by choosing the parameter $a$ sufficiently small.
By assumption~\ref{n0n1exists}, 
$n(\Omega)= n_0 + O(\Omega^{a_0/a})$ ($\Omega\rightarrow 0$) and
$n(\Omega)= n_1 + O((1 - \Omega)^{a_1/a})$ ($\Omega\rightarrow 1$).
Therefore, making the choice $a < \min(a_0,a_1)$, we obtain 
$dn/d\Omega|_{\Omega=0,1} =0$ (provided that the $O(\cdot)$-terms
are well-behaved). We are thus mimicking the features 
of an exact polytrope asymptotically as far as possible.

Let us now consider composite equations of state where $p=p(\rho)$ is
piecewise ${\mathcal C}^1$ on $(0,\infty)$.
In such situations
$p, m, r$ are continuous, but $dp/d\rho$ is not and neither is
$n(\Omega)$. It follows from the definitions that $\eta$, $q$ and hence
$\Omega$, $Q$ are continuous.  The mass density $\rho$ is continuous if
$p=p(\rho)$ is ${\mathcal C}^0$ on $(0,\infty)$, otherwise $\rho$
makes one or several jumps (a situation that corresponds to phase
transitions). The variables $u$ and $U$ are continuous if $\rho$ is,
but if $\rho$ jumps from some value  $\rho_1$ to $\rho_2$, then $u$
and $U$ make a jump according to  $u_1/u_2 = U_1(1-U_2)/(U_2(1-U_1))=
\rho_1/\rho_2$, which follows directly from the definitions of $u$ and
$U$. Solutions described by piecewise ${\mathcal C}^1$ equations of
state on $(0,\infty)$ can be obtained by viewing the above
transformation as a map between state spaces associated with smoothly
extended equations of state. Note that, in general, if one starts out
with a regular solution, then this solution has to be matched with a
non-regular one associated with the other (extended) equation of state.

In addition to the dynamical system it is of interest to consider the following
{\it auxiliary equations}:

\begin{subequations}\label{UQOmegaaux}
\begin{align}
\label{dr}
  \frac{dr}{d\lambda} &=  (1-U)(1-Q)r \\
\label{dm}
  \frac{dm}{d\lambda} &= U(1-Q)m \\
\label{r}
  r^2 &= \frac{1}{4\pi}\left(\frac{U}{1-U}\right)\left(\frac{Q}{1-Q}\right)
  \left(\frac{\Omega}{1-\Omega}\right)^{1/a}\rho^{-1}(\Omega) \\
\label{m}
  m^2 &= \frac{1}{4\pi}\left(\frac{U}{1-U}\right)\left(\frac{Q}{1-Q}\right)^3
\left(\frac{\Omega}{1-\Omega}\right)^{3/a}\rho^{-1}(\Omega) \\
\label{mr}
  \frac{m}{r} &= 
  \left(\frac{Q}{1-Q}\right)\left(\frac{\Omega}{1-\Omega}\right)^{1/a}
\end{align}
\end{subequations}

%%%%%%%%%%%%%%%%%%%%%%%%%%%%%%%%%%%%%%%%%%%%%%%%%%%%%%%%%%%%%%%%%%%%%%%%
%%%%%%%%%%%%%%%%%%%%%%%%%%%%%%%%%%%%%%%%%%%%%%%%%%%%%%%%%%%%%%%%%%%%%%%%
%%%%%%%%%%%%%%%%%%%%%%%%%%%%%%%%%%%%%%%%%%%%%%%%%%%%%%%%%%%%%%%%%%%%%%%%
%%%%%%%%%%%%%%%%%%%%%%%%%%%%%%%%%%%%%%%%%%%%%%%%%%%%%%%%%%%%%%%%%%%%%%%%
\section{Dynamical systems analysis}
\label{dynamicalsystemsanalysis}
%%%%%%%%%%%%%%%%%%%%%%%%%%%%%%%%%%%%%%%%%%%%%%%%%%%%%%%%%%%%%%%%%%%%%%%%

An important feature of the dynamical system~(\ref{UQOmega})
is that all orbits are strictly monotonically decreasing in $\Omega$ 
except if $\Omega = 0,1$, $Q=0$ or $U=1$. This fact will be of key importance
when we prove the following theorem in Appendix A:

\begin{theorem}\label{fix} 
All solutions converge to fixed points when $\lambda \rightarrow \pm \infty$ 
except when $n_0$ $(n_1)$ is equal to 5. 
In this latter case solutions also converge 
to a 1-parameter set of closed orbits 
$C_1$ $(C_2)$ when $\lambda \rightarrow \infty (-\infty)$.
\end{theorem}

\begin{figure}[htp]
	\centering
	\includegraphics[width=0.7\textwidth]{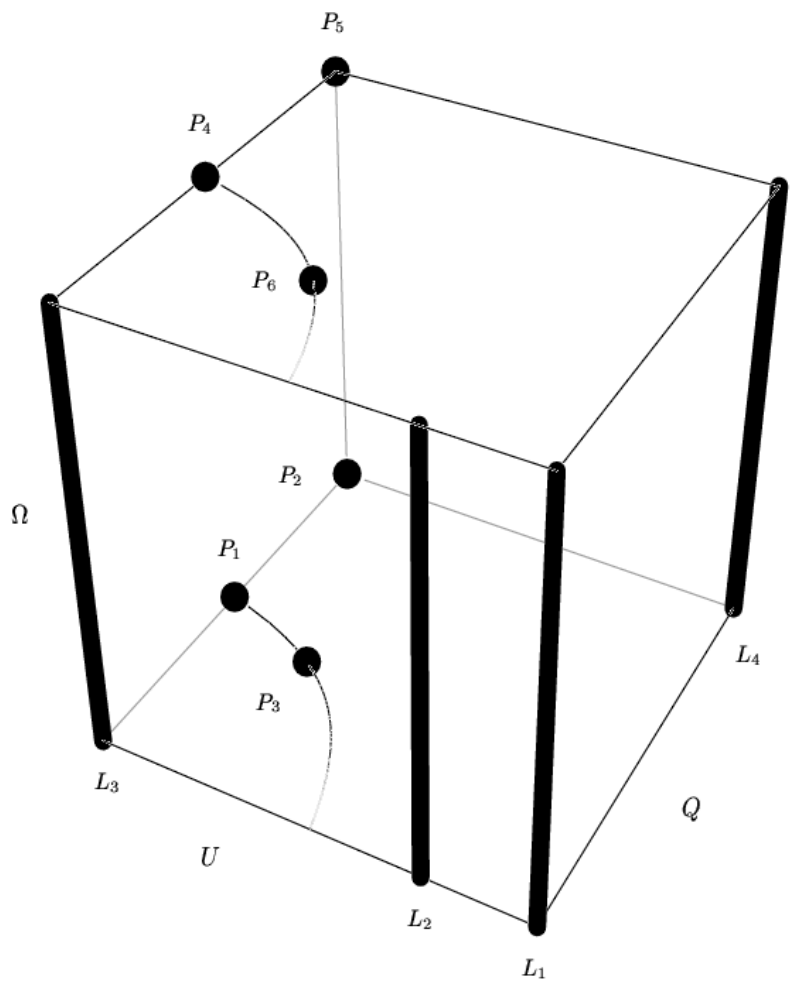}
	\caption{The compact state space, i.e., the unit cube $[0,1]^3$, of the    
		dynamical system~(\ref{UQOmega}). For any index-function $n(\eta)$ with
		$n_0>0$ and $n_1>0$ there exist at least $4$ fixed points
		($P_1, P_2, P_4, P_5$) and $4$ ``fixed lines'', i.e., 
		lines consisting of fixed points ($L_1,L_2, L_3, L_4$).
		In addition, for $n_0>3$ (and/or $n_1 >3$), we have the fixed point $P_3$ 
		(and/or $P_6$). The position of $P_3$ ($P_6$) depends on $n_0$ ($n_1$): 
		$P_3$ and $P_6$ move along the indicated curves
		as $n_i$ increases. In the picture we have chosen $n_0 =4$ and
		$n_1 =6$.}
    	\label{cube}
\end{figure}

Thus the fixed points and the periodic orbits $C_1, C_2$ describe 
the asymptotic features of the solutions. It is therefore natural to
begin with a qualitative analysis of the dynamical system by investigating
the fixed points. The state space, i.e., the unit cube $[0,1]^3$,
and the fixed points of the dynamical system~(\ref{UQOmega}) are depicted 
in Figure~\ref{cube}. 
Table~\ref{tab:UQcube} lists all fixed points together with
their local properties, i.e., the eigenvalues of
the Jacobian of~(\ref{UQOmega}) at the fixed points.

%\begin{figure}[htp]
%	\psfrag{L1}[cc][cc]{{\small $L_1$}}
%	\psfrag{L2}[cc][cc]{{\small $L_2$}}
%	\psfrag{L3}[cc][cc]{{\small $L_3$}}
%	\psfrag{L4}[cc][cc]{{\small $L_4$}}
%	\psfrag{L5}[cc][cc]{{\small $L_5$}}
%	\psfrag{P1}[cc][cc]{{\small $P_1$}}
%	\psfrag{P2}[cc][cc]{{\small $P_2$}}
%	\psfrag{P3}[cc][cc]{{\small $P_3$}}
%	\psfrag{P4}[cc][cc]{{\small $P_4$}}
%	\psfrag{P5}[cc][cc]{{\small $P_5$}}
%	\psfrag{P6}[cc][cc]{{\small $P_6$}}
%	\psfrag{O}[cc][cc]{{$\Omega$}}
%	\psfrag{Q}[cc][cc]{{$Q$}}
%	\psfrag{U}[cc][cc]{{$U$}}
%	\centering
%	\includegraphics[width=0.7\textwidth]{theCUBE.eps}
%	\caption{The compact state space, i.e., the unit cube $[0,1]^3$, of the    
%		dynamical system~(\ref{UQOmega}). For any index-function $n(\eta)$ with
%		$n_0>0$ and $n_1>0$ there exist at least $4$ fixed points
%		($P_1, P_2, P_4, P_5$) and $4$ ``fixed lines'', i.e., 
%		lines consisting of fixed points ($L_1,L_2, L_3, L_4$).
%		In addition, for $n_0>3$ (and/or $n_1 >3$), we have the fixed point $P_3$ 
%		(and/or $P_6$). The position of $P_3$ ($P_6$) depends on $n_0$ ($n_1$): 
%		$P_3$ and $P_6$ move along the indicated curves
%		as $n_i$ increases. In the picture we have chosen $n_0 =4$ and
%		$n_1 =6$.}
%    	\label{cube}
%\end{figure}

\begin{table}[ht]
  \begin{center}
    \begin{tabular}{|c|cccc|c|c|}
      \hline
      Fixed point & $U$ & $Q$ & $\Omega$ &   & Eigenvalues & Restrictions \\  \hline 
  	& & & & & & \\[-0.3cm]
      $L_1$ & 1 & 0 & $\Omega_0$ &   & $1\ , \ 1\ , \ 0$ & \\
      $L_2$ & $\sfrac{3}{4}$ & 0 & $\Omega_c$ &   &
      $-\sfrac{3}{4}\ , \ \sfrac{1}{2}\ , \ 0$ & \\
      $L_3$ & 0 & 0 & $\Omega_0$ &   & $3\ , \ -1\ , \ 0$ & \\
      $L_4$ & 1 & 1 & $\Omega_0$ &   & $0 \ , \ 0 \ , \ 0$ & \\
      $L_5$ & $U_0$ & 1 & 0 &  & $0 \ , \ -(1-U_0)\ , \ -a (1-U_0)$ & $n_0 =0$ \\
      $L_6$ & $U_0$ & 1 & 1 &   & $0 \ , \ -(1-U_0)\ , \ a (1- U_0)$ & $n_1 =0$  \\
      $P_1$ & 0 & $\sfrac{1}{2}$ & 0 &   &$-\sfrac{n_0 -3}{2}\ , \
      \sfrac{1}{2} \ , \ -\sfrac{a}{2}$ &  \\
      $P_2$ & 0 & 1 & 0 &  & $-n_0\ , \ -1\ , \ -a$ & \\
      $P_3$ & $\sfrac{n_0 -3}{2(n_0 -2)}$ & $\sfrac{2}{1+n_0}$ & 0 &  &
      $\sfrac{\beta}{4}\left(5-n_0 \pm
        i \sqrt{b}\right)\ , \ -a \beta $ 
      & $n_0>3$ \\     
      $P_4$ & 0 & $\sfrac{1}{2}$ & 1 &  & $-\sfrac{n_1 -3}{2}\ , \
      \sfrac{1}{2} \ , \ \sfrac{a}{2}$ &  \\   
      $P_5$ & 0 & 1 & 1 &  & $-n_1\ , \ -1\ , \ a$ & \\  
      $P_6$ & $\sfrac{n_1 -3}{2(n_1 -2)}$ & $\sfrac{2}{1+n_1}$ & 1 &  &
            $\sfrac{\beta}{4}\left(5-n_1 \pm
        i \sqrt{b}\right)\ , \ a \beta $
        & $n_1>3$ \\   
        \hline
    \end{tabular}
  \end{center}
    \caption{Local properties of the fixed points.
	In the table $n(0) = n_0$, $n(1) = n_1$,
	$b= -1-22n_i+7 n_i^2$,
	$\beta = \sfrac{n_i-1}{(n_i-2)(n_i+1)}$, where 
	$n_i = n_0$ for $P_3$ and $n_i = n_1$ for $P_6$; $b$ is negative
	for $n_i<(11+8\sqrt{2})/7$ and positive for $n_i>(11+8\sqrt{2})/7$.}
    \label{tab:UQcube}
\end{table}

Tables~\ref{emittors} and \ref{attractors} list from which 
fixed points (or periodic orbits) the orbits originate and 
in which fixed points (or periodic orbits) they
end. If the stable subspace of a hyperbolic fixed point $P$ is $m$-dimensional, 
this means that a $(m-1)$-parameter family of orbits ends in $P$.
Note that $L$ is a one-parameter set of fixed points, so that,
if the dimension of the stable subspaces of the fixed points on $L$ is $m$, 
then $L$ attracts a $m$-dimensional family of orbits.
Note that we restricted our attention to limit sets for orbits 
in the interior of the state space in Tables~\ref{emittors} 
and~\ref{attractors}.

\begin{table}[ht]
\begin{center}
    \begin{tabular}{|l|ccc|}
      \hline
	Range of $n_1$	& Fixed point & Unstable subspace & Dimension \\ \hline
		& & & \\[-0.2cm]
	\fbox{$n_1< 3$}\hspace*{1cm}
		& $P_4$ 	& $<-4 (n_1-2) e_1 + e_2, \,e_2, \,e_3>$ & 3 \\
		& $L_1$ 	& $<e_1, e_2>$ & 2 \\ 
		& $L_2$ 	& $<-\frac{3}{40}\, n(\Omega_c) e_1 +
				2 e_2 +
 				a (\Omega_c^2-\Omega_c) e_3>$ & 1 \\[0.2cm] \hline
		& & & \\[-0.1cm]
       \fbox{$n_1 = 3$}\hspace*{1cm}
		& $P_4$ 	& \multicolumn{2}{c|}{--- 1-dimensional unstable, 
					1-dimensional center subspace ---} \\
		& $L_1$ 	& $<e_1, e_2>$ & 2 \\ 
		& $L_2$ 	& $<-\frac{3}{40}\, n(\Omega_c) e_1 +
				2 e_2 +
 				a (\Omega_c^2-\Omega_c) e_3>$ & 1 \\[0.2cm] \hline
		& & & \\[-0.1cm]
	\fbox{$3<n_1< 5$} \hspace*{1cm}
		& $P_6$ 	& $<e_1, e_2, e_3>$ & 3 \\
		& $L_1$ 	& $<e_1, e_2>$ & 2 \\ 
		& $L_2$ 	& $<-\frac{3}{40}\, n(\Omega_c) e_1 +
				2 e_2 +
 				a (\Omega_c^2-\Omega_c) e_3>$ & 1 \\[0.2cm] \hline
		& & & \\[-0.1cm]
	\fbox{$n_1= 5$} \hspace*{1cm}
		& $P_6$ 	& $<e_1, e_2, e_3>$ & 3 \\
		& $L_1$ 	& $<e_1, e_2>$ & 2 \\ 
		& $L_2$ 	& $<-\frac{3}{40}\, n(\Omega_c) e_1 +
				2 e_2 +
 				a (\Omega_c^2-\Omega_c) e_3>$ & 1 \\[0.1cm]

		& $C_2$		& \multicolumn{2}{c|}{--- --- 1-parameter family of %
				periodic $\alpha$-limit orbits --- ---}\\[0.2cm] \hline
		& & & \\[-0.1cm]
	\fbox{$5<n_1$} \hspace*{1cm} 
		& $P_6$ 	& $<e_3>$ & 1 \\
		& $L_1$ 	& $<e_1, e_2>$ & 2 \\ 
		& $L_2$ 	& $<-\frac{3}{40}\, n(\Omega_c) e_1 +
				2 e_2 +
 				a (\Omega_c^2-\Omega_c) e_3>$ & 1 \\[0.2cm] \hline
    \end{tabular}
   \caption{\label{emittors}Sources for interior solutions.}
\end{center}
\end{table}
 
\begin{table}[ht]
\begin{center}
    \begin{tabular}{|l|ccc|}
	\hline
	Range of $n_0$ 	& Fixed point & Stable subspace & Dimension \\ \hline
		& & & \\[-0.2cm]
	\fbox{$n_0=0$}\hspace*{1cm}
		& $L_5$ 	& $<U_0 (3-4 U_0)e_1 +e_2,\,e_3>$ & 2 \\[0.2cm] \hline
		& & & \\[-0.1cm]
	\fbox{$0<n_0< 3$} \hspace*{1cm}
		& $P_2$         & $<e_1, e_2, e_3>$ & 3 \\[0.2cm] \hline
		& & & \\[-0.1cm]
        \fbox{$n_0=3$} \hspace*{1cm}
		& $P_2$         & $<e_1, e_2, e_3>$ & 3 \\[0.2cm] \hline
		& & & \\[-0.1cm] 
	\fbox{$3<n_0<5$} \hspace*{1cm} 
		& $P_1$ 	& $<-4 (n_0-2) e_1 + e_2, \,e_3>$ & 2 \\
		& $P_2$ 	& $<e_1, e_2, e_3>$ & 3 \\ 
		& $P_3$ 	& $<e_3>$	& 1 \\[0.2cm] \hline
		& & & \\[-0.1cm]
	\fbox{$n_0=5$} \hspace*{1cm} 
		& $P_1$ 	& $<-4 (n_0-2) e_1 + e_2, \,e_3>$ & 2 \\
		& $P_2$ 	& $<e_1, e_2, e_3>$ & 3 \\ 
		& $P_3$ 	& $<e_3>$	& 1 \\
		& $C_1$		& \multicolumn{2}{c|}{--- 1-parameter set of %
					periodic $\omega$-limit orbits ---}\\[0.2cm] \hline
		& & & \\[-0.1cm]
	\fbox{$5<n_0$} \hspace*{1cm} 
		& $P_1$ 	& $<-4 (n_0-2) e_1 + e_2, \,e_3>$ & 2 \\
		& $P_2$ 	& $<e_1, e_2, e_3>$ & 3 \\ 
		& $P_3$ 	& $<e_1, e_2, e_3>$ & 3 \\[0.2cm] \hline
    \end{tabular}
   \caption{\label{attractors}Attractors for interior solutions.}
\end{center}
\end{table}

Below we discuss asymptotic physical features associated with the
various fixed points and $C_1, C_2$. However, we already now mention
that the solutions with a regular center originate from $L_2$. Thus,
as is well known, there exists a 1-parameter set of regular solutions
parametrized by $\Omega_c$ (or equivalently by the central density 
$\rho_c$ for a given equation of state). 
We refer to this set of orbits as the regular set.

\begin{figure}[htp]
        \centering
	\includegraphics[height=0.28\textwidth]{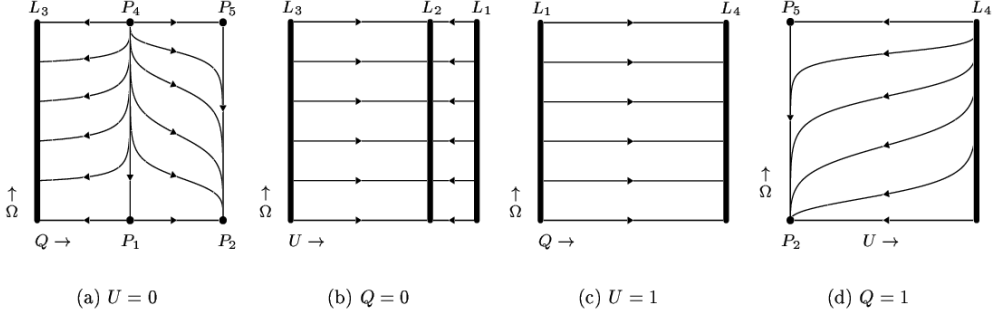}
	\caption{Orbits on the side faces. (a) is the $U=0$ plane, (b) the
		$Q=0$ plane, (c) the $U=1$ plane and (d) the $Q=1$ plane.
		For the latter we had to chose some index-function, the
		other three are independent of $n$.}
	\label{sidefaces}
\end{figure}

We then observe that all six faces of the cube are invariant subspaces.
Equations~(\ref{UQOmega}) can be solved explicitly on the four side faces
and we get solutions as shown in Figure~\ref{sidefaces}.

%\begin{figure}[htp]
%	\psfrag{L1}{{\scriptsize $L_1$}}
%	\psfrag{L2}{{\scriptsize $L_2$}}
%	\psfrag{L3}{{\scriptsize $L_3$}}
%	\psfrag{L4}{{\scriptsize $L_4$}}
%	\psfrag{L5}{{\scriptsize $L_5$}}
%	\psfrag{P1}{{\scriptsize $P_1$}}
%	\psfrag{P2}{{\scriptsize $P_2$}}
%	\psfrag{P3}{{\scriptsize $P_3$}}
%	\psfrag{P4}{{\scriptsize $P_4$}}
%	\psfrag{P5}{{\scriptsize $P_5$}}
%	\psfrag{P6}{{\scriptsize $P_6$}}
%	\psfrag{O}[cb][Bl]{{\scriptsize $\begin{array}{c}\uparrow\\ \Omega\end{array}$}}
%	\psfrag{Q}{{\scriptsize $Q\rightarrow$}}
%	\psfrag{U}{{\scriptsize $U\rightarrow$}}
%	\centering
%        \subfigure[$U=0$]{
%		\label{sideU0}
%		\includegraphics[height=0.2\textwidth]{Uis0.eps}}\quad
%     	\subfigure[$Q=0$]{
%		\label{sideQ0}
%		\includegraphics[height=0.2\textwidth]{Qis0.eps}}\quad
%	\subfigure[$U=1$]{
%		\label{sideU1}
%		\includegraphics[height=0.2\textwidth]{Uis1.eps}}\quad
%        \subfigure[$Q=1$]{
%		\label{sideQ1}
%		\includegraphics[height=0.2\textwidth]{Qis1.eps}}
%	\caption{Orbits on the side faces. (a) is the $U=0$ plane, (b) the
%		$Q=0$ plane, (c) the $U=1$ plane and (d) the $Q=1$ plane.
%		For the latter we had to chose some index-function, the
%		other three are independent of $n$.}
%    \label{sidefaces}
%\end{figure}

%\begin{figure}[htp]
%	\psfrag{L1}{{\small $L_1$}}
%	\psfrag{L2}{{\small $L_2$}}
%	\psfrag{L3}{{\small $L_3$}}
%	\psfrag{L4}{{\small $L_4$}}
%	\psfrag{L5}{{\small $L_5$}}
%	\psfrag{P1}{{\small $P_1$}}
%	\psfrag{P2}{{\small $P_2$}}
%	\centering
%        \subfigure[$n_0=0$]{
%		\label{base0}
%		\includegraphics[width=0.3\textwidth]{n0f.eps}}\quad
%     	\subfigure[$n_0=1$]{
%		\label{base1}
%		\includegraphics[width=0.3\textwidth]{n1f.eps}}\quad
%	\subfigure[$n_0=2$]{
%		\label{base2}
%		\includegraphics[width=0.3\textwidth]{n2f.eps}}\quad
%        \subfigure[$n_0=3$]{
%		\label{base3}
%		\includegraphics[width=0.3\textwidth]{n3f.eps}}\quad
%        \subfigure[$n_0=3.15$]{
%		\label{base315}
%		\includegraphics[width=0.3\textwidth]{n315f.eps}}\quad
%        \subfigure[$n_0=4$]{
%		\label{base4}
%		\includegraphics[width=0.3\textwidth]{n4f.eps}}\quad
%        \subfigure[$n_0=5$]{
%		\label{base5}
%		\includegraphics[width=0.3\textwidth]{n5f.eps}}\quad
%        \subfigure[$n_0=6$]{
%		\label{base6}
%		\includegraphics[width=0.3\textwidth]{n6f.eps}}\quad
%        \subfigure[$n_0=20$]{
%		\label{base20}
%		\includegraphics[width=0.3\textwidth]{n20f.eps}}\quad
%    \caption{Orbits in the polytropic subset $\Omega=0$ in terms of the 
%		variables $U$ and $Q$ for 
%		$n_0=0$ (a), 
%		$0<n_0\leq 3$ [(b), (c), (d)], 
%		$3<n_0\leq (11+8\sqrt{2})/7$ [(e)],
%		$(11+8\sqrt{2})/7 <n_0 < 5$ [(f)],
%		$n_0=5$ (g),
%		$n_0>5$ [(h), (i)].}
%    \label{base}
%\end{figure}

\begin{figure}[htp]
	\centering
        \subfigure[$n_0=0$]{
		\label{base0}
		\includegraphics[width=0.3\textwidth]{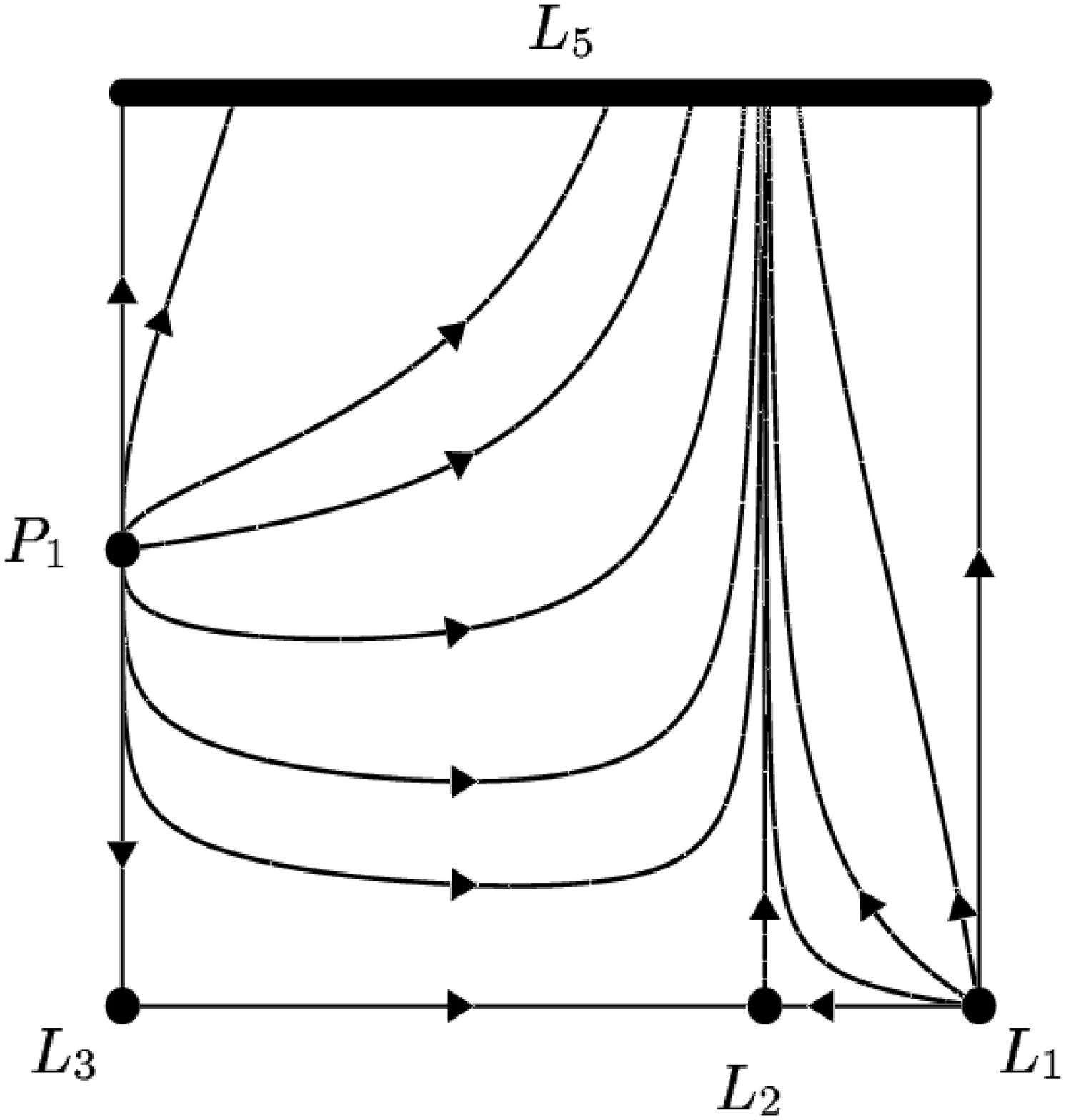}}\quad
     	\subfigure[$n_0=1$]{
		\label{base1}
		\includegraphics[width=0.3\textwidth]{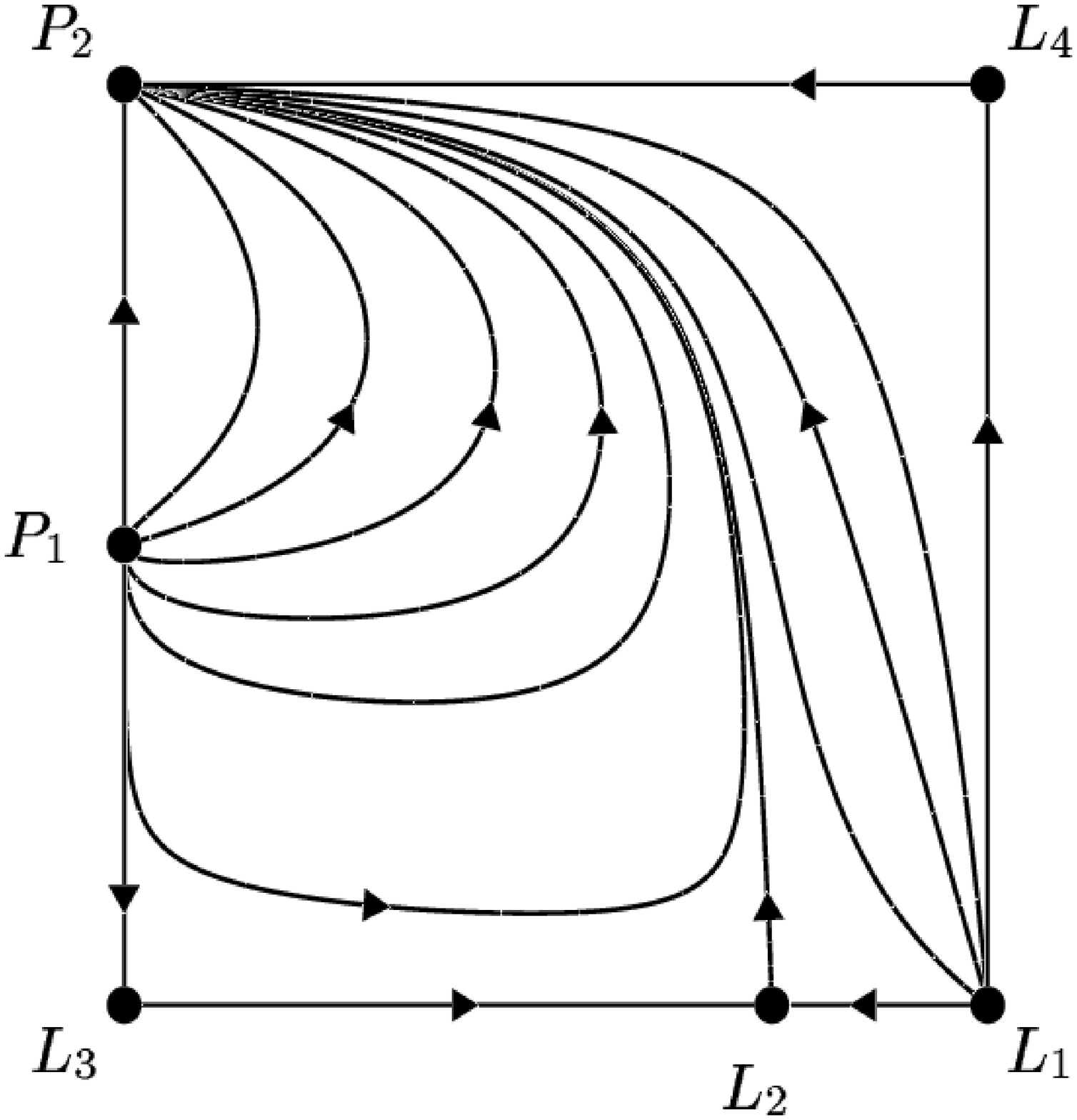}}\quad
	\subfigure[$n_0=2$]{
		\label{base2}
		\includegraphics[width=0.3\textwidth]{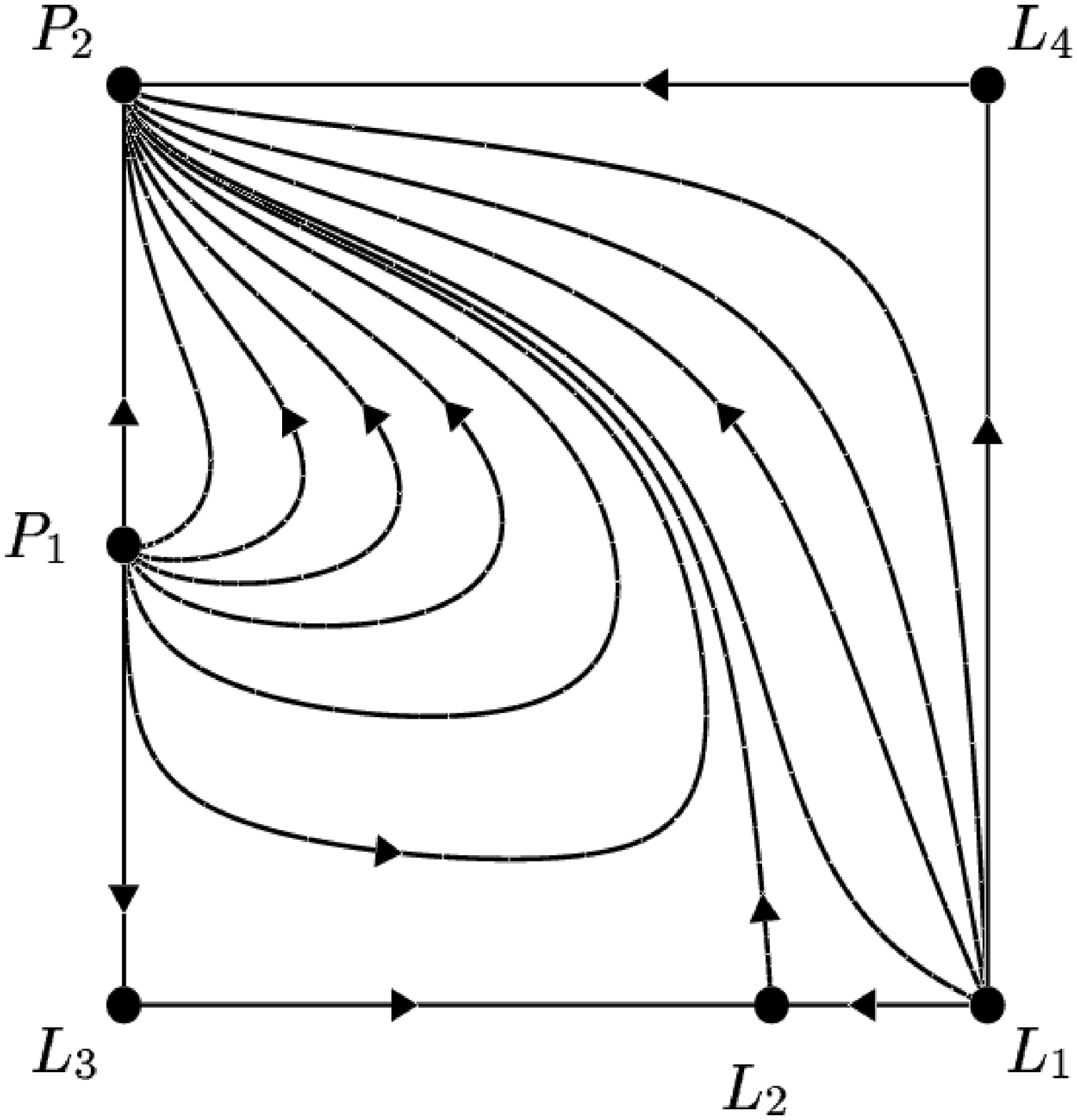}}\quad
        \subfigure[$n_0=3$]{
		\label{base3}
		\includegraphics[width=0.3\textwidth]{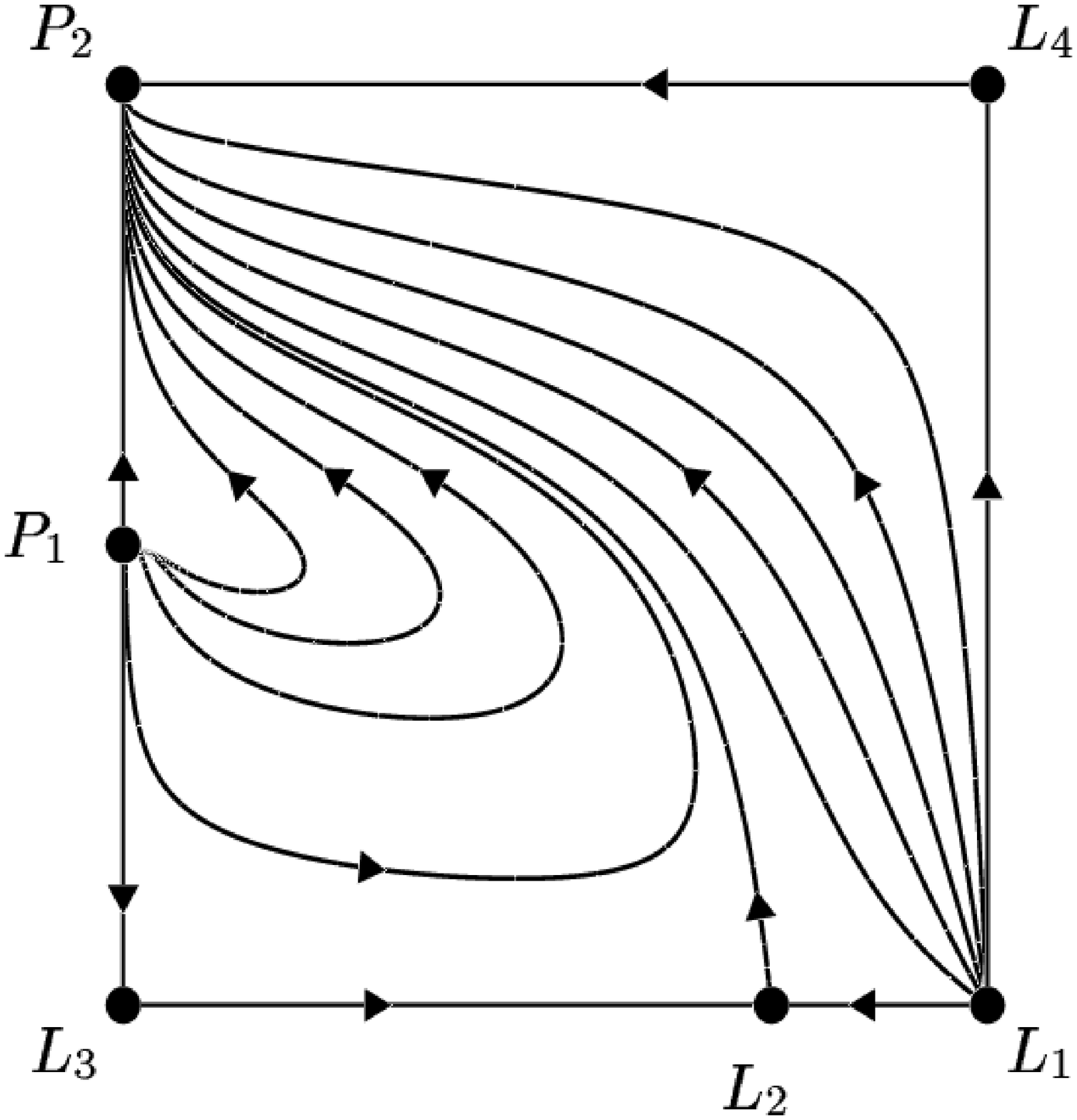}}\quad
        \subfigure[$n_0=3.15$]{
		\label{base315}
		\includegraphics[width=0.3\textwidth]{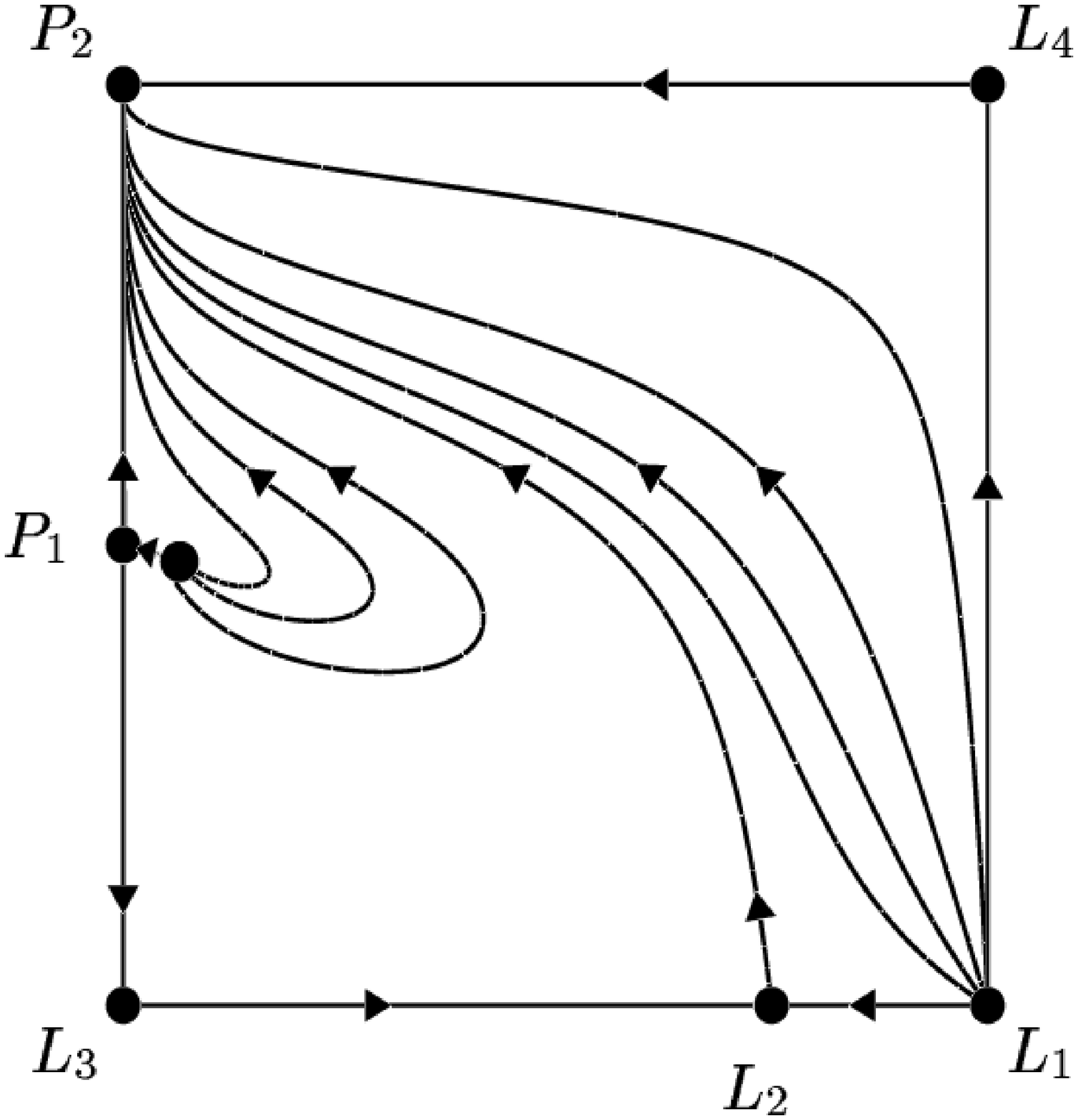}}\quad
        \subfigure[$n_0=4$]{
		\label{base4}
		\includegraphics[width=0.3\textwidth]{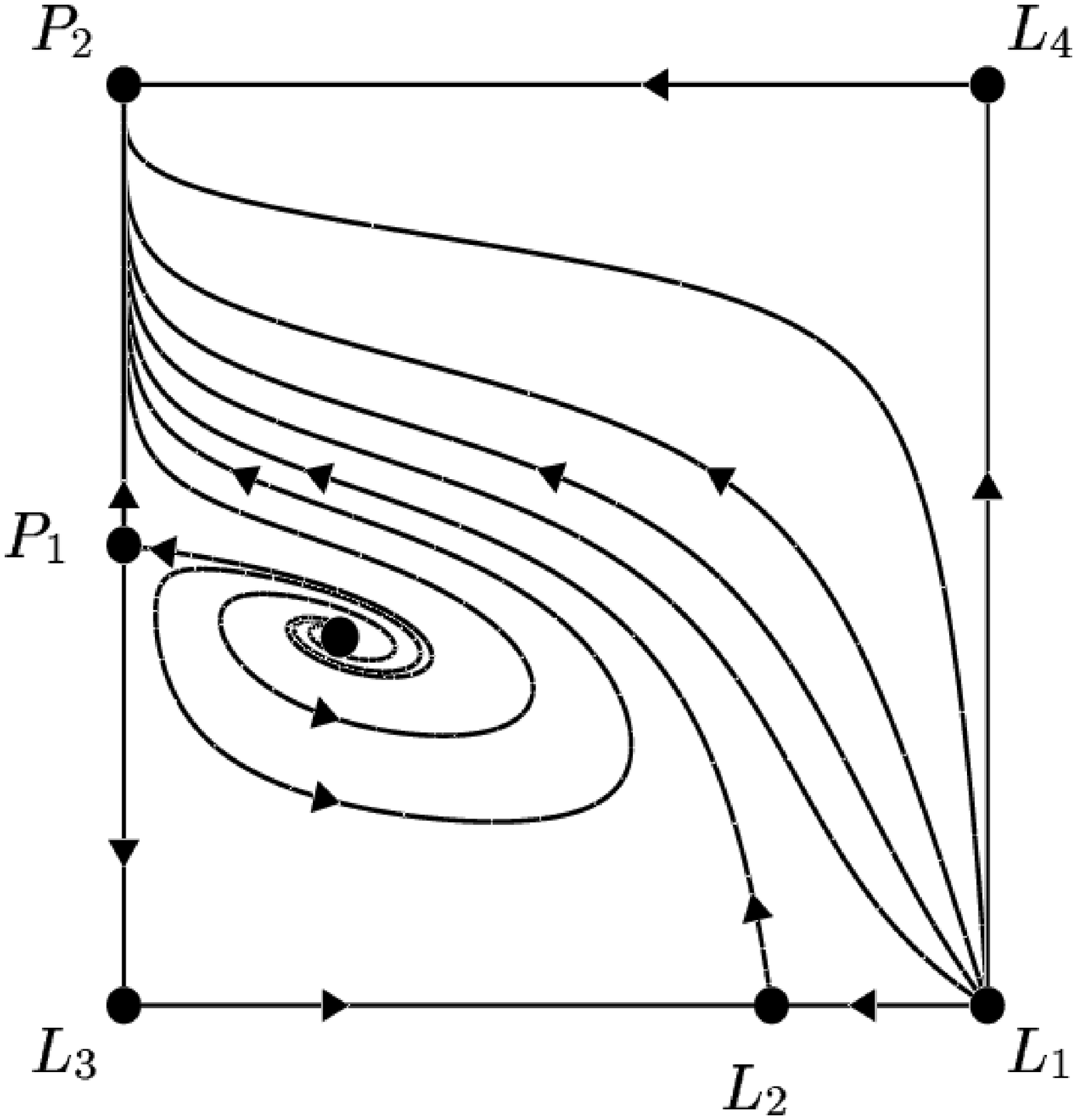}}\quad
        \subfigure[$n_0=5$]{
		\label{base5}
		\includegraphics[width=0.3\textwidth]{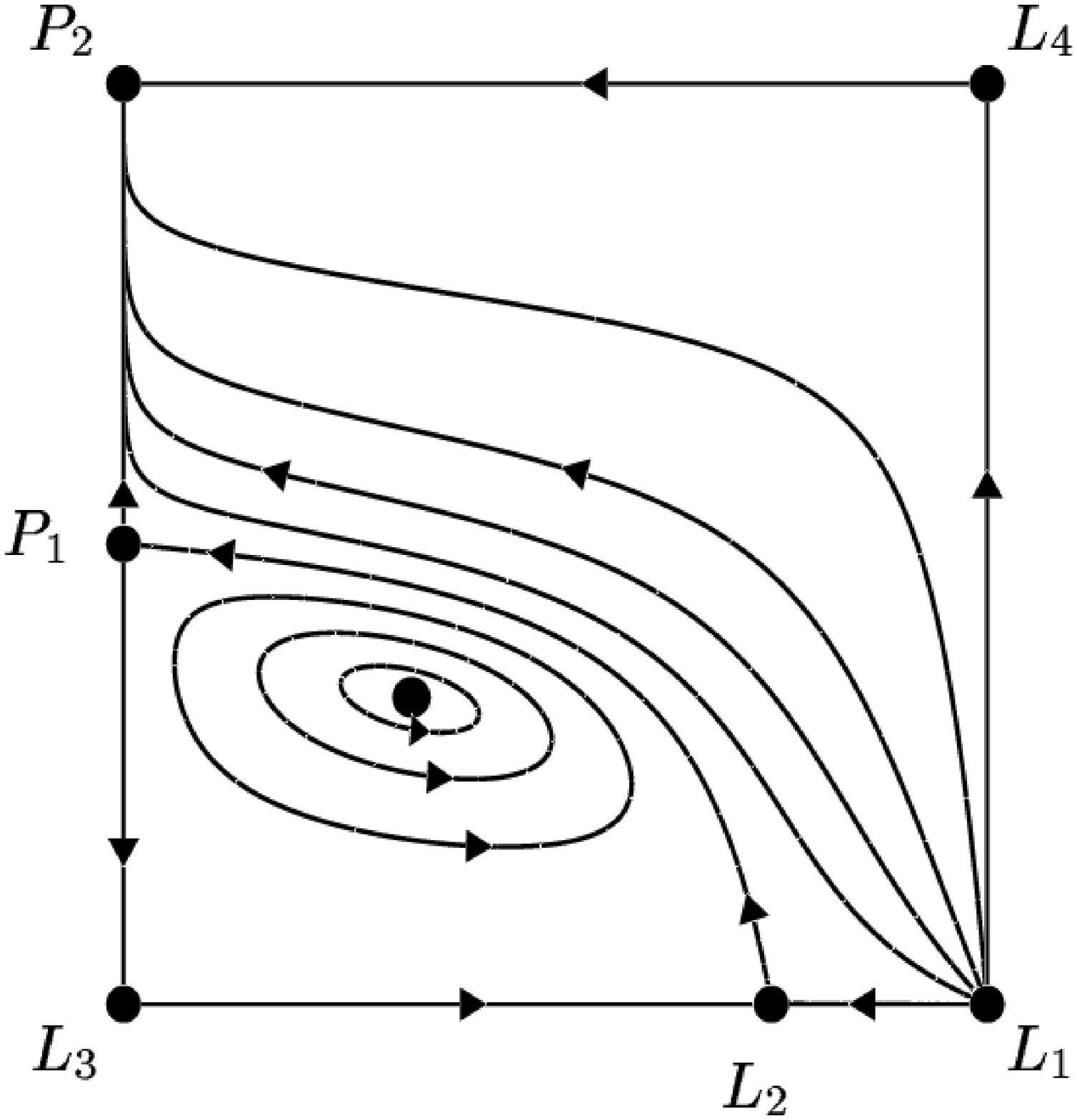}}\quad
        \subfigure[$n_0=6$]{
		\label{base6}
		\includegraphics[width=0.3\textwidth]{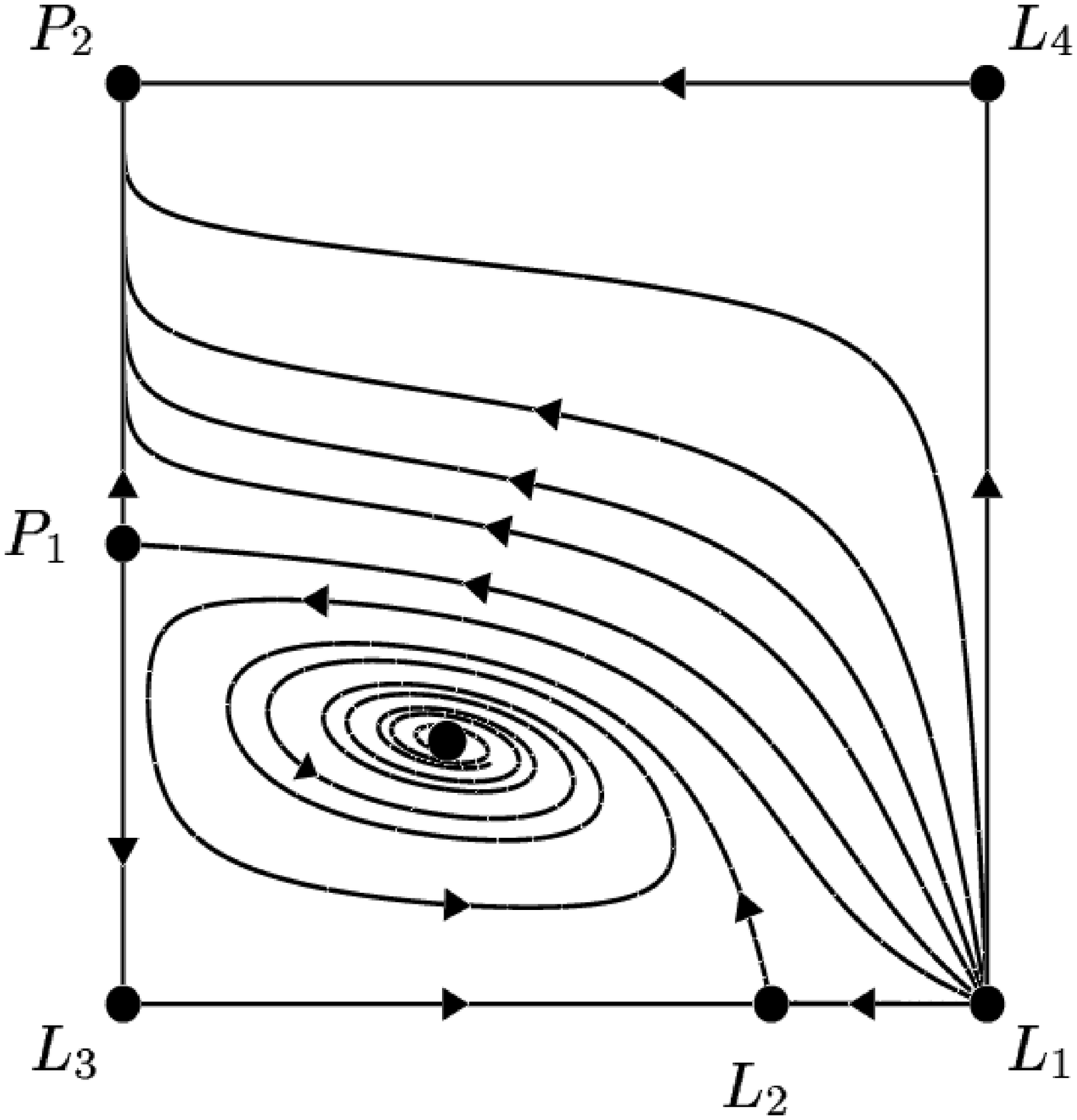}}\quad
        \subfigure[$n_0=20$]{
		\label{base20}
		\includegraphics[width=0.3\textwidth]{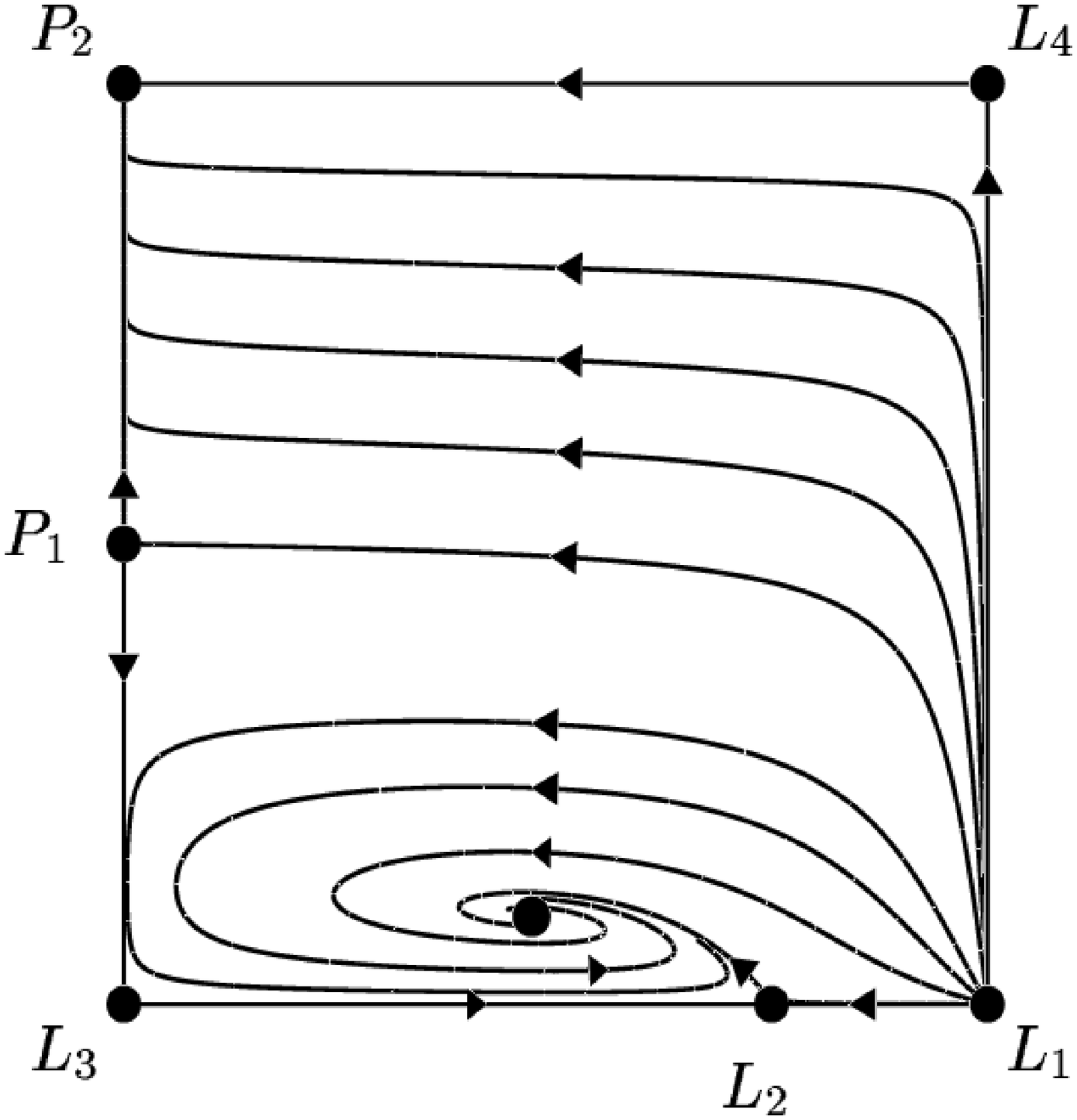}}\quad
    \caption{Orbits in the polytropic subset $\Omega=0$ in terms of the 
		variables $U$ and $Q$ for 
		$n_0=0$ (a), 
		$0<n_0\leq 3$ [(b), (c), (d)], 
		$3<n_0\leq (11+8\sqrt{2})/7$ [(e)],
		$(11+8\sqrt{2})/7 <n_0 < 5$ [(f)],
		$n_0=5$ (g),
		$n_0>5$ [(h), (i)].}
    \label{base}
\end{figure}

%\begin{figure}[htp]
%                \centering
%		\includegraphics[width=0.96\textwidth]{compress3.ps}
%    \caption{Orbits in the polytropic subset $\Omega=0$ in terms of the 
%		variables $U$ and $Q$ for 
%		$n_0=0$ (a), 
%		$0<n_0\leq 3$ [(b), (c), (d)], 
%		$3<n_0\leq (11+8\sqrt{2})/7$ [(e)],
%		$(11+8\sqrt{2})/7 <n_0 < 5$ [(f)],
%		$n_0=5$ (g),
%		$n_0>5$ [(h), (i)].}
%    \label{base}
%\end{figure}

The $\Omega=0$ plane (and the $\Omega=1$ plane, since it is described
by the same equations) display more complicated features. Note that
the equations for $U$ and $Q$ decouple in equation~(\ref{UQOmega})
when one has a polytropic equation of state. This leads to a
2-dimensional system which may be identified with the $\Omega=0$
subset (or equivalently the $\Omega=1$ subset). Hence we  denote the
$\Omega=0$ plane as the ``polytropic subset''. The orbit structure for
the polytropic subset is given in Figure~\ref{base}.

The polytropic subset is of key importance to subsequent discussions
(and highly interesting in itself). It is clear that there exist
bifurcations when $n_0 =0,3,5$ (and similarly for $n_1$), which is related to
the fact that some fixed points become non-hyperbolic. It is worth noticing
that the case $n_0=0$ (corresponding to an incompressible fluid) is
exactly solvable, as is $n_0=5$, which is
particularly useful. Apart from the fixed points, we observe a family of
closed orbits $C_1$ centered around $P_3$. The orbits for the $n_0=5$
case can be characterized as the sets $\{U,Q\}$ satisfying
\begin{equation}\label{n05mon}
  \Phi = U(4-7Q) - 3(1 - 2Q) - C U^{-1/2}Q^{-3/2}(1-U)^{3/2}(1-Q)^{5/2} = 0,
\end{equation}
where the parameter $C$ is required to be $C \geq -1/4$. This can be
shown by observing that $\frac{d \Phi}{d \lambda} = 0$. The closed
orbits are characterized by $C$ ranging between  $-1/4 < C < 0$; the
fixed point $P_3$ is represented by $C=-1/4$ while $C=0$
characterizes the solution with regular center, i.e., the orbit
connecting $L_2$ and $P_1$. The remaining orbits correspond to
positive values of $C$ (compare with Figure~\ref{base}). It is shown
in Appendix A that $C_1$ $(C_2)$ act as $\omega$-limit
($\alpha$-limit) sets for orbits coming from the interior of the state
space.

Center manifold analysis shows that no interior orbit ends at $P_1$ 
when $n_0=3$ and that $P_4$ is a source for a 2-parameter family of
orbits when $n_1=3$. For details, see Appendix A.

\begin{figure}[htp]
	\psfrag{L1}{{$L_1$}}
	\psfrag{L2}{{$L_2$}}
	\psfrag{L3}{{$L_3$}}
	\psfrag{L4}{{$L_4$}}
	\psfrag{L5}{{$L_5$}}
	\psfrag{P1}{{$P_1$}}
	\psfrag{P2}{{$P_2$}}
	\psfrag{P3}{{$P_3$}}
	\psfrag{P4}{{$P_4$}}
	\psfrag{P5}{{$P_5$}}
	\psfrag{P6}{{$P_6$}}
	\psfrag{Q}[cb][Bl]{{$\begin{array}{c}\uparrow\\ Q\end{array}$}}
	\psfrag{U}{{$U\rightarrow$}}
	\centering
        \subfigure[$\frac{d\ln m}{d\lambda}$]{
		\label{acquiremass}
		\includegraphics[height=0.4\textwidth]{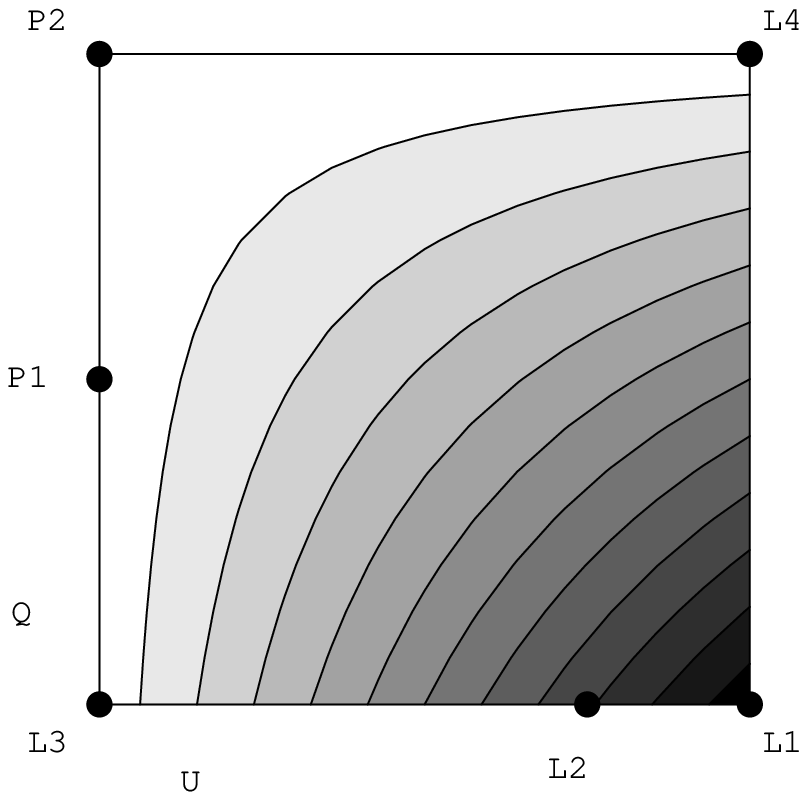}}\qquad
     	\subfigure[$\frac{d\ln r}{d\lambda}$]{
		\label{acquireradius}
		\includegraphics[height=0.4\textwidth]{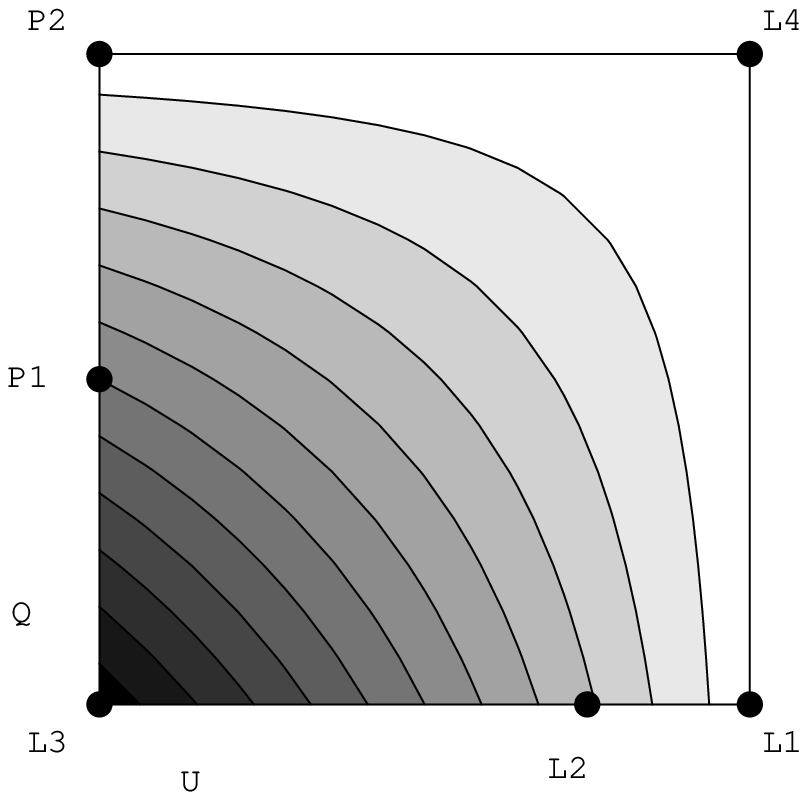}}
	\caption{Subfigure (a) is a contour plot of 
		$\frac{d\ln m}{d\lambda}$, (b) 
		corresponds to $\frac{d\ln r}{d\lambda}$. 
		The black regions are values of the functions
		close to one (their maximum value) while the white regions 
		correspond to zero (their minimum value).}  
  	\label{acquire}
\end{figure}

To understand the solutions' mass and radius properties within the
state space picture, it is useful to consider
equations~(\ref{dr}) and~(\ref{dm}). Note that they do not involve
$\Omega$, and it thus follows that 2D contour plots 
(corresponding to any section $\Omega=\mathrm{const}$) 
yield a picture of where solutions gain
radius and mass in the present setting (see Figure~\ref{acquire}).
It is easily seen that no mass is acquired near the side faces $U=0$ and
$Q=1$, which is in accord with the fact that the fixed points $P_1$ and
$P_2$ are attractors for solutions with finite mass (see below for details).
In contrast, $\frac{d\ln r}{d\lambda}$ does not vanish near $P_1$, but only
near $P_2$ (and $L_5$, if present), the only 
attractor for solutions with finite radii. Clearly, 
orbits spiraling into $P_3$ must acquire both infinite mass
and radius. Moreover, as $U(\lambda)$ and $Q(\lambda)$ become
asymptotically constant in this case, we can specify
the asymptotic growth of $m(r)$ as a function of $r$:
$m(r) \sim r^{(n_0-3)/(n_0-1)}$, see~(\ref{centralsol}) below.

Now recall that an orbit stands for
a solution $(m(r), p(r), \rho(r))$ ($r>0$), via 
(\ref{uqom}) and that $d\ln r = (1-U)(1-Q) d\lambda$. 
The Newtonian potential $v(r)$ is related to $\eta(r)$ 
via $v(r)-v_S = -\eta(r)$, where $v_S = v|_{p=0}$. 
For models with finite radius (with surface $\{p=0\}$) 
$v_S$ is the surface potential. At the surface the interior solution 
is joined to an exterior vacuum solution (standard junction).
Note that the exterior solution is not represented in the interior of the cube. 
For solutions with infinite radii, $p>0$ everywhere and $v_S=0$.

We can interpret the different solutions  of (\ref{UQOmega}) in terms
of familiar variables by using the relation between $\eta$ (i.e,
$\Omega(\lambda)$) and the Newtonian potential and the variable 
transformations~(\ref{uqom}). 
We first analyze the
orbits in the neighborhood of fixed points from which orbits originate
(Table~\ref{emittors}), and then turn to the solutions in the
neighborhood of fixed points where orbits end
(Table~\ref{attractors}).

$L_1$: The 2-parameter family of orbits that comes from $L_1$
corresponds to solutions with the following asymptotic expressions 
for the potential $v(r)$ and the mass $m(r)$: 
\begin{subequations}\label{negmasssols}
\begin{align}
v(r) & = v_0 + \frac{1}{2} \, C\, \delta r^2 + v_S + O(\delta r^3)\\
m(r) &=  R\, C\, \delta r \,(1+\delta r) + O(\delta r^3)\, .
\end{align}
\end{subequations}
Here, $v_0=-(\frac{\Omega_0}{1-\Omega_0})^{1/a}$ and $C>0$ are the two
arbitrary constants that characterize the various orbits, and 
$\delta r= \frac{r-R}{R}>0$, 
where $R=\sqrt{\frac{1}{4\pi}\, C\, \rho^{-1}(\eta)|_{(-v_0)}}\,$.
The orbits that come from $L_1$ 
can be viewed as originating from the negative mass
cube, as discussed in Appendix~\ref{B}. A brief investigation of the 
system~(\ref{UQOmeganeg}) and its relatively simple structure 
(Table~\ref{tab:UQcubeneg}) shows that there exists a 2-parameter family of
solutions with a negative mass singularity, for which $v(r) = -M/r +
v_{const.}$ (where $M<0, v_{const} = \mbox{const}$) when
$r\rightarrow 0$.
For $r>0$, these solutions pick up positive mass monotonically. 
Eventually, when they have reached zero mass, they 
leave the negative mass cube and enter %into 
the positive mass cube through $L_1$. This happens at the radius $r=R$
(compare with~(\ref{negmasssols})).  

$L_2$: Each fixed point on $L_2$ gives rise to exactly one 
orbit that moves into the interior of the cube, 
so that the line $L_2$ as a whole generates a 1-parameter
family of orbits parametrized by $\Omega_c$. 
These orbits correspond to the regular solutions of~(\ref{mpeq}), i.e.,
\begin{subequations}\label{regsols}
\begin{align}
v(r) & = v_c + \frac{1}{2} \, \frac{4\pi}{3}\, \rho_c \,r^2 + v_S + O(r^4)\\
m(r) &= \frac{4\pi}{3}\,r^3\, \rho_c \,
	(1 + \frac{4\pi}{10} \, n_c\, \frac{\rho_c}{v_c}\, r^2) + O(r^7)
\end{align}
\end{subequations}
when $r\rightarrow 0$.
Here, $v_c = -(\frac{\Omega_c}{1-\Omega_c})^{1/a}$ serves as the free parameter,
$\rho_c = \rho(\eta)|_{(-v_c)}$ is the central density, and $n_c$ is given as 
$n_c = n(\eta)|_{(-v_c)}$.
The regular solutions are certainly the most important perfect fluid 
solutions as they serve as stellar models.

$P_4$: This point is a source for interior solutions as long as
$n_1\leq 3$. The 2-parameter family of orbits that originates from 
$P_4$ corresponds to solutions asymptotically described by
\begin{subequations}\label{posmasssing}
\begin{align}
\label{posmasssinggeneric} 
v(r) & =  -\frac{M}{r}\,(1 + T \, r^{3-n_1}) + v_{const} + v_S + O(r^\varepsilon)\, ,
&&{\rm when}\quad n_1 \neq 2,3\:,\\ \label{posmassn2} 
v(r) & = -\frac{M}{r}\,(1 - 4\pi\rho_+ M \, r \log r) + v_{const} + v_S + 
O(r\log r)\, ,
&&{\rm when}\quad n_1=2\:,\\
v(r) & = \frac{\rm const}{r \sqrt{-\log r}} + v_{const} + v_S +O(\cdot)\, , 
&&{\rm when}\quad n_1=3\:, 
\end{align}
\end{subequations}
in the regime $r\rightarrow 0$,
i.e., the solutions possess a positive mass singularity at $r=0$. 
The constants that characterize this 2-parameter family of solutions 
when $n_1\neq2,3$ are $M>0$ and $v_{const}\,$; $T$ is given by 
$T= -4\pi \rho_+ M^{n_1-1} (n_1-2)^{-1} (n_1-3)^{-1}$. The $\varepsilon$ 
appearing in the order-term in~(\ref{posmasssing}) satisfies
$\varepsilon>0$ for $n_1 < \min(2+a_1, \frac{5}{2})$ 
(for $a_1$ see Section~\ref{equationsofstate}), 
and otherwise $2-n_1< \varepsilon < 0$,
which means that $O(r^\varepsilon)$ hides additional singular terms 
in this case. Like the term $T \, r^{3-n_1}$, these terms depend solely
on the free parameter $M$.

For $n_1=2$, on the space $<e_1,e_2>$, the linearization of~(\ref{UQOmega})
has a Jordan normal form and the potential looks like in 
expression~(\ref{posmassn2}).
For $n_1=3$ we have a center manifold and
the potential's leading term is as given above.
In this case the mass behaves like $m(r) \rightarrow \frac{\rm const}{\sqrt{-\log r}}$ 
for $r\rightarrow 0$.
This case represents a transition between~(\ref{posmasssinggeneric}) and the 
solutions~(\ref{strangesol}) generated by $P_6$ for $n_1>3$.

$P_6$: There exists a family of orbits that originates from $P_6$ when 
$n_1 > 3$, but one 
special orbit plays the main role. Its potential looks like
\begin{equation}\label{strangesol}
v(r) = -C \, r^{-2/(n_1-1)}\, (1 + O(r^\alpha)) + v_S 
\end{equation}
as $r\rightarrow 0$. 
Here, $C^{n_1-1} = \frac{1}{4 \pi\rho_+} \frac{2 (n_1-3)}{(n_1-1)^2}$ and
$\alpha=\frac{2 a_1}{(n_1-1)}$. 
For $3< n_1 < 5$, (\ref{strangesol}) is the central orbit of 
a 2-parameter family of orbits asymptotically approaching (\ref{strangesol}).
In the case $(11+8\sqrt{2})/7<n_1<5$, we observe a superposition of 
(\ref{strangesol}) and dying oscillatory modes. 
For $n_1 \geq 5$ (\ref{strangesol}) is the only solution generated by $P_6$.
For further details we refer to the treatment of the attractive
counterpart of $P_6$, the fixed point $P_3$.
	
We now turn to the various end points of the orbits:

$L_5$: When $n_0=0$ the 1-parameter set of fixed points 
$L_5 = \{(U_0,1,0)\}$ 
attracts a 2-parameter family of orbits.
The corresponding solutions can easily be shown to possess finite radii $R$ and masses $M$
($R$ and $M$ are of course different for the different solutions). 
To first order the potential has the form
\begin{equation}\label{l5}
v(r) = -C\, \delta r + v_S\:,  
\end{equation} 
where $\delta r = \frac{R-r}{R}$.
The different orbits are characterized by the two parameters $U_0$ and $C>0$.
At $r=R$ the density is non-vanishing, $\rho = \rho_-$,
so that solutions of this type can be viewed as asymptotically 
incompressible towards low pressures. The radius $R$ is given by 
$R= \sqrt{\frac{1}{4\pi \rho_-}\, \frac{U_0}{1-U_0} \,C}$,
the total mass $M$ of a solution is given by $M= R\, C\,$, whereby
the surface potential is $v_S = M/R =C $.

$P_1$: Recall that $n_0>3$ is necessary for these solutions to exist.
The 1-parameter family of orbits that converge to $P_1$ corresponds to 
solutions with infinite radii and possess the 
following potential, mass, and density as $r\rightarrow \infty$:
\begin{subequations}\label{infsolswithfinitemass}
\begin{align}
v(r) & = -\frac{M}{r}\, ( 1 - \frac{4}{n_0-3}\,C\,r^{3-n_0}) + O(\cdot) \\
m(r) & = M \, (1-4 C\,\frac{n_0-2}{n_0-3}\,r^{3-n_0}) + O(\cdot) \\
\rho(r) & = \rho_- M^{n_0} r^{-n_0} (1+ O(r^{3-n_0}))\, .
\end{align}
\end{subequations}
Here, $C$ denotes $C= \frac{\pi\rho_-}{n_0-2}\, M^{n_0-1}$ and $M>0$. 
The solutions~(\ref{infsolswithfinitemass}) thus have
finite masses $M$ (different for different solutions) although they have 
infinite radii. For the special case of an exact polytrope with
$n(\eta)\equiv 5$ the associated regular
solutions are given by
$v(r)=-M /\sqrt{\frac{4\pi}{3}\, \rho_- M^4+r^2}$.

$P_2$: When $n_0>0$, the fixed point $P_2$ acts as an attractor
for a 2-parameter family of orbits.
Since these orbits correspond to solutions with finite masses and radii, 
they play the most prominent role in our considerations.
\begin{subequations}\label{finitesols}
\begin{align}
v(r) & = -C\, \delta r - C\, \delta r^2 + v_S + O(\delta r^{2+\alpha}) \\
m(r) & = R\, C \,\left(\,
1- \frac{C^{n_0}\, D^{-1}}{n_0 + 1}\, \delta r^{n_0+1} \,\right) + O(\cdot) \\
\rho(r) & = \frac{1}{4\pi R^2}\, C^{n_0+1}\,D^{-1}\,\delta r^{n_0} +O(\cdot)\, .
\end{align}
\end{subequations}
As before, $\delta r$ stands for $\delta r = \frac{R-r}{R}$. The constants
$C>0$, $D>0$ are arbitrary and $\alpha= \min\{1, n_0\}$.
Combining the earlier asymptotic expression for $\rho(\eta)$ 
with the above equations, one can show that the radii and masses of the 
solutions that end at $P_2$ are given by
$R^2=\frac{1}{4\pi \rho_-}\,\frac{C}{D}\, ;\, M^2=\frac{1}{4\pi\rho_-}\,\frac{C^3}{D}$.

$P_3$: For $3<n_0\leq 5$, the fixed point $P_3$ attracts a single 
orbit that corresponds to a solution with
infinite mass and radius. For $r\rightarrow \infty$ we have 
\begin{subequations}\label{centralsol}
\begin{align}
v(r) & = - K r^{-2/(n_0-1)} \,(1+ O(r^{-\alpha})) \\
m(r) & = \frac{2}{n_0-1}\, K\, r^{(n_0-3)/(n_0-1)} \,(1+ O(r^{-\alpha}))\:,
\end{align}
\end{subequations}
where $\alpha = \frac{2 a_0}{n_0-1}$ and 
$K = \frac{1}{4\pi\rho_-}\, \frac{2(n_0-3)}{(n_0-1)^2}$.
For $n_0>5$ the solution~(\ref{centralsol}) becomes
the center of a 2-parameter family of orbits that oscillate
around~(\ref{centralsol}) and converge to it
when $r\rightarrow \infty$.
\begin{subequations}\label{osci}
\begin{align}
v(r) & = - K r^{-2/(n_0-1)} \,
\big(1+ H r^{-\alpha_1} \: \mbox{osc}(\alpha_2 \log r) + O(\cdot)\,\big)\:, \\
\nonumber
& \quad \mbox{where}\quad \mbox{osc}(x) = (k_1 \alpha_2 - k_2 \alpha_1) \,\cos x + 
(k_1 \alpha_1 + k_2 \alpha_2)\,\sin x \\[3pt]
m(r) & = \frac{2}{n_0-1}\, K\, r^{(n_0-3)/(n_0-1)} \,
\big(1 + H r^{-\alpha_1}\: \overline{\mbox{osc}}(\alpha_2 \log r) + O(\cdot)\,\big)\:,\\
\nonumber
& \quad \mbox{where}\quad \overline{\mbox{osc}}(x) = 
(1-\alpha_1)\,\mbox{osc}(x) + \alpha_2 \,\mbox{osc}^\prime(x)
\end{align}
\end{subequations}
In~(\ref{osci}) $K$ is again defined as in~(\ref{centralsol}); furthermore,
$\alpha_1 = \frac{n_0-5}{2(n_0-1)}\,$, $\alpha_2 =\frac{\sqrt{b}}{2(n_0-1)}\,$,
and $H = 16\,\frac{(n_0-2)^2}{(n_0-1)^2}\,\frac{1}{\alpha_1^2+\alpha_2^2}$.
The constants $k_1$ and $k_2$ parameterize the various solutions.
Note that $v(r)$ and $m(r)$ are only of the form~(\ref{osci}),
if the condition $4 a_0 > (n_0-5)$ is satisfied (whereby $\alpha>\alpha_1$);
else the oscillations are dying out more rapidly than
the $O(r^{-\alpha})$ term in~(\ref{centralsol}).

When $n_0 = 5$ $(n_1 = 5)$ the periodic orbits $C_1$ $(C_2)$ constitute limit 
sets for interior orbits and it is possible to obtain asymptotic expressions. 
We refrain from giving such expressions, but to obtain them one uses the known 
exact solutions for polytropes with $n=5$, whose asymptotic expressions directly 
yield the desired expressions for more general equations of state since 
$dn(\Omega)/d\Omega=0$ when $\Omega \rightarrow 0$.

\begin{remark}
We ignored those fixed points which are not mentioned in the list since 
we are only interested in interior orbits. 
Only $L_4$ requires a closer investigation, since
the linearization of~(\ref{UQOmega}) at $L_4$ is zero.
However, considerations in Appendix A show that
no interior orbits converge to $L_4$ for $\lambda\rightarrow\pm\infty$.
\end{remark}

\begin{remark}
Finally note that the orbits on the side face $U=0$ 
(see Figure~\ref{sidefaces}) correspond to the vacuum solutions 
$v(r) = -\frac{M}{r} + C + v_S$ ($M>0$), where
$C=0$ corresponds to the orbit $P_4 \rightarrow P_1$, while $C<0$ $(C>0)$
characterizes the orbits on the left (right) hand side.
The orbits on the side face $Q=0$ correspond to solutions 
with constant potential. The orbits on $U=1$ and
$Q=1$ are unphysical since $d\ln r = 0$ in these cases.
\end{remark}

%%%%%%%%%%%%%%%%%%%%%%%%%%%%%%%%%%%%%%%%%%%%%%%%%%%%%%%%%%%%%%%%%%%%%%%%
%%%%%%%%%%%%%%%%%%%%%%%%%%%%%%%%%%%%%%%%%%%%%%%%%%%%%%%%%%%%%%%%%%%%%%%%
%%%%%%%%%%%%%%%%%%%%%%%%%%%%%%%%%%%%%%%%%%%%%%%%%%%%%%%%%%%%%%%%%%%%%%%%
%%%%%%%%%%%%%%%%%%%%%%%%%%%%%%%%%%%%%%%%%%%%%%%%%%%%%%%%%%%%%%%%%%%%%%%%
\section{Examples}
\label{examples}
%%%%%%%%%%%%%%%%%%%%%%%%%%%%%%%%%%%%%%%%%%%%%%%%%%%%%%%%%%%%%%%%%%%%%%%%

In this section we give some examples and discuss some features
concerning the relationship between the equation of state and various
models' radii and masses. We will come back to this issue in the next
section where we give some general theorems.

%%%%%%%%%%%%%%%%%%%%%%%%%%%%%%%%%%%%%%%%%%%%%%%%%%%
\subsection{Mass-radius features}
\label{massradiusfeatures}
%%%%%%%%%%%%%%%%%%%%%%%%%%%%%%%%%%%%%%%%%%%%%%%%%%%

Recall from the discussion of the fixed point $P_2$ that it acts
as an attractor for a 2-parameter family of orbits 
and thereby generates a 2-parameter family of finite perfect fluid solutions
(cf.~(\ref{finitesols})).
An orbit of this family can be uniquely characterized by
two parameters $C>0$ and $D>0$, where
$C=\lim_{\lambda\rightarrow\infty} \frac{\Omega^{1/a}}{1-Q}$ and
$D=\lim_{\lambda\rightarrow\infty} \frac{\Omega^{n/a}}{U}$.
The associated physical variables display the asymptotic
behavior~(\ref{finitesols}) as $r\rightarrow R$, and
for the finite radii $R$ and total masses $M$ the relations
\begin{equation}\label{universalMR}
R^2 =\frac{1}{4\pi \rho_-}\,\frac{C}{D} \quad,\quad
M^2 = \frac{1}{4\pi\rho_-}\,\frac{C^3}{D}\:
\end{equation}
hold.
We may thus refer to~(\ref{universalMR}) as a universal $(M,R)$-relation.

Consider a 1-parameter family of solutions, described by a parameter
$s$  (e.g., the regular solutions with $s = \rho_c$, the central
density).  We then have $C(s)$, $D(s)$, and hence 
$R(s)=R(C(s),D(s))\,$;$\,M(s)=M(C(s),D(s))$, whereby we obtain a curve in a $(M,R)$-diagram
for the particular family under consideration. This can be visualized
as making a particular 1-dimensional cut through the graphs $R(C,D)$
and $M(C,D)$, defined by~(\ref{universalMR}).

Equation~(\ref{universalMR}) can be brought into a more
illustrative form by introducing cylindrical coordinates
around $P_2$, i.e., set $U=\varepsilon \sin\phi$, 
$1-Q= \varepsilon \cos\phi$, $\Omega=h^a$ (with $\varepsilon$ small).
Then the asymptotic expansion around $P_2$ yields,
\begin{equation}\label{universalMRcyl}
r^2 = \frac{1}{4 \pi \rho_-}\, h^{1-n_0} \, \tan\phi + O(\varepsilon^a)\, , \quad
m^2 = \frac{1}{4 \pi \rho_-}\,\frac{1}{\varepsilon^2} h^{3-n_0} \, \tan\phi \, 
(1+ \tan^2\phi) + O(\varepsilon^a)
\end{equation}
as $\lambda\rightarrow\infty$, and, accordingly, 
\begin{equation}\label{mrcyl}
M=(4\pi \rho_-)^{-1/(n_0-1)}\:
\lim_{\lambda\rightarrow\infty}
\big[\varepsilon^{-1} (\tan\phi)^{1/(n_0-1)} (\cos\phi)^{-1}\big]
\:R^{(3-n_0)/(1-n_0)}\:.
\end{equation}

Equation~(\ref{mrcyl}) and related formulas are useful in e.g.,
numerical computations: take a certain small fixed $\varepsilon$ and
follow an orbit attracted by $P_2$ to its intersection with the small
cylinder. Then the values of $\phi$ and $h$ of the point of
intersection yield $M$ and $R$.

\begin{figure}[htp]
	\psfrag{t1}{{\small Small $\rho_c$}}
	\psfrag{t2}[cc][cc]{{\small Large $\rho_c$}}
	\psfrag{t4}[rc][rc]{{\small Small $\rho_c$}}
	\psfrag{t3}[rc][rc]{{\small Large $\rho_c$}}
	\psfrag{t5}{{\small Small $\rho_c$}}
	\psfrag{t6}{{\small Large $\rho_c$}}
	\psfrag{t7}{{\small Small $\rho_c$}}
	\psfrag{t8}[cc][cc]{{\small Large $\rho_c$}}
	\psfrag{t9}[cc][cc]{{\small Small $\rho_c$}}
	\psfrag{t10}[cc][cc]{{\small Large $\rho_c$}}
	\psfrag{M}{{\small ${\mathcal M}$}}
	\psfrag{R}[cc][cc]{{\small ${\mathcal R}$}}
	\centering
	\subfigure[$n=0.25$]{
		\label{MR025}
		\includegraphics[width=0.3\textwidth]{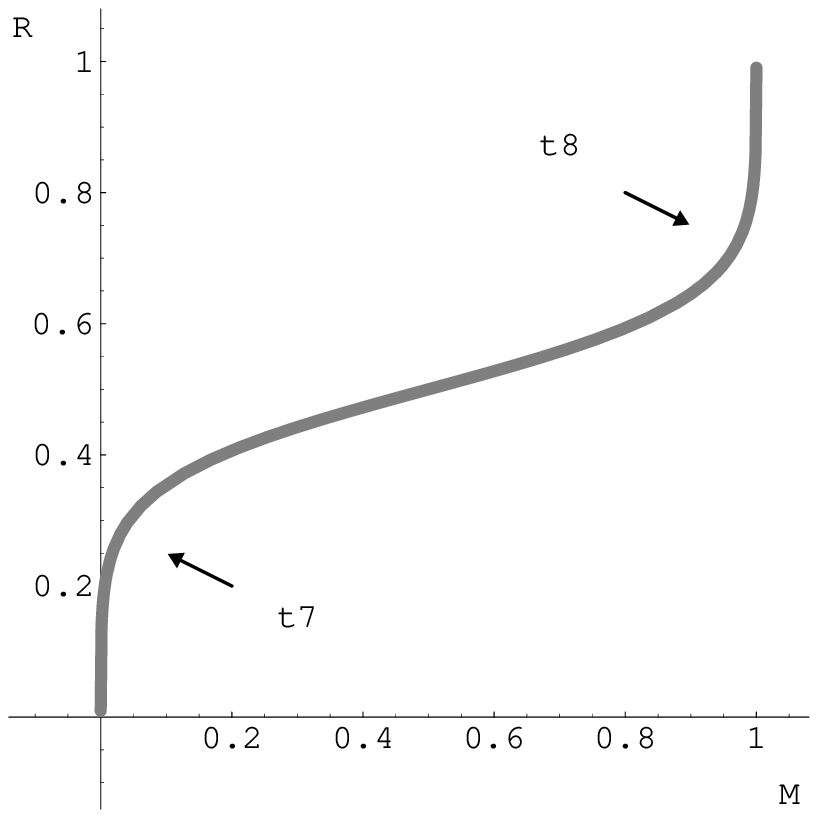}}\quad
        \subfigure[$n=1$]{
		\label{MR1}
		\includegraphics[width=0.3\textwidth]{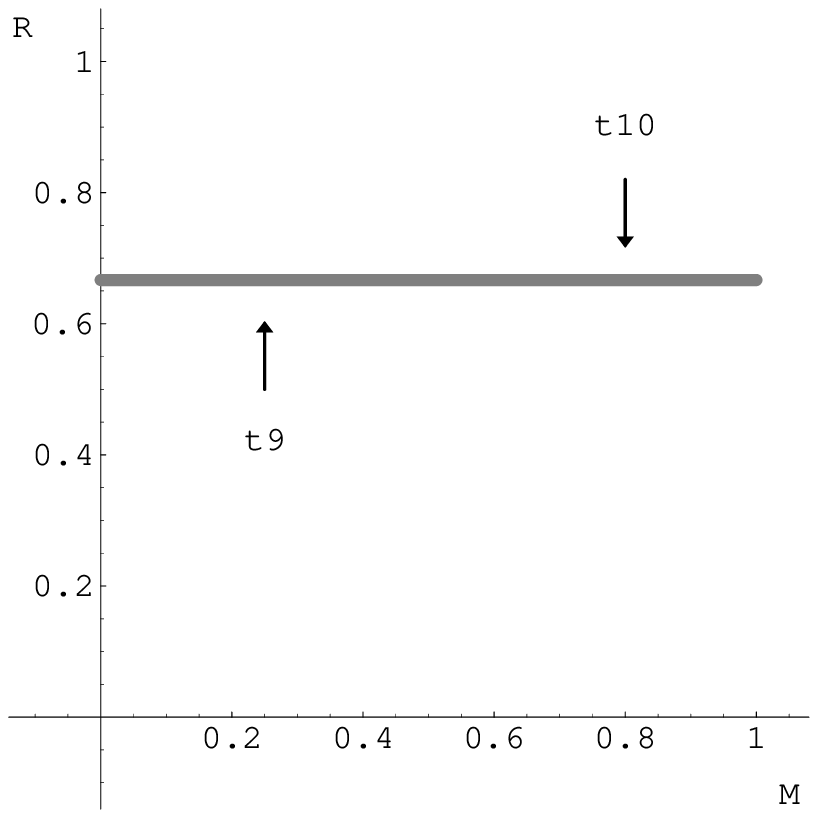}}\quad
     	\subfigure[$n=1.5$]{
		\label{MR1point5}
		\includegraphics[width=0.3\textwidth]{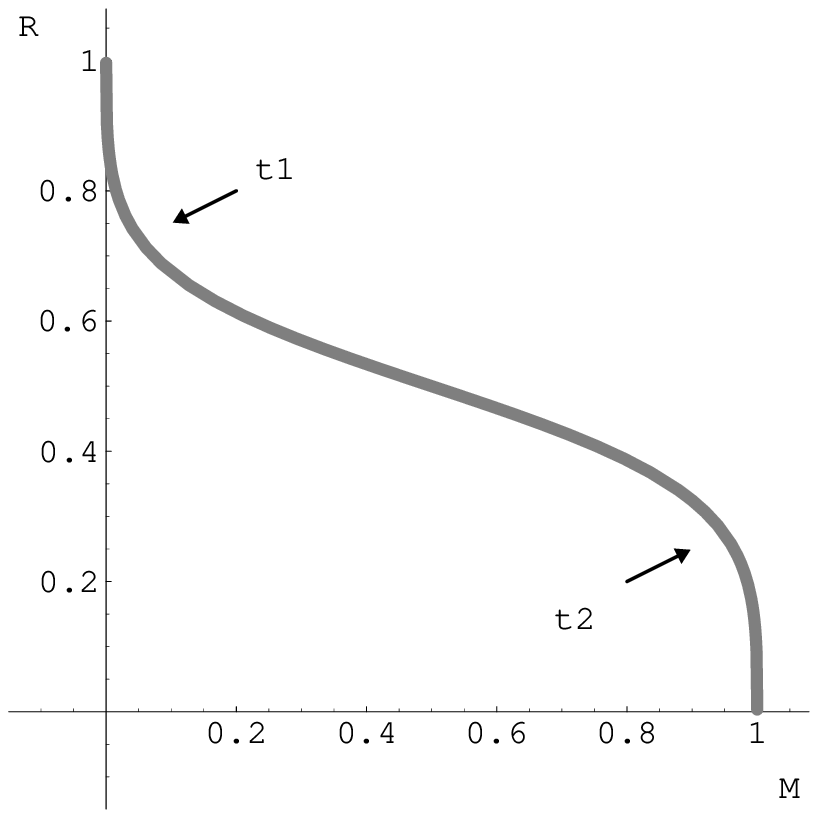}}\quad
	 \subfigure[$n=3$]{
		\label{MR3}
		\includegraphics[width=0.3\textwidth]{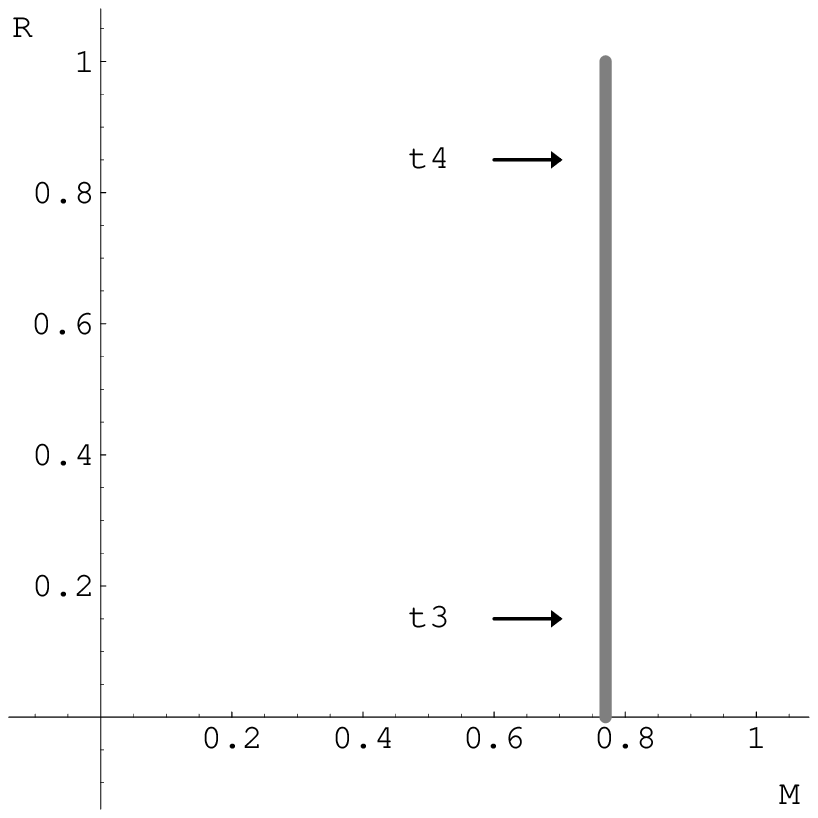}}\qquad
     	\subfigure[$n=4$]{
		\label{MR4}
		\includegraphics[width=0.3\textwidth]{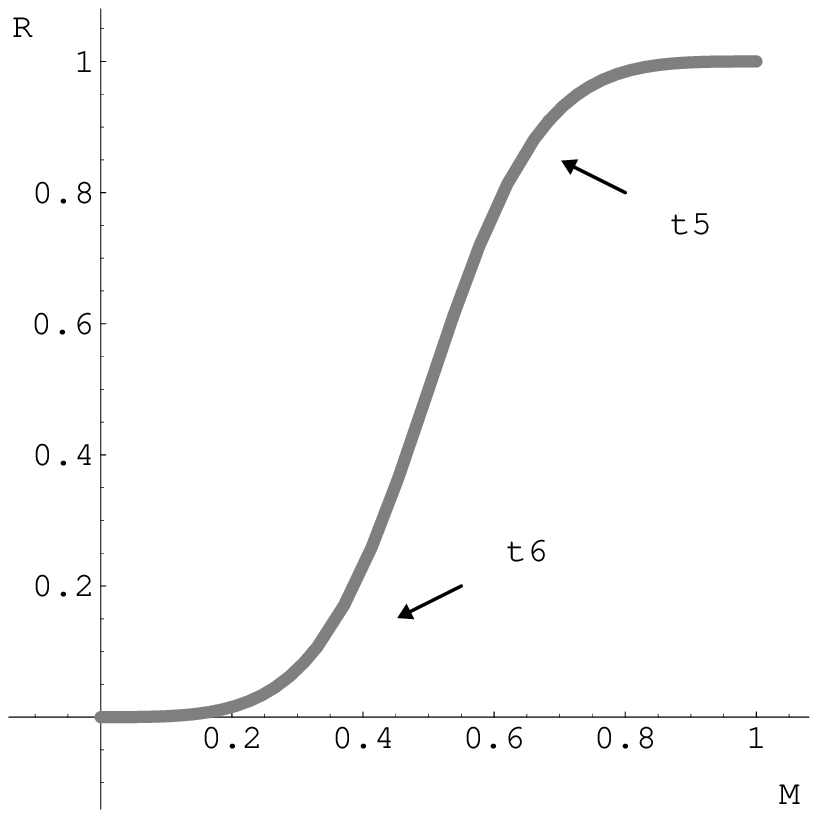}}\qquad
	 \caption{The mass-radius diagram for polytropic equations of state
		with different polytropic indices $n$.
		We use bounded variables, ${\mathcal M}= (M/\tilde{M})(1+M/\tilde{M})^{-1}$ 
		and ${\mathcal R} = (R/\tilde{R})(1+R/\tilde{R})^{-1}$, 
		where $\tilde{M}$ and $\tilde{R}$ are typical values.} 
 	 \label{polyMR}
\end{figure}

Note that~(\ref{universalMRcyl}) and~(\ref{mrcyl}) 
automatically yield the standard relations
(see, e.g.,~\cite{book:KippWeig1994}) for the polytropes $n(\eta)\equiv n$:
Since~(\ref{UQOmega}) decouples for polytropic equations of state, 
all corresponding 
regular orbits intersect the small cylinder at the same value of $\phi$.
Moreover, (\ref{Omegaeq}) can be integrated, and
expressed in $\eta$ we obtain $\eta(\lambda) = \eta_c
\exp\big(-\int_{-\infty}^\lambda Q_{\mathrm{p}}(\lambda) 
(1-U_{\mathrm{p}}(\lambda)) d\lambda\big)$,
where $(U_{\mathrm{p}},Q_{\mathrm{p}})$ is the solution 
of~(\ref{Ueq}) and (\ref{Qeq}).
At some value $\lambda=\tilde{\lambda}$ the orbit intersects the cylinder, where
$\tilde{\lambda}$ is independent of $\eta_c$, whence we conclude
that $h\propto \eta_c$. Therefore, for the polytropes,
\begin{equation}\label{MRpoly}
R^2= const\: \eta_c^{1-n} \qquad M^2= const\: \eta_c^{3-n}
\qquad
M= const \:R^{(3-n)/(1-n)},
\end{equation}
where the constants only depend 
on $\rho_-$ and on $n$ (but not on $\rho_c$).
$(M,R)$-diagrams for some polytropes
are displayed in Figure~\ref{polyMR}. 

%%%%%%%%%%%%%%%%%%%%%%%%%%%%%%%%%%%%%%%%%%%%%%%%%%%
\subsection{Chandrasekhar's equation of state}
\label{chandrasekhar}
%%%%%%%%%%%%%%%%%%%%%%%%%%%%%%%%%%%%%%%%%%%%%%%%%%%

The equation of state of a completely degenerate, ideal Fermi gas
reads (see e.g.,~\cite{Shapiro/Teukolsky:1983})
\begin{equation}\label{chandraeqofstate}
\rho(x) = c_\rho x^3 \quad,\quad 
p(x) = c_p \big( \,x \sqrt{1+x^2}\: (2 x^2/3-1)+\log(x+ \sqrt{1+x^2})\, \big)\:,
\end{equation}
where $c_\rho$ and $c_p$ are constants.
The equation of state is given in implicit form and the parameter $x$ is the 
``dimensionless Fermi momentum''.
In the non-relativistic limit, $x \ll 1$, the pressure 
$p(x) \propto x^5\,$, and in the ultra-relativistic limit, $x \gg 1$,
$p(x) \propto x^4$. Hence the equation of state~(\ref{chandraeqofstate})
behaves like a polytrope with $n_0 = 3/2$ ($n_1 = 3$) for low (large) densities.

In terms of $\eta$ the equation of
state~(\ref{chandraeqofstate}) and the polytropic index function 
are represented by simple expressions.
For convenience, we use
a dimensionless variable $\eta$, which is just a rescaled
version of our original $\eta$-variable according to $\eta/c_\eta$,
where $c_\eta = (8/3)(c_p/c_\rho)$. We obtain 
\begin{equation}\label{chandranice}
\rho(\eta) \,=\, \rho_- \,\big( \eta \,(1 + \frac{\eta}{2}) \big)^{3/2}\:,
\end{equation}
and
\begin{equation}\label{chandraindex}
n(\eta) \,=\, 3\; \frac{\eta +1}{\eta +2}\:.
\end{equation}
This latter expression makes it evident that we deal with an asymptotically 
polytropic equation of state for which $n_0 =3/2$, $n_1 =3$, and that
$a_0 = a_1 =1\,$.

Clearly, as is well known, all stellar models with a completely degenerate ideal 
Fermi gas have finite radii and masses since $n_0 = 3/2$ 
(compare with the theorems in the next section). 

\begin{figure}[htp]
	\psfrag{L1}{{\small $L_1$}}
	\psfrag{L2}{{\small $L_2$}}
	\psfrag{L3}{{\small $L_3$}}
	\psfrag{L4}{{\small $L_4$}}
	\psfrag{L5}{{\small $L_5$}}
	\psfrag{P1}{{\small $P_1$}}
	\psfrag{P2}{{\small $P_2$}}
	\psfrag{P3}{{\small $P_3$}}
	\psfrag{P4}{{\small $P_4$}}
	\psfrag{P5}{{\small $P_5$}}
	\psfrag{P6}{{\small $P_6$}}
	\psfrag{O}[cb][Bl]{{$\begin{array}{c}\uparrow\\ \Omega\end{array}$}}
	\psfrag{Q}[cb][Bl]{{$\begin{array}{c}\uparrow\\ Q\end{array}$}}
	\psfrag{Q1}{{$Q\rightarrow$}}
	\psfrag{U}{{$U\rightarrow$}}
	\centering
        \subfigure[Projection to $\Omega=0$]{
		\label{chandradown}
		\includegraphics[width=0.3\textwidth]{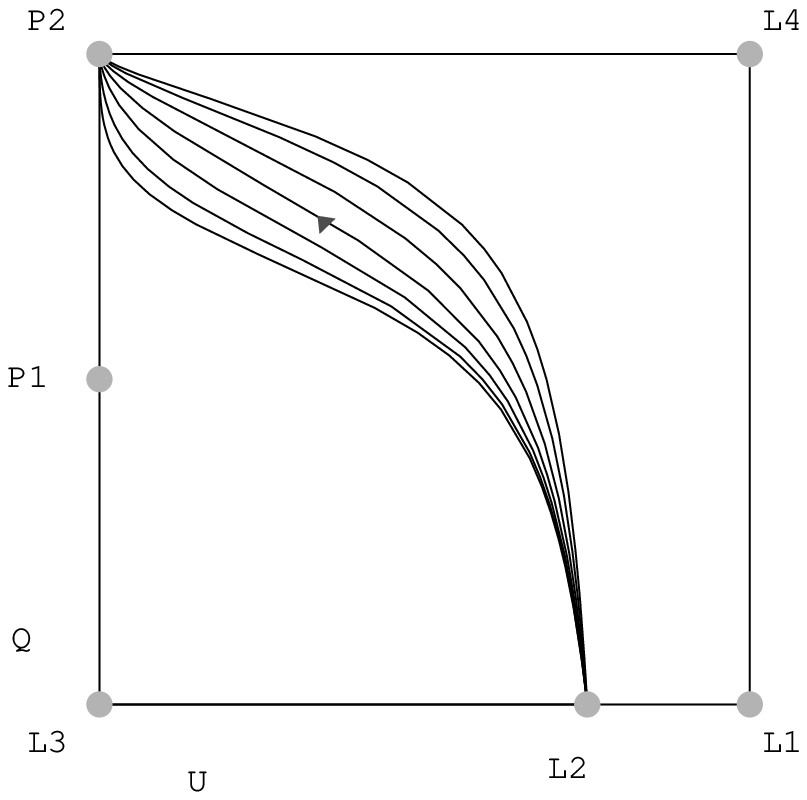}}\qquad
     	\subfigure[Projection to $U=0$]{
		\label{chandraside}
		\includegraphics[width=0.3\textwidth]{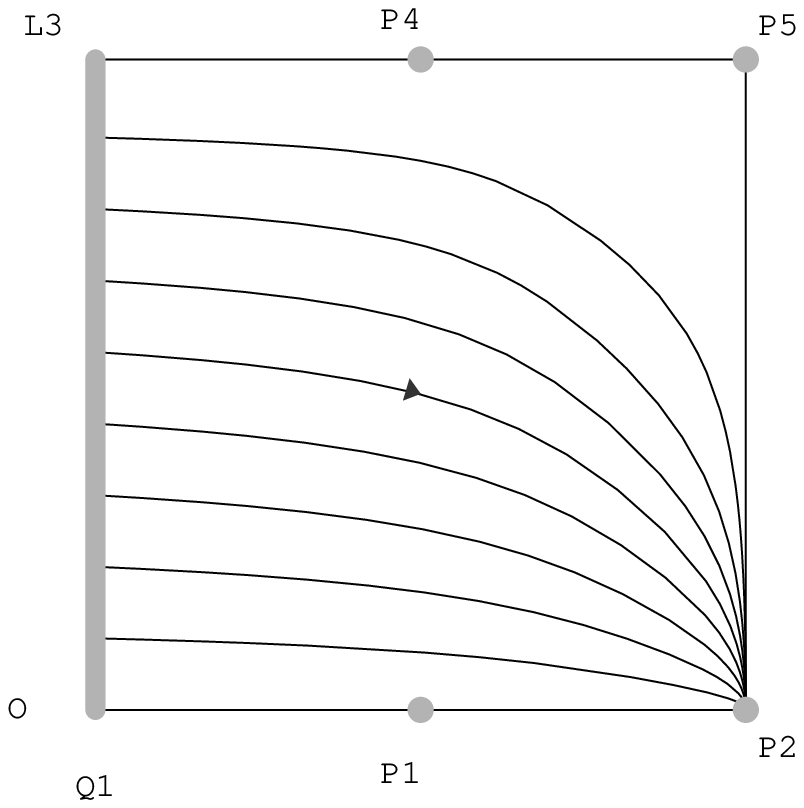}}
	\caption{Solutions associated with Chandrasekhar's equation of state 
	projected down to $\Omega=0$. The topmost orbit coincides with the 
	polytropic orbit for $n=3/2$, the lowermost orbit coincides with the polytropic 
	orbit for $n=3$.}
    \label{chandraorbits}
\end{figure}

The compact state space $(U,Q,\Omega)$ provides a picture of the
solution space of a given equation of state. In
Figure~\ref{chandraorbits}, projections onto the $\Omega = 0$ and 
$U =0$ planes of the regular ``Chandrasekhar solutions'' are shown. 
In Figure~\ref{chandradown}, for
small central densities the orbits coincide with the $n=3/2$
polytropic orbits. With increased central density the orbits bend
down until, in the limit of infinite densities, the orbits approach
the $n=3$ polytropic solution.

\begin{figure}[htp]
	\psfrag{t11}{{\small Small $\rho_c$}}
	\psfrag{t12}[rc][rc]{{\small Large $\rho_c$}}
	\psfrag{M}{{\small ${\mathcal M}$}}
	\psfrag{R}[cc][cc]{{\small ${\mathcal R}$}}
	\centering
	\includegraphics[width=0.4\textwidth]{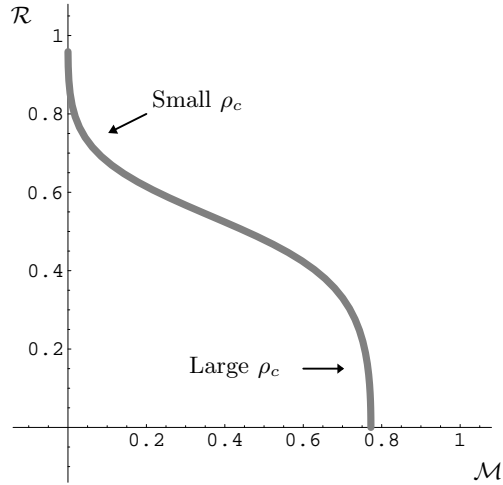}
	\caption{The mass-radius diagram for Chandrasekhar's equation of state.
		We use bounded variables, ${\mathcal M} = (M/\tilde{M})(1+M/\tilde{M})^{-1}$ 
		and ${\mathcal R} = (R/\tilde{R})(1+R/\tilde{R})^{-1}$, 
		where $\tilde{M}$ and $\tilde{R}$ are typical values.} 
    \label{chandraMR}
\end{figure}

Also the corresponding mass-radius diagram ``interpolates''
between the polytropic ends. For small central densities we observe
that $M$ and $R$ behave like for a polytrope with index $n=3/2$, while
for high central densities the mass approaches the 
Chandrasekhar limit and thereby becomes independent of $\rho_c$,
analogous to a polytrope with index $n=3$.
For the mass-radius diagram see Figure~\ref{chandraMR}.

%%%%%%%%%%%%%%%%%%%%%%%%%%%%%%%%%%%%%%%%%%%%%%%%%%%
\subsection{Composite equations of state}
\label{compositeequationsofstate}
%%%%%%%%%%%%%%%%%%%%%%%%%%%%%%%%%%%%%%%%%%%%%%%%%%%

Since the equation of state only enters the system of equations
through the function $n(\Omega)$, one can view the solutions as a
"changing" polytrope. Figure~\ref{base} gives an intuitive idea of
how solutions behave in the compact state space.  It is possible to
approximate an equation of state by discretizing $n(\Omega)$ and thus
to view it as a composite equation of state.

We will now give some mass-radius diagrams for simple composite equations of
state: consider ${\mathcal C}^0$ equations of state that correspond to 
functions $n(\Omega)$ described by two constant values, $n_0, n_1$, 
and a jump from $n_0$ to $n_1$ at a given 
$\Omega=\Omega_{\mathrm{j}}$. 
In this present case $U,Q,\Omega$ are continuous
(compare with the remark at the end of Section~\ref{dynamicalsystemsformulation})
and the map between the state spaces associated with the polytropes $n_0,n_1$ 
can be viewed as gluing the state spaces together at $\Omega_{\mathrm{j}}$. 
Hence a regular solution that corresponds to a composite equation of state
follows the polytropic flow associated with the index $n_1$ until
$\Omega$ has reached $\Omega_{\mathrm{j}}$. At this point
the polytropic flow associated with $n_0$ takes over.

\begin{figure}[htp]
	\centering
     	\subfigure[$n_0=4$, $n_1=7$]{
		\label{cp47}
		\includegraphics[width=0.3\textwidth]{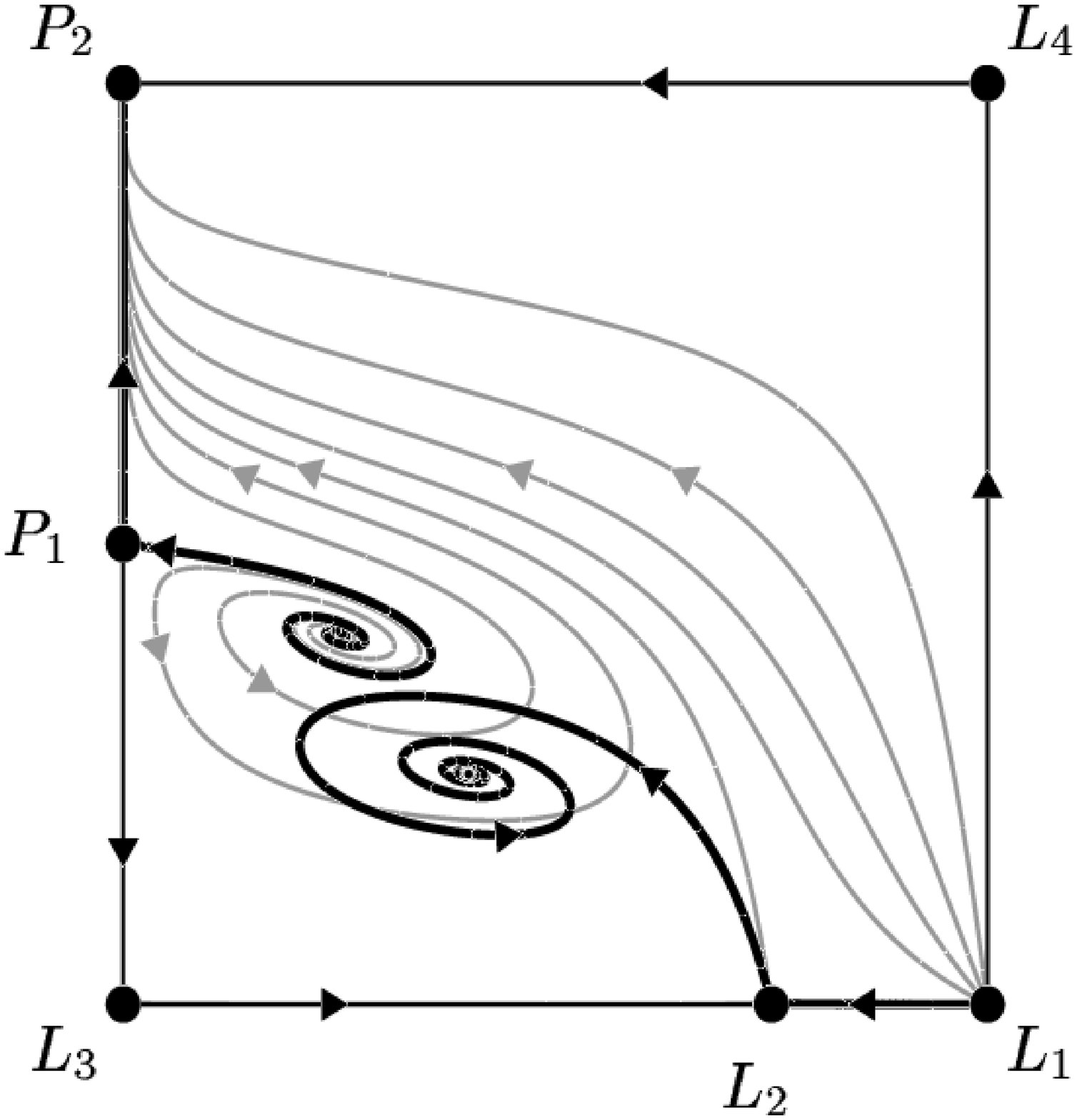}}\quad
        \subfigure[$n_0=4$, $n_1=6.2$]{
		\label{cp462}
		\includegraphics[width=0.3\textwidth]{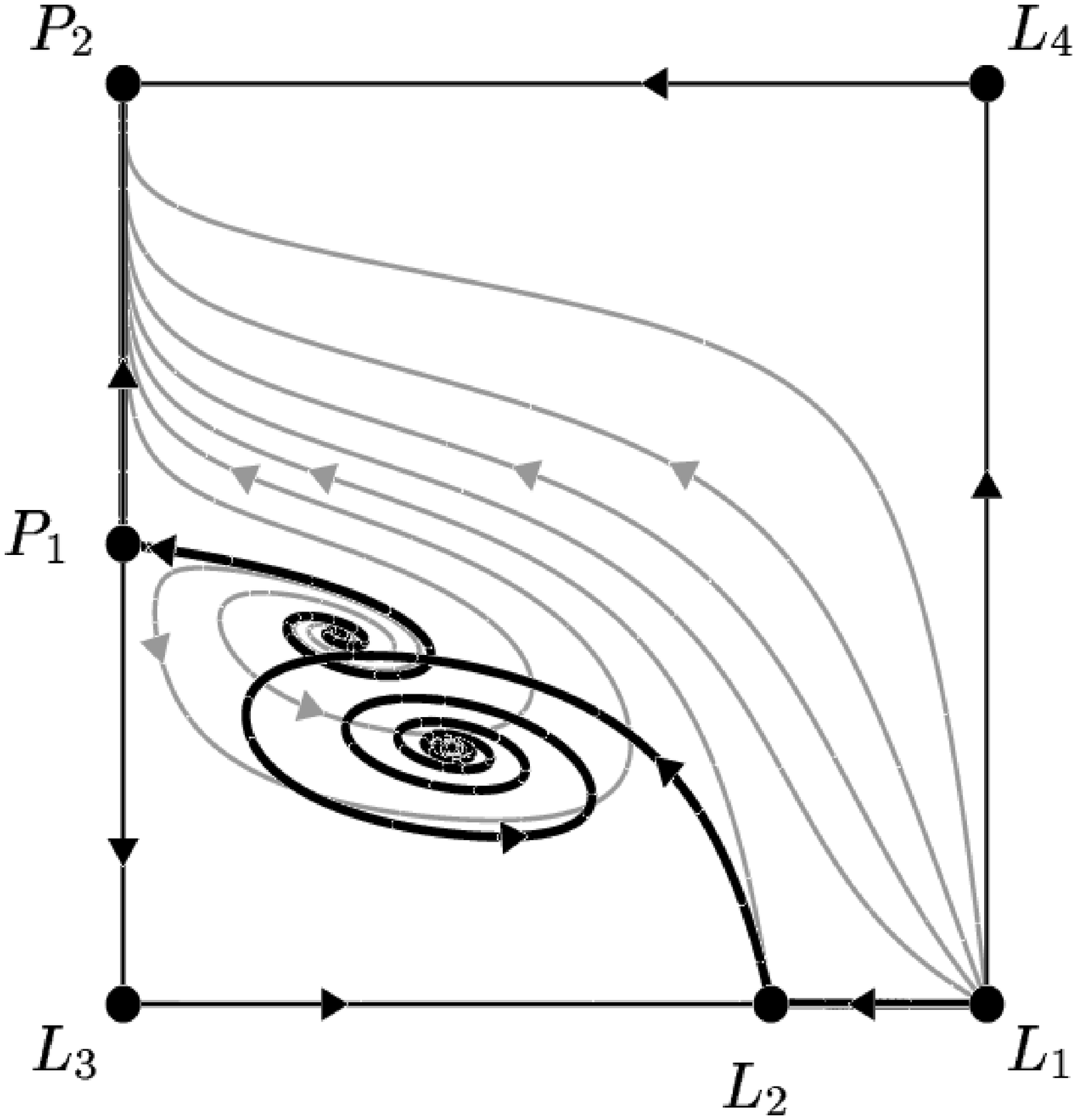}}\quad
    	\subfigure[$n_0=4$, $n_1=5.88$]{
		\label{cp4588}
		\includegraphics[width=0.3\textwidth]{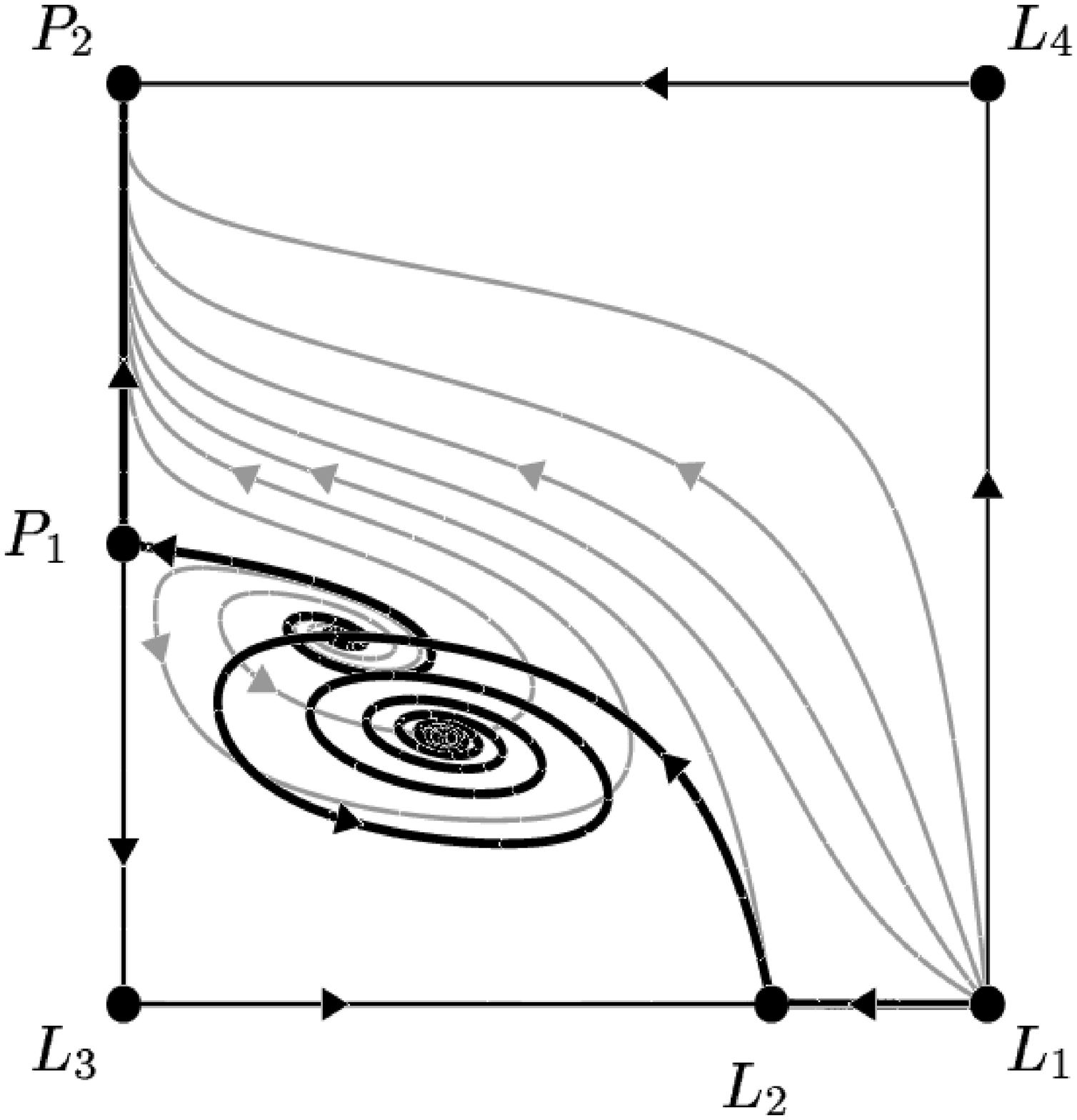}}
	\caption{Composite equation of 
		state $n(\eta) = n_0$ for $\eta \leq \eta_{\mathrm{j}}$ and
		$n(\eta) = n_1$ for $\eta>\eta_{\mathrm{j}}$. For
		(a), (b), and (c), $n_0 = 4$ while $n_1$ varies.
		The orbits are projected down to $\Omega=0$. 
		A regular solution starts from $L_2$ and proceeds to some point
		(depending on $\eta_c$) on the $n_1$-orbit (where $\Omega$ 
		($\eta$) has reached $\Omega_{\mathrm{j}}$ ($\eta_{\mathrm{j}}$)). 
		Here the orbit is matched with the
		$n_0$-flow, which is subsequently followed.
		A regular solution with infinite radius and finite mass
		is constructed by gluing the regular orbit, 
		which originates from $L_2$, to 
		the orbit that ends at $P_1$.} 
    \label{compositeflow}
\end{figure}

Figures~\ref{compositeflow} are projections to the $\Omega=0$ plane. 
Regular solutions proceed along the spiral ($n_1$-flow)
up to some point, at which they switch to
the $n_0$-flow. If this point lies
on the curve that ends at $P_1$, then the solution has infinite radius
but finite mass (see~(\ref{infsolswithfinitemass})). 

%\begin{figure}[htp]
%	\psfrag{L1}{{\small $L_1$}}
%	\psfrag{L2}{{\small $L_2$}}
%	\psfrag{L3}{{\small $L_3$}}
%	\psfrag{L4}{{\small $L_4$}}
%	\psfrag{L5}{{\small $L_5$}}
%	\psfrag{P1}{{\small $P_1$}}
%	\psfrag{P2}{{\small $P_2$}}
%	\psfrag{P3}{{\small $P_3$}}
%	\psfrag{P4}{{\small $P_4$}}
%	\psfrag{P5}{{\small $P_5$}}
%	\psfrag{P6}{{\small $P_6$}}
%	\centering
%     	\subfigure[$n_0=4$, $n_1=7$]{
%		\label{cp47}
%		\includegraphics[width=0.3\textwidth]{cp47.eps}}\quad
%        \subfigure[$n_0=4$, $n_1=6.2$]{
%		\label{cp462}
%		\includegraphics[width=0.3\textwidth]{cp462.eps}}\quad
%    	\subfigure[$n_0=4$, $n_1=5.88$]{
%		\label{cp4588}
%		\includegraphics[width=0.3\textwidth]{cp4589.eps}}
%	\caption{Composite equation of 
%		state $n(\eta) = n_0$ for $\eta \leq \eta_{\mathrm{j}}$ and
%		$n(\eta) = n_1$ for $\eta>\eta_{\mathrm{j}}$. For
%		(a), (b), and (c), $n_0 = 4$ while $n_1$ varies.
%		The orbits are projected down to $\Omega=0$. 
%		A regular solution starts from $L_2$ and proceeds to some point
%		(depending on $\eta_c$) on the $n_1$-orbit (where $\Omega$ 
%		($\eta$) has reached $\Omega_{\mathrm{j}}$ ($\eta_{\mathrm{j}}$)). 
%		Here the orbit is matched with the
%		$n_0$-flow, which is subsequently followed.
%		A regular solution with infinite radius and finite mass
%		is constructed by gluing the regular orbit, 
%		which originates from $L_2$, to 
%		the orbit that ends at $P_1$.} 
%    \label{compositeflow}
%\end{figure}

\begin{figure}[htp]
	\centering
     	\subfigure[$n_0=6$, $n_1=8$]{
		\label{cp68}
		\includegraphics[width=0.3\textwidth]{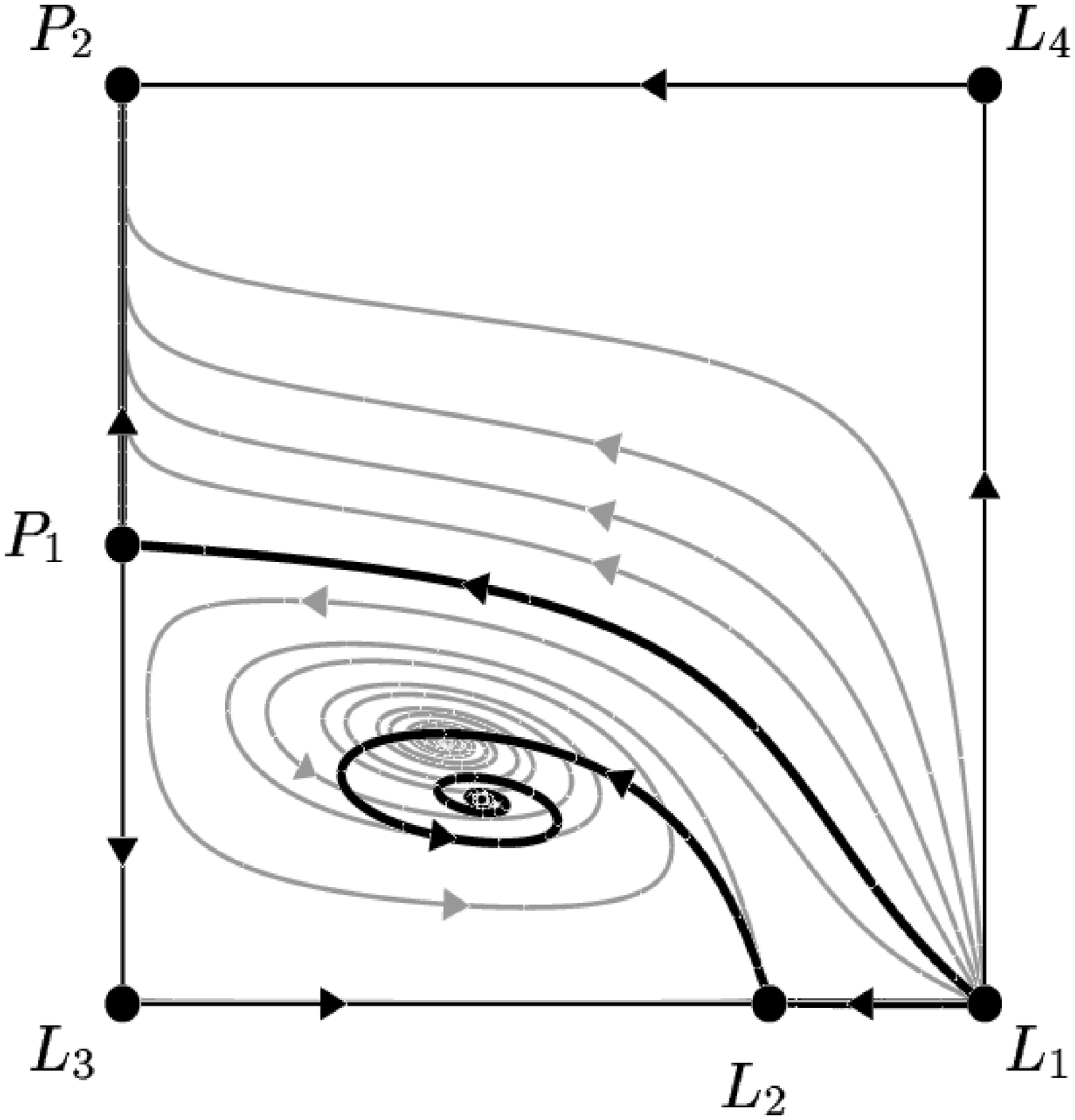}}\quad
        \subfigure[$n_0=6$, $n_1=4$]{
		\label{cp64}
		\includegraphics[width=0.3\textwidth]{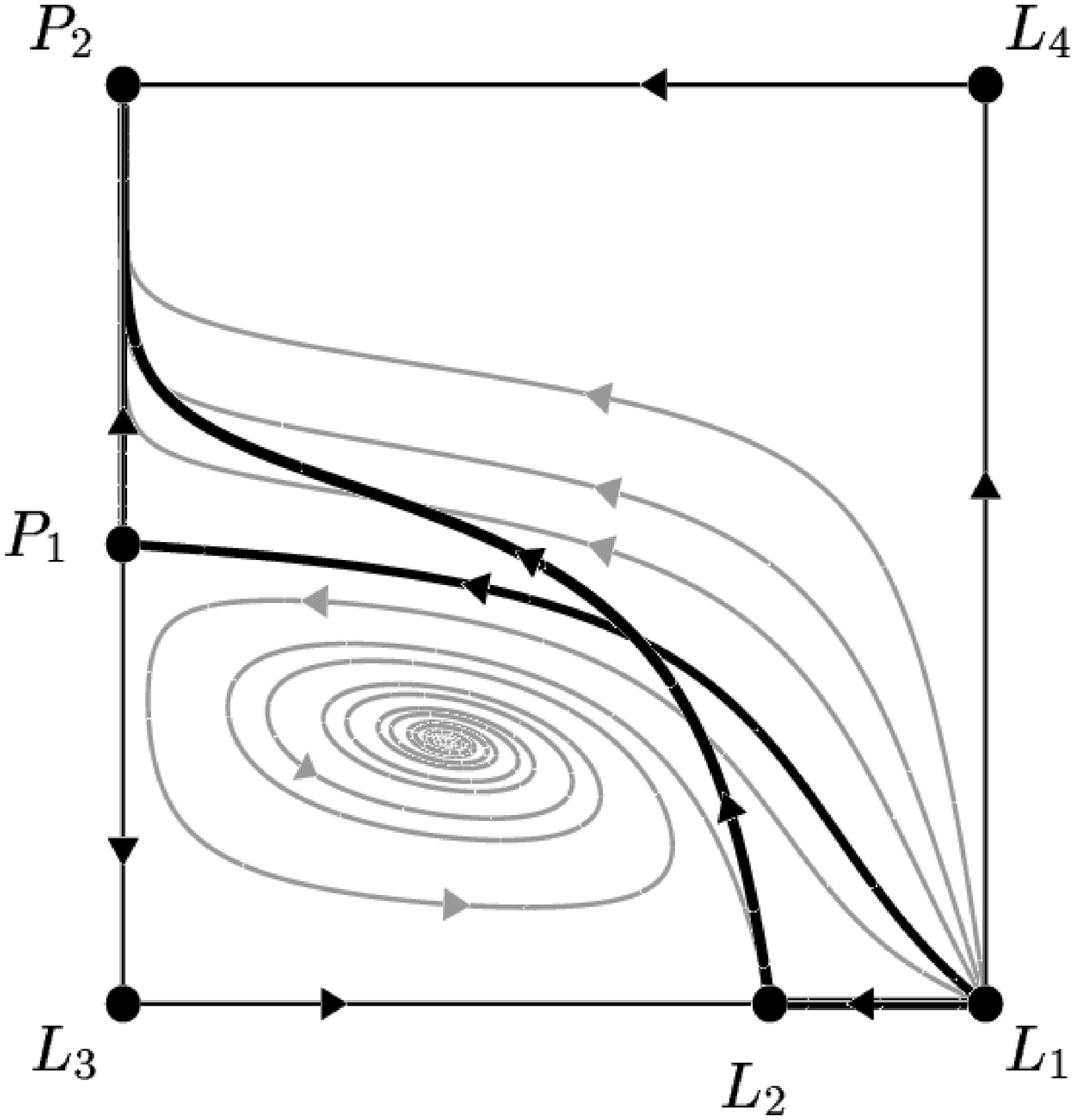}}
	\caption{Composite equation of state 
		$n(\eta) = n_0$ for $\eta \leq \eta_{\mathrm{j}}$ and
		$n(\eta) = n_1$ for $\eta>\eta_{\mathrm{j}}$. $n_0 = 6$ for
		(a) and (b), $n_1$ varies.
		Orbits projected down to $\Omega=0$. 
		Compare with Figure~\ref{compositeflow}.}
    \label{compositeflow2}
\end{figure}

%\begin{figure}[htp]
%	\centering
%	\includegraphics[width=0.6\textwidth]{figure9.ps}
%	\caption{Composite equation of state 
%		$n(\eta) = n_0$ for $\eta \leq \eta_{\mathrm{j}}$ and
%		$n(\eta) = n_1$ for $\eta>\eta_{\mathrm{j}}$. $n_0 = 6$ for
%		(a) and (b), $n_1$ varies.
%		Orbits projected down to $\Omega=0$. 
%		Compare with Figure~\ref{compositeflow}.}
%    \label{compositeflow2}
%\end{figure}

We observe that when $n_0$ is chosen between $3 < n_0 < 5$ 
and $n_1 > 5$, a subset of 
the following possible scenarios occur:
\begin{itemize}
\item 
The two spirals do not intersect and thus every regular solution has a finite mass and radius. 
See, e.g., Figure~8(a).
\item 
Finitely many solutions with infinite radius - but with 
finite mass - are embedded into a continuum of finite solutions.
Compare with, e.g., Figure~8(b).
\item
A continuum of finite solutions contains
one solution with infinite radius and mass. This solution
represents the accumulation point for an infinite number
of solutions with infinite radii and finite masses.
This happens if the (projection of the) $n_1$-spiral passes through $P_3$.
See, e.g., Figure~8(c).
\item
Infinitely many solutions with infinite radii and finite masses
are embedded into a continuum of finite solutions, but without
an accumulation.
This is the case if $P_6$ possesses $(U,Q)$ coordinates lying
on the $n_0$-spiral.
\item
For a certain special choice of $n_0, n_1$ the last two scenarios
can occur simultaneously.
\end{itemize}

Using relations describing the polytropic spirals like~(\ref{censpiral}), 
it is possible to give a qualitative answer to the question  
which central values yield infinite solutions.

For composite equations of state with $n_0>5$ 
the following is possible:
\begin{itemize}
\item 
Every regular solution has an infinite radius and mass (Figure~9(a)).
\item 
There exists one solution
that has infinite radius, but finite mass.
This solution separates a continuum of solutions with infinite radii
and masses from a continuum of solutions with finite radii and masses.
See, e.g., Figure~9(b).
\end{itemize}

\begin{figure}[htp]
	\psfrag{v}[cc][cc]{{\scriptsize $\hat{\eta}_c$}}
	\psfrag{M}[lc][cc][1][-90]{{\scriptsize ${\mathcal M}$}}
	\psfrag{R}[lc][cc][1][-90]{{\scriptsize ${\mathcal R}$}}
	\centering
        \subfigure[$n_0=4$, $n_1=6.2$]{
		\label{462R}
		\includegraphics[width=0.4\textwidth]{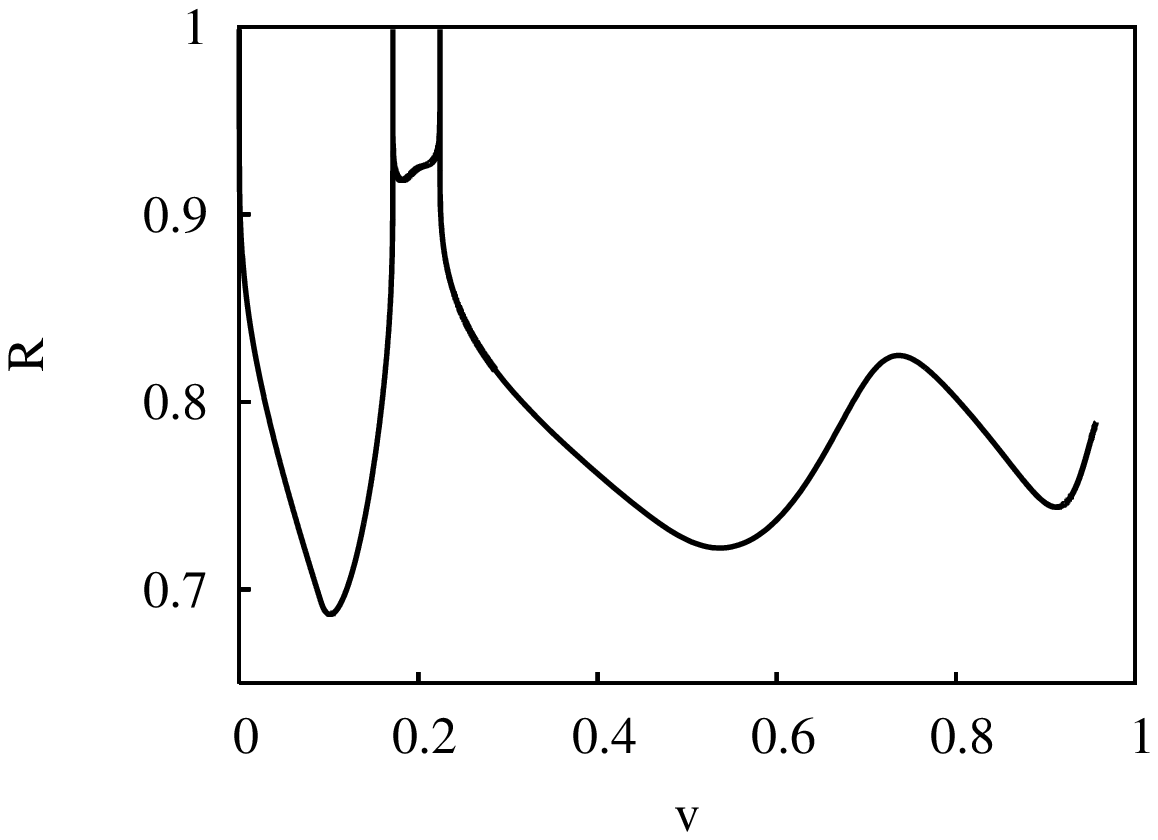}}\quad
     	\subfigure[$n_0=4$, $n_1=6.2$]{
		\label{462M}
		\includegraphics[width=0.4\textwidth]{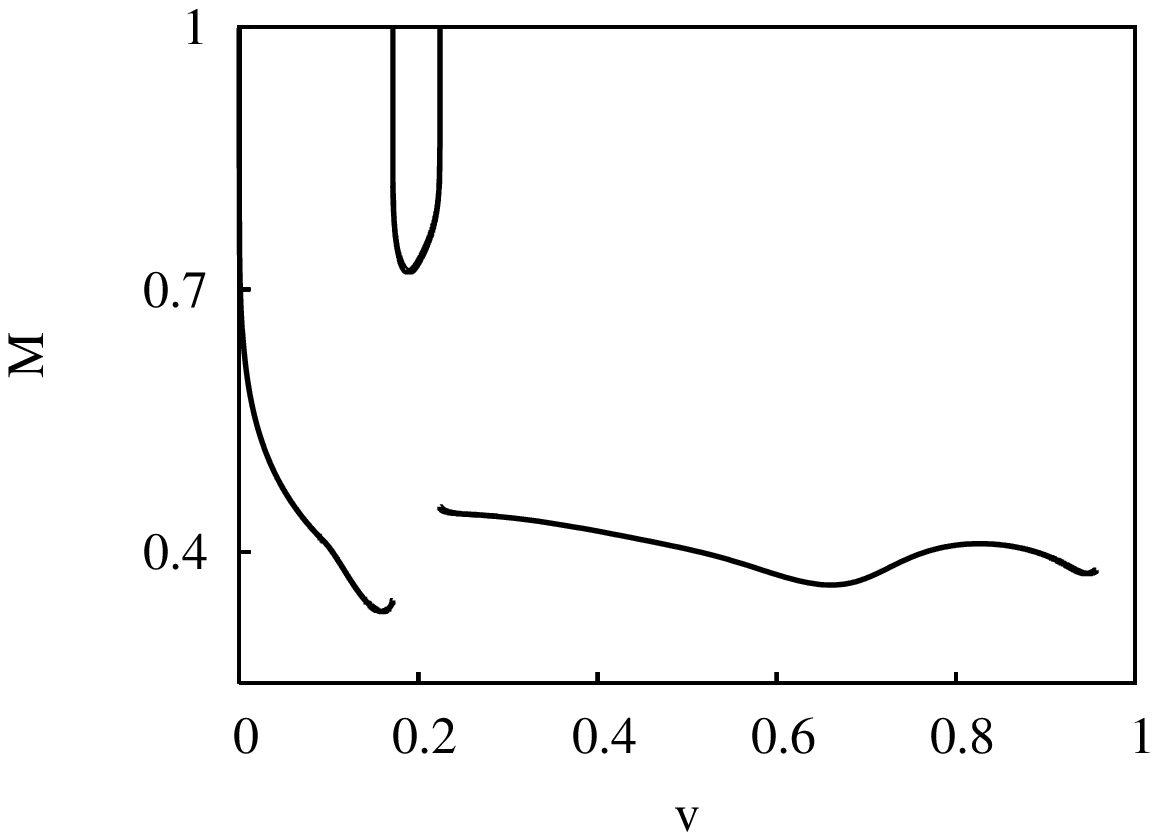}}\quad
	\subfigure[$n_0=4$, $n_1=5.88$]{
		\label{4588RL}
		\includegraphics[width=0.4\textwidth]{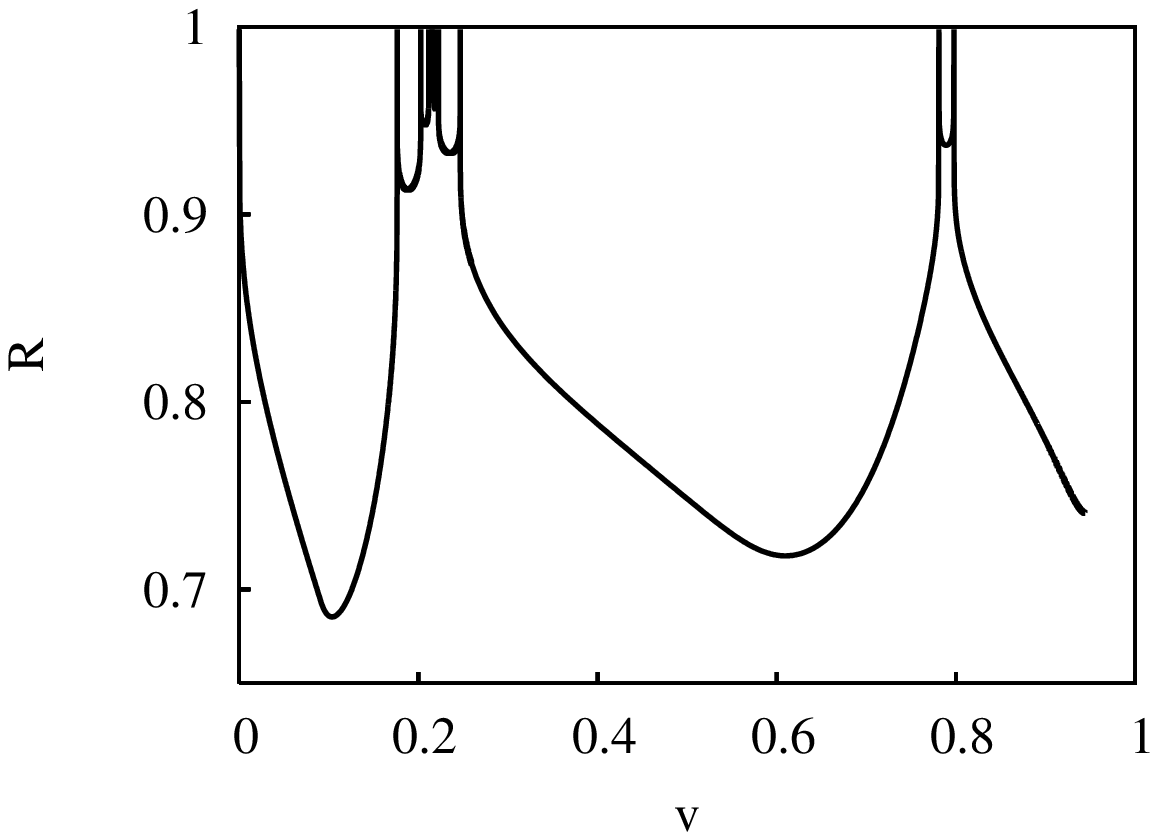}}\quad
	\subfigure[$n_0=4$, $n_1=5.88$]{
		\label{4588RS}
		\includegraphics[width=0.4\textwidth]{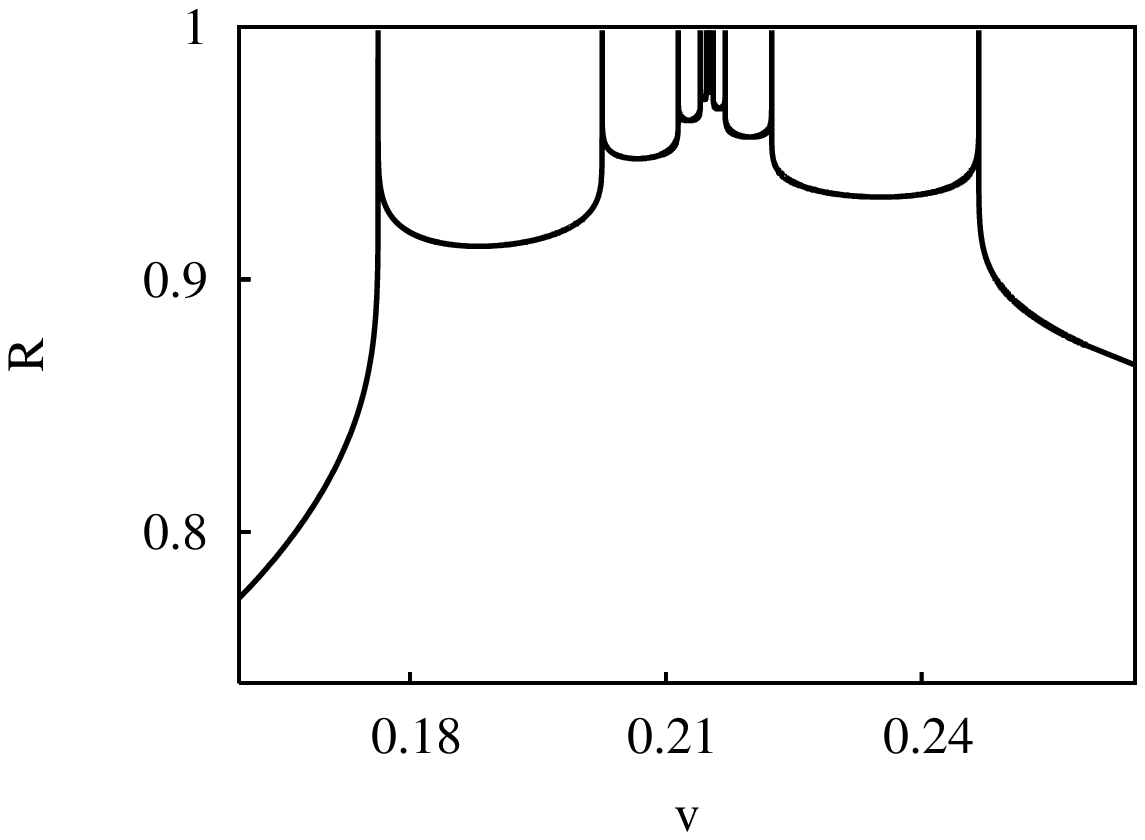}}\quad
	\subfigure[$n_0=4$, $n_1=5.88$]{
		\label{4588RXS}
		\includegraphics[width=0.4\textwidth]{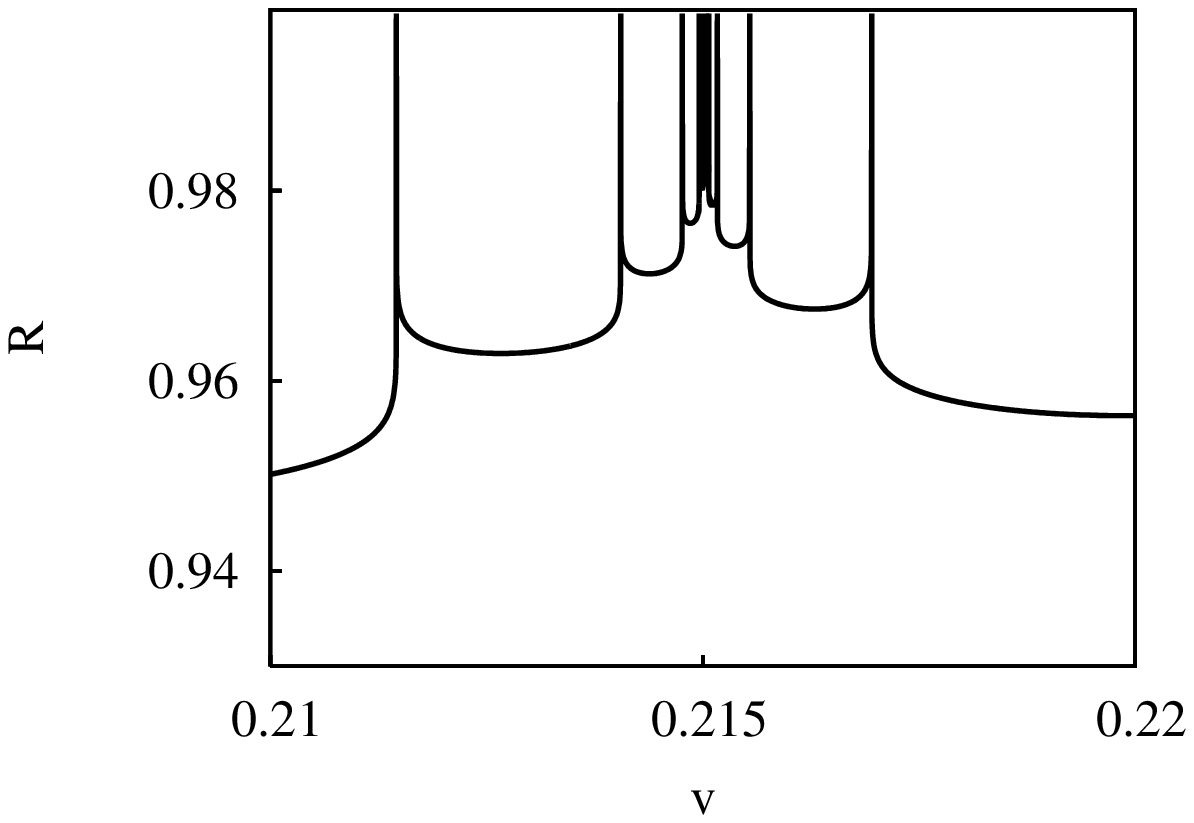}}\quad
	\subfigure[$n_0=4$, $n_1=5.88$]{
		\label{4588MS}
		\includegraphics[width=0.4\textwidth]{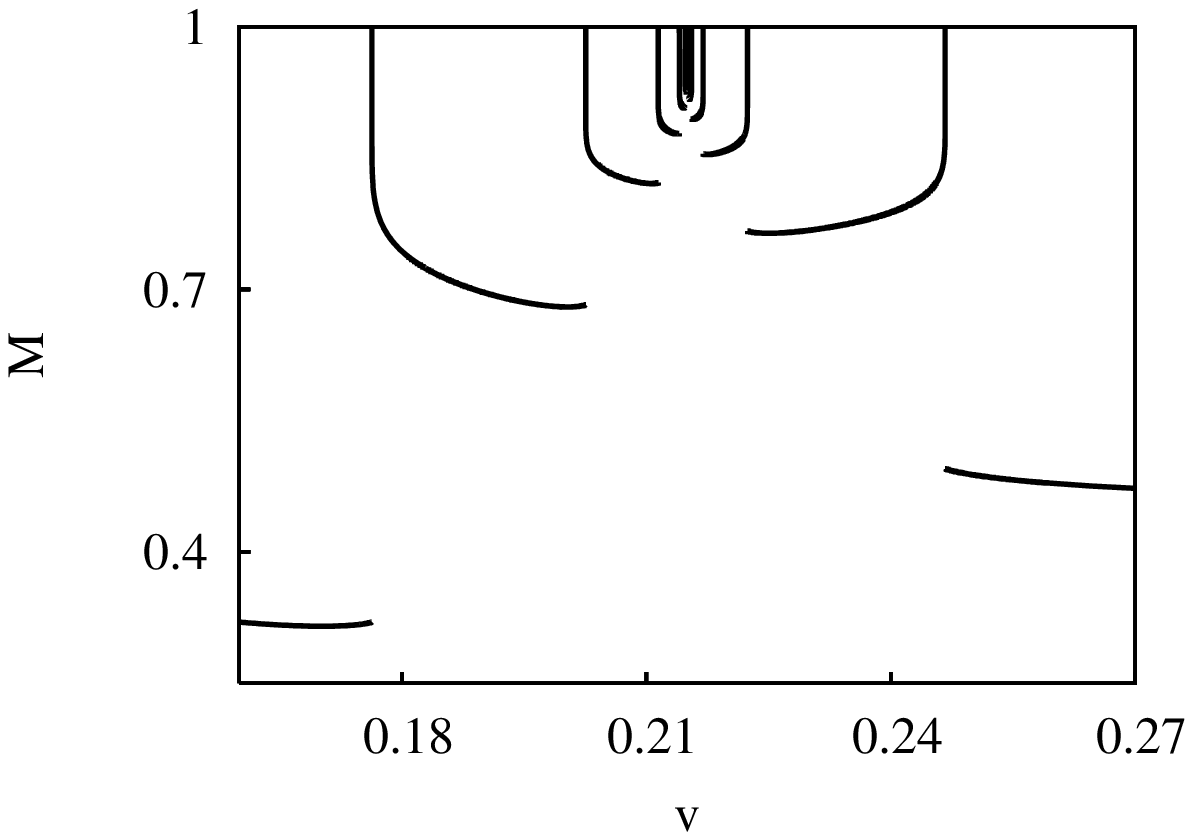}}\quad
	\caption{Composite equation of 
		state $n(\eta) = n_0$ for $\eta \leq \eta_{\mathrm{j}}$ and
		$n(\eta) = n_1$ for $\eta>\eta_{\mathrm{j}}$. $n_0 = 4$ and $n_1 = 6.2$ or
		$n_1 \approx 5.88$.
		The diagrams display the radius $R$ and the total mass $M$
		of regular solutions as functions of the central value $\eta_c$.
		For $n_1 =6.2$, $R$ becomes infinite for two values of $\eta_c$, 
		while $M$ is finite, although the mass can become arbitrarily large. 
		For $n_1 \approx 5.88$, we observe an infinite number of solutions
		with infinite radii and finite masses. These solutions
		accumulate at a certain value of $\eta_c$, where
		both $R$ and $M$ are infinite. By comparison with Figure~8(b)
		and~8(c) the qualitative features of the $R$- and $M$-diagrams
		can easily be explained. In the diagrams bounded quantities are used:
		$\hat{\eta}_c = (\eta_c/10)(1+\eta_c/10)^{-1}$,
		${\mathcal M} = \log(1+M) (1+\log(1+M))^{-1}$, and
		${\mathcal R} = \log(1+R) (1+\log(1+R))^{-1}$. Note that
	        ${\mathcal R}=1$ corresponds to an infinite radius.}
	\label{compositeMandRdiag}
\end{figure}

\begin{figure}[htp]
	\psfrag{M}[cc][cc]{{\scriptsize ${\mathcal M}$}}
	\psfrag{R}[lc][cc][1][-90]{{\scriptsize ${\mathcal R}$}}
	\centering
        \subfigure[$n_0=4$, $n_1=6.2$]{
		\label{462MR}
		\includegraphics[width=0.45\textwidth]{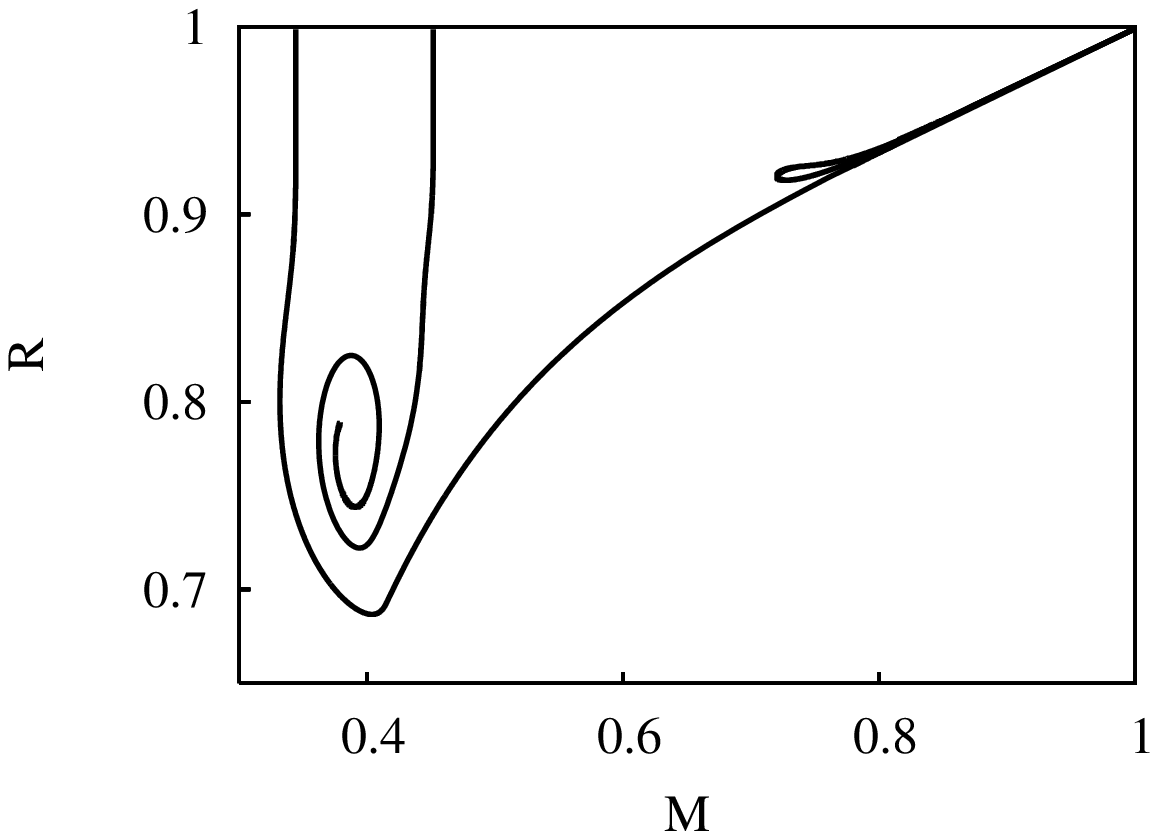}}\quad
	\subfigure[$n_0=4$, $n_1=5.88$]{
		\label{4588MR}
		\includegraphics[width=0.45\textwidth]{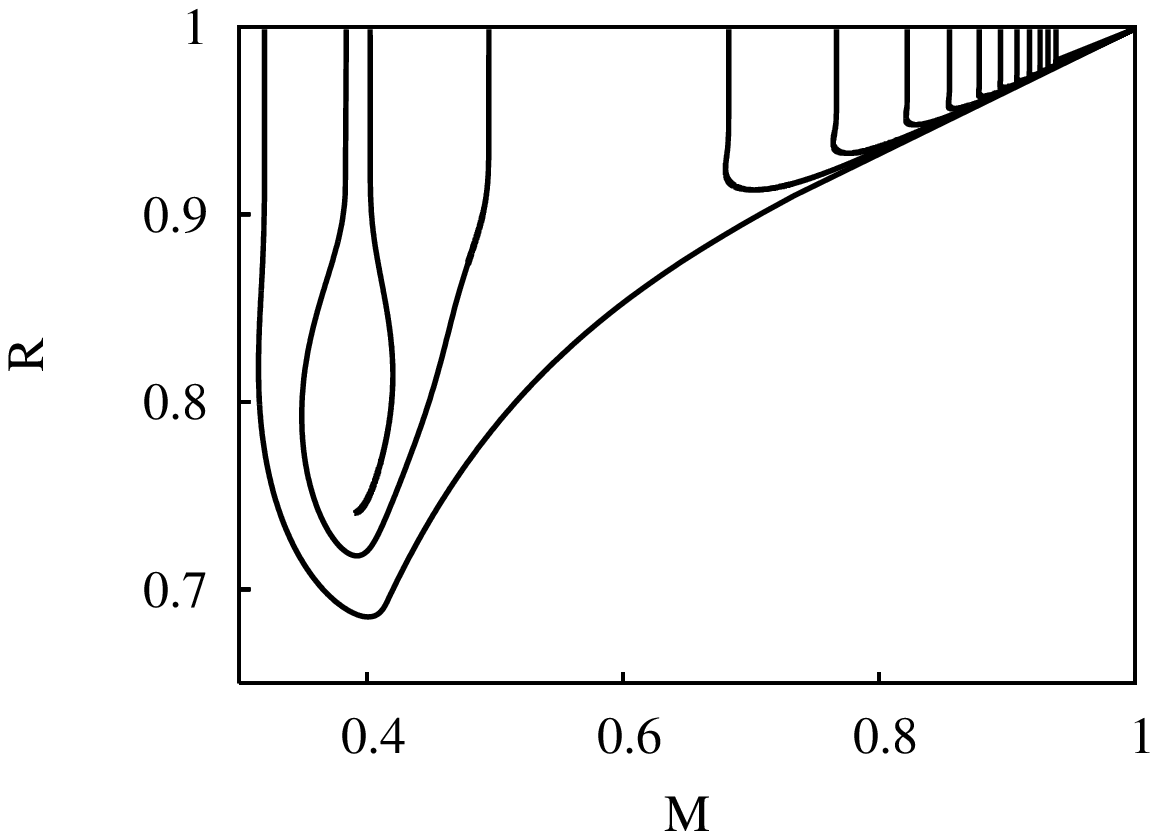}}\quad
	\caption{Mass-radius diagrams for composite equations of 
		state, i.e., $n(\eta) = n_0$ for $\eta \leq \eta_{\mathrm{j}}$ and
		$n(\eta) = n_1$ for $\eta>\eta_{\mathrm{j}}$. 
		For $n_0 = 4$, $n_1 = 6.2$ the diagram can be understood as
		follows: For small $\eta_c$
		$(M,R)$ is located in the right upper corner of
		the diagram. With increasing $\eta_c$ a point 
		$(k(M),k(R))\approx (0.35,1)$ is reached, then there is a
		jump to a small loop graph in the right upper corner.
		Having passed through the loop there is a jump to a 
		point $(k(M),k(R))\approx (0.45,1)$, 
		and subsequently the curve goes down until 
		it ends up in a $(M,R)$-spiral (not completely 
		represented in the diagram).
		For $n_1 \approx 5.88$ the $(M,R)$-diagram becomes
		more complicated. As is only indicated, 
		in the upper right corner an infinite	
		number of small graphs accumulate. 
		In the diagrams bounded quantities are used:
		${\mathcal M} = \log(1+M) (1+\log(1+M))^{-1}$ and
		${\mathcal R} = \log(1+R) (1+\log(1+R))^{-1}$.}
	\label{compositeMRdiag}
\end{figure}

As seen from Figures~\ref{compositeMandRdiag} and~\ref{compositeMRdiag}, 
the mass-radius diagrams can become quite complicated 
already for this simple class of equations of state.
However, by juxtaposing Figure~\ref{compositeMandRdiag} 
and Figure~\ref{compositeflow} the qualitative 
features of the mass-radius diagrams can easily be explained.

Note that it is easy to construct smooth equations of state
displaying similar features. One could, e.g., study 
equations of state of the Chandrasekhar type~(\ref{chandraindex})
with general constants. One can also choose to use some suitable 
regularization at the jump to obtain solutions, with a smooth equation of state, 
that display similar features.

%%%%%%%%%%%%%%%%%%%%%%%%%%%%%%%%%%%%%%%%%%%%%%%%%%%%%%%%%%%%%%%%%%%%%%%%
%%%%%%%%%%%%%%%%%%%%%%%%%%%%%%%%%%%%%%%%%%%%%%%%%%%%%%%%%%%%%%%%%%%%%%%%
%%%%%%%%%%%%%%%%%%%%%%%%%%%%%%%%%%%%%%%%%%%%%%%%%%%%%%%%%%%%%%%%%%%%%%%%
%%%%%%%%%%%%%%%%%%%%%%%%%%%%%%%%%%%%%%%%%%%%%%%%%%%%%%%%%%%%%%%%%%%%%%%%
\section{Qualitative results}
\label{qualitativeresults}
%%%%%%%%%%%%%%%%%%%%%%%%%%%%%%%%%%%%%%%%%%%%%%%%%%%%%%%%%%%%%%%%%%%%%%%%

We start this section by deriving a number of inequalities that give 
information about the solution space for various equations of state.

For regular solutions it is clear that $\rho(r) \leq {\bar\rho}(r)$, 
where the average density ${\bar \rho}$ is defined by
${\bar \rho} = 3m/(4\pi r^3)$, if $dp/d\rho \geq 0$ (or equivalently
$n(\Omega) \geq 0$). In  our approach this inequality translates to the
simple statement that $U\leq3/4$. This follows from the fact that
$u=3\rho/{\bar \rho}$. An alternative proof is obtained by first
noting that $dU/d\lambda = -3n(\Omega)Q/64$ on the $U=3/4$ surface,
and that $dU/d\lambda<0$ if $U>3/4$ and $n(\Omega) \geq 0$. We then
note that the eigenvector for the transversally hyperbolic set $L_2$
has the components $(-3n(\Omega_c)/40,2)$ in the $(U,Q)$-plane. Hence
regular orbits start on the inside of the surface or on it if
$n(\Omega) = 0$; in the latter case they continue on the surface as
long as $n(\Omega)=0$. Thus, because of the inequality, it follows
that they cannot pass into the region where  $U>3/4$.

Another inequality follows from that $\rho_c \geq {\bar \rho} \geq
\rho$, when $dp/d\rho \geq 0$ ($n(\Omega) \geq 0$), and the definition
of $U$. It is given by $\rho(\Omega)/\rho_c \leq
U/(3(1-U))$. Combining this with  equation (\ref{UQOmegaaux}) yields
\begin{subequations}\label{ineq}
\begin{align}
\label{rad}
  r^2 & \geq \frac{3}{4\pi}\left(\frac{Q}{1-Q}\right)
  \left(\frac{\Omega}{1-\Omega}\right)^{1/a}\rho_c^{-1} \\
\label{mass}
  m^2 & \geq \frac{3}{4\pi}\left(\frac{Q}{1-Q}\right)^3\left
(\frac{\Omega}{1-\Omega}\right)^{3/a}\rho_c^{-1}\ .
\end{align}
\end{subequations}

As another example, which will be used in proofs of theorems below,
consider a polytrope, characterized by $n$, and another equation of state,
characterized by the index function $n(\Omega)$, such that $n > n(\Omega)$ 
(or $n < n(\Omega)$) for $0 \leq \Omega \leq \Omega_{\rm max}$. 
Now consider a 1-parameter set of solutions of the polytrope,
e.g, the regular solution set for which 
$0 \leq \Omega_c \leq \Omega_{\rm max}$. Since the 
equation for $\Omega$ decouples in the polytropic case, 
it follows that the polytropic invariant set is described by a particular 
surface $\Psi(U,Q) = 0$ in the state space
$U,Q,\Omega$, and that
$d\Psi/d\lambda = (\partial \Psi/\partial U)(dU/d\lambda) +  (\partial
\Psi/\partial Q)(dQ/d\lambda) = 0$ on the surface.
We restrict ourselves to subsets that satisfy $dU/d\lambda <0$ 
and $dQ/d\lambda > 0$ in the interior state space. The regular
subsets for polytropes with $n\leq 5$ are particularly interesting
subsets of this type  (this will be shown below; an intuitive picture
of what types of sets that satisfy this requirement can be obtained by
considering Figure~\ref{base}). From the above it follows
that  $(\partial \Psi/\partial U)$ and $(\partial \Psi/\partial Q)$
have the same (non-zero) sign. We can always choose $\Psi$ so
that the sign of $(\partial \Psi/\partial U)$ is positive.  

Let us now
consider the described "polytropic surface" $\Psi(U,Q,\Omega) = \Psi(U,Q) = 0$
in the state space of the other equation of state. An evaluation of
the derivative of $\Psi$ {\it on the surface\/} yields
\begin{eqnarray}\label{polineq}
\nonumber
\left(\frac{d\Psi}{d\lambda}\right)_{n(\Omega)} & = & 
 \left(\frac{d\Psi}{d\lambda}\right)_{n(\Omega)} - 
 \left(\frac{d\Psi}{d\lambda}\right)_n = \\ 
& =  & \left(\frac{\partial \Psi}{\partial U}\right)
 \left(
 \left(\frac{dU}{d\lambda}\right)_{n(\Omega)} - 
 \left(\frac{dU}{d\lambda}\right)_n
 \right) + 
 \left(\frac{\partial \Psi}{\partial Q}\right)
 \left(
 \left(\frac{dQ}{d\lambda}\right)_{n(\Omega)} - 
 \left(\frac{dQ}{d\lambda}\right)_n
 \right) = \\ \nonumber
& =  & \left(\frac{\partial \Psi}{\partial U}\right)
 U(1-U)^2\,Q(n - n(\Omega))\, ,
\end{eqnarray}
where we have made use of a pointwise comparison of the two flows and 
where the one associated with $n$ is zero on the surface. 
The above quantity is positive (negative if $n < n(\Omega)$)
in the interior state space and acts as a "semi-permeable membrane"
for solutions with $0 \leq \Omega \leq \Omega_{\rm max}$. Note that
if $\Omega_{\rm max} = 1$, this surface will provide a restriction in
the state space for all solutions. As an example, we now consider
$n=5$ with $\Omega_{\rm max} = 1$, which plays a key role in several 
of the theorems concerning models with finite radii below.

As mentioned earlier (see~(\ref{n05mon})), 
the regular subset for $n=5$ is characterized by 
\begin{equation}\label{regn05}
  U = \frac{3(1 - 2Q)}{4 - 7Q}\ .
\end{equation}
Now define
\begin{equation}\label{Phidef}
  \Phi(U,Q,\Omega) = (4-7Q)U - 3(1-2Q)\ .
\end{equation}
We now consider the surface $\Phi(U,Q,\Omega) = 0$ inside 
the state space, and note that the surface has an outward pointing normal.
The flow through the surface is given by the derivative of $\Phi$ 
on the surface. Equation (\ref{UQOmega}) yields
\begin{equation}\label{dPhi} 
  \left(\frac{d\Phi}{d\lambda}\right)_{\rm surface} = 
  3 (1-2 Q) (1-U)^2\,Q (5 - n(\Omega))\ .
\end{equation}
Note that $Q \leq 1/2$ on the surface, and hence that the sign in the
interior state space is determined by $5 - n(\Omega)$, in accordance
with (\ref{polineq}). When $n(\Omega) < 5$ $(> 5)$ the derivative is
positive (negative) and the vector field is pointing "outwards"
("inwards"), as can be intuitively deduced by considering
Figure~\ref{base} when one superimposes the regular orbit for $n=5$ onto
the figures for the other polytropic indices. We also note that by
continuity the equilibrium point $P_3$ $(P_6)$ is inside the
intersection of the surface with the $\Omega = 0$ $(\Omega = 1)$
plane. Thus for example, it follows from this inequality, together
with the eigenvector properties associated with $L_2$, that the regular surface
of $n(\Omega) <5$ $(>5)$ is "outside" ("inside") the
regular "$n=5\,$-surface" (once orbits are outside (inside), which they are in the 
neighborhood of $L_2$, they are prevented by the inequality to pass through the
surface).

Moreover, since it follows from equation~(\ref{uqomeq}) that
$dU/d\lambda<0, dQ/d\lambda > 0$, if $n(\Omega) \leq 5$, outside the
regular $n=5$-surface, the above "outside" statement is also true for
$n$ if $5 \geq n > n(\Omega)$.

Let us now list a number of theorems, relating equations of state to
properties concerning total masses and radii of various solutions
(primarily the regular ones). 
For the proofs (which are mostly given in Appendix~\ref{A})
we can make use of methods provided by dynamical systems theory.
This is a considerable advantage compared to the cumbersomeness
of several proofs in the literature; for theorems
of a similar type as the theorems below see, 
e.g., \cite{art:Heinzle2002}, \cite{Rendall/Schmidt:1991}, 
\cite{art:Makino1984}, \cite{Makino:1998}, \cite{art:Makino2000}.
In addition to our unified approach to already known statements we will 
also prove new results.

\begin{theorem}\label{n0leq3} (Finiteness of perfect fluid solutions).
Consider an asymptotically polytropic equation of state 
$\rho(p) = K  p^{n_0/(n_0+1)}\, ( 1 + O(p^{a_0/(n_0 + 1)}) )$ 
($p \rightarrow 0$) with asymptotic index $n_0 \leq 3$.
Then any corresponding perfect fluid solution, either regular
or non-regular, has finite mass and radius. 
\end{theorem}

\proof 
We distinguish two cases, $n_0<3$ and $n_0=3$.  For $n_0<3$ the
proof is trivial. According to Theorem~\ref{fix} only fixed points
can be attractors on $\Omega=0$ in this case. Combining this with
Table~\ref{attractors}, it follows that there exists a unique
attractor for $n_0<3$, namely $L_5$ for $n_0=0$ and $P_2$ for
$0<n_0<3$. These fixed points correspond to solutions with finite
radii and masses (compare with~(\ref{l5}) and~(\ref{finitesols})).

In the case $n_0=3$ the argument must necessarily also involve $P_1$, which
is no longer hyperbolic, but possesses a 1-dimensional center subspace.
Therefore it cannot be excluded that $P_1$ is an endpoint of a family of 
orbits that comes from the interior of the state space. 
However, in Appendix~\ref{appendix:attractorsinthestatespace} 
it is shown that this is not the case.
Consequently, it follows that all orbits end at $P_2$, and hence 
Theorem~\ref{n0leq3} is established.
\proofend

\begin{theorem}\label{nless5} 
Consider an asymptotically polytropic equation of state $\rho(\eta)$
with index-function $n(\eta)\leq 5$ $\forall\eta$, $n(\eta)<5$
for arbitrarily small $\eta$.
Then the regular perfect fluid solutions, and
the solutions that possess a negative mass singularity, have
finite masses and radii.
\end{theorem}

\proof
Equation~(\ref{Phidef}) defines the surface of polytropic 
regular solution of index 5. It follows from the properties of
the eigenvectors of the transversally hyperbolic line $L_2$ and 
from~(\ref{dPhi}) that this surface shields the regular solutions, 
for any $n(\Omega) \leq 5$,
from the fixed point $P_3$ (when it exists). Taken together with 
Theorem~\ref{fix} and the properties of $P_1$,
it follows that $P_2$ if $n_0\neq 0$, and $L_5$, if $n_0=0$, 
are the only limit sets when $\lambda \rightarrow \infty$ for the 
regular solutions, and thus that they have finite radii and masses.
\proofend

\begin{theorem}\label{n0bigger3}
For any $n_0>3$ there exist
asymptotically polytropic equations of state (having $n_0$ as
their asymptotic index), such that some of the regular 
perfect fluid solutions possess infinite radii.
\end{theorem}

\proof
When $n_1 = 5 + \epsilon$, for some small $\epsilon$, the regular subset 
for $n_1$ intersects the surface associated with orbits that end at $P_1$ 
for an equation of state with polytropic index $n_0$. Hence one can 
obtain an example displaying the desired feature 
by constructing a composite equation of state. 
For further discussion, and a picture of solutions exhibiting this type of 
property, see Section~\ref{compositeequationsofstate} and 
Figure~\ref{compositeflow}.
\proofend

Hence Theorem~\ref{n0leq3} is sharp. From Theorem~\ref{nless5} we know that
an equation of state that yields regular solutions with infinite radii has to 
satisfy $n(\Omega) > 5$ in some range for $\Omega$. However, this condition 
is not sufficient, since there exists some equations of state that satisfy
$n(\Omega) > 5$ in some range for $\Omega$, for which all regular solutions 
have finite radii. We refer to the examples given above for composite
equations of state.

The dynamical systems formulation is also a powerful tool to
investigate the qualitative properties of the mass-radius diagram.
A spiral structure of the $(M,R)$-diagram has been observed 
in several situations and its existence has been proved
rigorously in some cases (see, e.g.,~\cite{art:Makino2000}).
The following theorem formulates a criterion when
such a structure occurs and gives information about
its concrete form.

\begin{theorem}\label{spiralthm} (Spiral structure of the $(M,R)$-diagram).
Consider an asymptotically polytropic equation of state
with asymptotic indices $0< n_0 \leq 3$ and $n_1$.
Then the mass-radius relation for high central pressures 
possesses a spiral structure, if and only if $n_1 > 5$
with the spiral given by
\begin{equation}\label{spirals}
\left(\begin{array}{c}
R(\eta_c) \\
M(\eta_c)
\end{array}\right) \:=\:
\left(\begin{array}{c}
R_O \\
M_O
\end{array}\right) 
\:+\:
\left(\frac{1}{\eta_c}\right)^{\gamma_1} \: {\mathcal B} \: 
{\mathcal J}(\gamma_2 \log\frac{1}{\eta}\,)\: b 
\:+\:
o(\left(\frac{1}{\eta_c}\right)^{\gamma_1}) \:,
\end{equation}
where $R_O$ and $M_O$ are constants, ${\mathcal B}$ is a non-singular
matrix, and $b$ a non-zero vector.
The matrix ${\mathcal J}(\varphi) \in \mathrm{SO}(2)$ describes a rotation by an angle
$\varphi$, and the constants $\gamma_1$ and $\gamma_2$ are given by
\begin{equation}\label{gamma12}
\gamma_1 = (n_1 -5) \,\frac{1}{4} \qquad 
\gamma_2 = \frac{1}{4}\, \sqrt{b}\:.
\end{equation}
\end{theorem}
Recall from Table~\ref{tab:UQcube} that $\sqrt{b} = \sqrt{-1 -22 n_1 + 7 n_1^2}$.

\proof
See Appendix~\ref{A}. 
\proofend

\begin{theorem}\label{MR2nd} (Polytropic behavior of the $(M,R)$-diagram).
Consider an asymptotically polytropic equation of state with asymptotic indices
$0<n_0$, and $0<n_1<5$. To first order, 
the mass-radius relation for high central
pressures is approximated by the mass-radius relation for an exact polytrope
with polytropic index $n_1$, i.e.,
\begin{equation}\label{highdenMR}
R(\eta_c) = R_{\mathrm{p}:n_1}(\eta_c)\, \zeta[n(\eta)] \qquad
M(\eta_c) = M_{\mathrm{p}:n_1}(\eta_c)\, \zeta[n(\eta)] \:,
\end{equation}
where $R_{\mathrm{p}:n_1}(\eta_c)$ and
$M_{\mathrm{p}:n_1}(\eta_c)$ are the radius and 
the mass for an exact polytrope
with index $n_1$ (compare with~(\ref{MRpoly})), i.e.,
\begin{equation}\label{MRpoly2}
R_{\mathrm{p}:n_1}(\eta_c) = c_R(n_1) \, \eta_c^{1-n_1} \qquad
M_{\mathrm{p}:n_1}(\eta_c) = c_M(n_1) \, \eta_c^{3-n_1}\:,
\end{equation}
and $\zeta$ is a functional acting on the index-function $n(\eta)$, given by
\begin{equation}\label{zeta}
\zeta[n(\eta)] = \exp\left(\frac{1}{2} \,
\int\limits_0^1 \frac{n(\eta)-n_0}{\eta} \, d\eta + \frac{1}{2}\,
\int\limits_1^\infty \frac{n(\eta)-n_1}{\eta} \, d\eta \right)\:.
\end{equation}
\end{theorem}

\proof
See Appendix~\ref{A}.
\proofend

%%%%%%%%%%%%%%%%%%%%%%%%%%%%%%%%%%%%%%%%%%%%%%%%%%%%%%%%%%%%%%%%%%%%%%%%
%%%%%%%%%%%%%%%%%%%%%%%%%%%%%%%%%%%%%%%%%%%%%%%%%%%%%%%%%%%%%%%%%%%%%%%%
%%%%%%%%%%%%%%%%%%%%%%%%%%%%%%%%%%%%%%%%%%%%%%%%%%%%%%%%%%%%%%%%%%%%%%%%
%%%%%%%%%%%%%%%%%%%%%%%%%%%%%%%%%%%%%%%%%%%%%%%%%%%%%%%%%%%%%%%%%%%%%%%%
\section{Concluding remarks}
\label{concludingremarks}
%%%%%%%%%%%%%%%%%%%%%%%%%%%%%%%%%%%%%%%%%%%%%%%%%%%%%%%%%%%%%%%%%%%%%%%%

In this paper we studied static spherically symmetric perfect 
fluid models in Newtonian gravity with asymptotically polytropic barotropic 
equations of state. We developed a dynamical systems formulation, 
which yielded both a pictorial representation of the solution spaces,
and a tool making it possible to derive theorems.

The framework was also used for numerical purposes, and
exhibited certain advantages, in particular in connection with
bifurcation phenomena. Such bifurcations cause numerical 
problems, which can be circumvented by use of approximate 
solutions near fixed points.

In the paper, the approach is to some extent tailored to the family of
equations of state that are considered. However, the underlying 
principles have a broader range of applicability -- in the present 
context of Newtonian perfect fluids, but also in other areas.

Let us first comment on the connection between the choice of 
variables and the asymptotic features of the equation of state. 
The present choice of the variable 
$\omega=\eta^a$, in~(\ref{uqomeq}),
is adapted to the asymptotic features, $n(\eta) = n_0 + O(\eta^{a_0})$,
$n(\eta) = n_1 + O(\eta^{-a_1})$, when $\eta \rightarrow 0$ and
$\eta\rightarrow \infty$.
However, one can similarly adapt 
to equations of state with quite general asymptotic behavior: 
assume that $n(\eta)$ behaves like
$n(\eta)=n_0 + \nu_0(\eta)$ and $n(\eta)=n_1 + \nu_1(\eta)$
(where $\nu_0(\eta)$ and $\nu_1(\eta)$ can go to zero
in different, arbitrary ways).
Choose $\omega = f(\eta)$, where $f(\eta)$ can be 
any smooth, monotonically 
increasing function adapted to the asymptotics of $n(\eta)$, i.e.,
a suitable choice of $f(\eta)$ is one where $f(\eta)$ asymptotically 
satisfies $f(\eta) = \nu_0(\eta)^k$ ($k<1$) and $f(\eta)=\nu_1(\eta)^{-k}$.
By this choice $n(\omega)$ becomes a differentiable function.
Using the variables $(u,q,\omega)$ we obtain again~(\ref{uqomeq}),
where the $d\omega/d\xi$ equation now
reads $d\omega/d\xi = -q ( \eta \,\frac{df}{d\eta})(f^{-1}(\omega))$.
Thus the system is similar to~(\ref{uqomeq}),
and allows similar treatment. 

Static spherically symmetric perfect fluid models 
are also of interest in General Relativity.
Relativistic stellar models with a variety of equations of state
have been studied in some earlier papers, see, e.g,~\cite{Rendall/Schmidt:1991}, 
\cite{Makino:1998}, \cite{art:Makino2000}, \cite{art:grstar_pol}. 
Several of the new ideas presented in this paper should
also be useful in the general relativistic context, 
and might yield new results in that area. Note that in the low pressure limit 
the relativistic equations approximate the Newtonian equations, and thus the 
fixed point analysis of a relativistic dynamical system
partly resemble the one presented in this paper, 
at least in the low pressure region (compare with~\cite{art:grstar_pol}).
Moreover, it might be possible to derive relativistic inequalities 
by adapting the idea of using certain Newtonian polytropic solutions to divide 
state spaces into tractable regions.

It is also possible to go beyond the perfect fluid assumption.
In a future paper we will consider the Vlasov-Poisson
equations for a self-gravitating collisionless gas.
The corresponding dynamical systems formulation 
displays many similarities with the present situation, 
although new features also occur.

Finally, it might be fruitful to go beyond 
the static and spherically symmetric assumptions and use the 
dynamical systems approach as a starting point for perturbation 
theory.

%%%%%%%%%%%%%%%%%%%%%%%%%%%%%%%%%%%%%%%%%%%%%%%%%%%%%%%%%%%%%%%%%%%%%%%%
%%%%%%%%%%%%%%%%%%%%%%%%%%%%%%%%%%%%%%%%%%%%%%%%%%%%%%%%%%%%%%%%%%%%%%%%
%%%%%%%%%%%%%%%%%%%%%%%%%%%%%%%%%%%%%%%%%%%%%%%%%%%%%%%%%%%%%%%%%%%%%%%%
%%%%%%%%%%%%%%%%%%%%%%%%%%%%%%%%%%%%%%%%%%%%%%%%%%%%%%%%%%%%%%%%%%%%%%%%

%%%%%%%%%%%%%%%%%%%%%%%%%%%%%%%%%%%%%%%%%%%%%%%%%%%%%%%%%%%%%%%%%%%%%%%%
%%%%%%%%%%%%%%%%%%%%%%%%%%%%%%%%%%%%%%%%%%%%%%%%%%%%%%%%%%%%%%%%%%%%%%%%
%%%%%%%%%%%%%%%%%%%%%%%%%%%%%%%%%%%%%%%%%%%%%%%%%%%%%%%%%%%%%%%%%%%%%%%%
%%%%%%%%%%%%%%%%%%%%%%%%%%%%%%%%%%%%%%%%%%%%%%%%%%%%%%%%%%%%%%%%%%%%%%%%
\subsection*{Acknowledgements} 
The authors would like to thank Alan Rendall and 
Robert Beig (J.M.H.) for helpful discussions. 
This work was supported by the DOC-program of the
{\it Austrian Academy of Sciences} (J.M.H.) 
and the {\it Swedish Research Council} (C.U.).
%%%%%%%%%%%%%%%%%%%%%%%%%%%%%%%%%%%%%%%%%%%%%%%%%%%%%%%%%%%%%%%%%%%%%%%

%%%%%%%%%%%%%%%%%%%%%%%%%%%%%%%%%%%%%%%%%%%%%%%%%%%%%%%%%%%%%%%%%%%%%%%%
%%%%%%%%%%%%%%%%%%%%%%%%%%%%%%%%%%%%%%%%%%%%%%%%%%%%%%%%%%%%%%%%%%%%%%%%
%%%%%%%%%%%%%%%%%%%%%%%%%%%%%%%%%%%%%%%%%%%%%%%%%%%%%%%%%%%%%%%%%%%%%%%%
%%%%%%%%%%%%%%%%%%%%%%%%%%%%%%%%%%%%%%%%%%%%%%%%%%%%%%%%%%%%%%%%%%%%%%%%

%%%%%%%%%%%%%%%%%%%%%%%%%%%%%%%%%%%%%%%%%%%%%%%%%%%%%%%%%%%%%%%%%%%%%%%%
%%%%%%%%%%%%%%%%%%%%%%%%%%%%%%%%%%%%%%%%%%%%%%%%%%%%%%%%%%%%%%%%%%%%%%%%
%%%%%%%%%%%%%%%%%%%%%%%%%%%%%%%%%%%%%%%%%%%%%%%%%%%%%%%%%%%%%%%%%%%%%%%%
%%%%%%%%%%%%%%%%%%%%%%%%%%%%%%%%%%%%%%%%%%%%%%%%%%%%%%%%%%%%%%%%%%%%%%%%

%%%%%%%%%%%%%%%%
\begin{appendix}
%%%%%%%%%%%%%%%%
%%%%%%%%%%%%%%%%%%%%%%%%%%%%%%%%%%%%%%%%%%%%%%%%%%%%%%%%%%%%%%%%%%%%%%%%
%%%%%%%%%%%%%%%%%%%%%%%%%%%%%%%%%%%%%%%%%%%%%%%%%%%%%%%%%%%%%%%%%%%%%%%%
%%%%%%%%%%%%%%%%%%%%%%%%%%%%%%%%%%%%%%%%%%%%%%%%%%%%%%%%%%%%%%%%%%%%%%%%
%%%%%%%%%%%%%%%%%%%%%%%%%%%%%%%%%%%%%%%%%%%%%%%%%%%%%%%%%%%%%%%%%%%%%%%%
\section{Proof of theorems}
\label{A}
%%%%%%%%%%%%%%%%%%%%%%%%%%%%%%%%%%%%%%%%%%%%%%%%%%%%%%%%%%%%%%%%%%%%%%%

%%%%%%%%%%%%%%%%%%%%%%%%%%%%%%%%%%%%%%%%%%%%%%%%%%%
\subsection{Dynamical systems}
\label{appendix:dynamicalsystems}
%%%%%%%%%%%%%%%%%%%%%%%%%%%%%%%%%%%%%%%%%%%%%%%%%%%

Let us briefly recall some main tools from the theory of
dynamical systems (see e.g., \cite{art:Crawford1991}, \cite{book:Carr1981}).

\begin{definition}($\omega$-limit set).
Consider a dynamical system on $\mathbb{R}^m$. The $\omega$-limit set
$\omega(x)$ of
a point $x\in\mathbb{R}^m$ is defined as the set of all 
accumulation points of the future orbit of $x$.
\end{definition}

\begin{remark}
Correspondingly, the $\alpha$-limit set $\alpha(x)$ is defined as
the set of accumulation points of the past orbit.
The $\omega$-limit of a set $S\subseteq \mathbb{R}^m$ is
$\omega(S) = \bigcup_{x\in S} \omega(x)$.
The $\omega$-limit sets ($\alpha$-limit sets) characterize
the future (the past) asymptotic behavior of the dynamical system.
The simplest examples for limit sets are fixed points and periodic orbits.
An orbit whose $\omega$- and $\alpha$-limit set is a fixed point is
called {\it heteroclinic}; a homoclinic orbit originates from and ends in 
one and the same fixed point. 
\end{remark}

\begin{theorem}\label{LaSalle}(LaSalle principle).
Consider a dynamical system $\dot{y} = f(y)$ 
on $\mathbb{R}^m$ and the closed, bounded, 
future invariant set $S\subseteq\mathbb{R}^m$.
If $Z: S\rightarrow \mathbb{R}$ is a ${\mathcal C}^1$ function which is 
monotonic along orbits in $S$, i.e., $\dot{Z} \leq 0$ ($\dot{Z} \geq 0$), 
then
\begin{equation}\label{omegalimitLa}
\omega(S) \subseteq 
\{s \in S\:|\:  \dot{Z} = 0\} \:.
\end{equation}
\end{theorem}

Replacing "future invariant" by "past invariant" results in the analogous
statement for the $\alpha$-limit set.

\begin{theorem}\label{monprin}(Monotonicity 
principle~\cite{book:WainwrightEllis1997}).
Consider a dynamical system on $\mathbb{R}^m$ and the invariant 
set $S\subseteq\mathbb{R}^m$.
If $Z: S\rightarrow \mathbb{R}$ is a ${\mathcal C}^1$ function which is 
strictly decreasing along orbits in $S$, then
\begin{subequations}\label{omegalimitmon}
\begin{align}
\omega(S) &\subseteq 
\{s \in \bar{S}\backslash S\:|\:\lim\limits_{x\rightarrow s} Z(x) \neq 
\sup\limits_{S} Z\} \\
\alpha(S) &\subseteq 
\{s \in \bar{S}\backslash S\:|\:\lim\limits_{x\rightarrow s} Z(x) \neq 
\inf\limits_{S} Z\} \:.
\end{align}
\end{subequations}
\end{theorem}

The monotonicity principle gives information about the global asymptotic
behavior of the dynamical system. Locally in the neighborhood of a fixed
point, the flow of the dynamical system 
is determined by the stability features of the fixed point.
If the fixed point is hyperbolic, i.e., if the linearization of
the system at the fixed point is a matrix possessing eigenvalues
with non-vanishing real parts, then the {\it Hartman-Grobman theorem} applies:
In a neighborhood of a hyperbolic fixed point the full nonlinear
dynamical system and the linearized system are topologically equivalent.
Non-hyperbolic fixed points are treated in center manifold theory.
We give a brief introduction (see, e.g., \cite{book:Carr1981} or~\cite{art:Crawford1991}):

Consider an autonomous dynamical system $\dot{y} = F(y)$,
where $y(t) \in \mathbb{R}^m$. Assume that there exists a fixed point $P$
and assume that the linearization of the system at $P$  
is described by a matrix which can be diagonalized. 
Accordingly, the phase space $\mathbb{R}^m$
can be decomposed into the direct sum $E^s\oplus E^u\oplus E^c$ of
the stable subspace $E^s$, the unstable subspace $E^u$ and the center subspace $E^c$.
Moreover, without loss of generality, we can choose variables such that 
the dynamical system can be written as
\begin{subequations}\label{fullnonlinear} 
\begin{align}
\dot{x}_1 & = A_1 x_1 + N_1(x_1,x_2) \\
\dot{x}_2 & = A_2 x_2 + N_2(x_1,x_2) \:.
\end{align}
\end{subequations}
Here, $x_1 \in E^c$ is an $m_c$-dimensional vector, $x_2\in E^s\oplus E^u$ is
($m_u+m_s$)-dimensional, the $A_i$ are matrices representing the linearized system, and 
the $N_i$ denote the nonlinear terms.

A center manifold $M^c$ is an invariant manifold in the phase space $\mathbb{R}^m$,
i.e., a submanifold of $\mathbb{R}^m$, which is invariant under the flow $\phi_t$ 
of the dynamical system. Moreover, $M^c$ must contain $P$ and is 
required to be tangent to $E^c$.
In a neighborhood of $P$ one can describe $M^c$ as the graph of a function
$h: E^c \rightarrow E^s\oplus E^u$, i.e., $(x_1, h(x_1)) \in M^c$.
The function $h$ is a solution to the differential equations
\begin{equation}\label{hequation}
\partial_{x_1} h(x_1) \:[ A_1 x_1 + N_1(x_1,h(x_1))]=
A_2 h(x_1) + N_2(x_1,h(x_1)),
\end{equation}
which also satisfies
$h(0)=0$ and $(\partial_{x_1} h)(0) = 0$ (tangency conditions). Recall that
$P=(0,0)$ in the adapted variables we use.
The Shoshitaishvili theorem (reduction theorem, \cite{art:Crawford1991},
\cite{art:Rendall/Tod1999}) 
generalizes the Hartman-Grobman theorem for the present situation:

\begin{theorem}\label{shoshitaishvili}(Shoshitaishvili theorem).
Consider the dynamical system~(\ref{fullnonlinear}) in a
neighborhood of the fixed point $P=(0,0)$. 
Then the flow of the full nonlinear system 
and the flow of the reduced system
\begin{subequations}\label{reduced} 
\begin{align}
\dot{x}_1 & = A_1 x_1 + N_1(x_1,h(x_1)) \\
\dot{x}_2 & = A_2 x_2 \:.
\end{align}
\end{subequations}
are locally equivalent, i.e., there exists a local 
homeomorphism $\Psi$ on phase space, such that
$\phi_t^{\mathrm{full}} = \Psi^{-1} \circ \phi_t^{\mathrm{reduced}} \circ \Psi$.
Here, $h$ is given by~(\ref{hequation}).
\end{theorem}

Equation~(\ref{hequation}) is in general a nonlinear partial differential equation
and cannot be solved in closed form. However, it holds that if
a function $\tilde{h}$ solves~(\ref{hequation}) up to $O(x_1^n)$ as well as 
the tangency conditions, 
then $h(x_1) = \tilde{h}(x_1) + O(x_1^n)$.
Hence, one can solve~(\ref{hequation}) approximately by a formal power series.

%%%%%%%%%%%%%%%%%%%%%%%%%%%%%%%%%%%%%%%%%%%%%%%%%%%
\subsection{Attractors in the state space}
\label{appendix:attractorsinthestatespace}
%%%%%%%%%%%%%%%%%%%%%%%%%%%%%%%%%%%%%%%%%%%%%%%%%%%

In this subsection we prove
that the global asymptotic behavior of the dynamical system~(\ref{UQOmega})
is fully determined by the fixed points of Table~\ref{tab:UQcube} and the closed
orbits $C_1, C_2$ (see Section~\ref{dynamicalsystemsanalysis}).

{\bf Theorem~\ref{fix}.} 
{\it 
All solutions converge to fixed points when $\lambda\rightarrow \pm\infty$ 
except when $n_0$ ($n_1$) is equal to 5. In this latter case solutions also
converge to the 1-parameter set of closed orbits $C_1$ ($C_2$) when
$\lambda\rightarrow \infty (-\infty)$.
}

\proof
All orbits of the dynamical system~(\ref{UQOmega}) are strictly monotonically 
decreasing in $\Omega$ except when $\Omega=0$, $\Omega=1$, $Q=0$, and $U=1$. 
It follows from the monotonicity principle that the $\omega$-limit set and
the $\alpha$-limit set of the state space reside on these
invariant subspaces.

More precisely, if $x$ is an interior point of the state space, then
\begin{equation}\label{alphaomega}
\alpha(x) \subseteq \{\Omega =1\}\cup\{Q=0\} \quad\mbox{and}\quad 
\omega(x) \subseteq \{\Omega=0\}\:.
\end{equation}
To see this, note that $U$ is a monotonically decreasing function
in a large part of the state space including $\{3/4<U<1\}\times[0,1]^2$.
This excludes attracting sets in this region and prevents interior orbits 
of having $\omega$-limits on the $\{U=1\}$ side face. Considering the known orbits and 
the local structure of the fixed points on $\{Q=0\}$, and making use of the Shoshitaishvili
theorem~\ref{shoshitaishvili}, it becomes clear that the 
fixed points act as $\alpha$-limit sets only, and not as $\omega$-limits.
Finally note that $Q$ is a monotonically increasing function 
when $0<Q<1$ and $1/2 < U$, ensuring in particular that $L_4$ cannot contain
an $\alpha$-limit of any interior orbit and thus leaving 
$\{\Omega =1\}\cup\{Q=0\}$ as the only set containing $\alpha$-limits.
Note that if $x$ lies on any of the side faces, then $\alpha(x), \omega(x)$ are
fixed points on the side faces themselves, as follows from the known solution structure.

We now come to the remaining point and investigate what type of
attractors can be found in the $\{\Omega=0\}$ plane
(and the identical $\{\Omega=1\}$ plane).
For $n_0 \leq 3$ the function
\begin{equation}\label{Z}
Z = \left(\frac{U}{1-U}\right) \left(\frac{Q}{1-Q}\right)^3 \ 
\end{equation}
is monotonic, which is seen from
\begin{equation}\label{dZ}
\frac{dZ}{d\lambda} = \big( 2U(1 - Q) + (3 - n(\Omega)) Q (1-U) \big) Z\ .
\end{equation}
This excludes attractors 
in the interior of the $\Omega = 0$ set when $n_0 \leq 3$.
Consequently, combined with the solution structure on the boundary it 
follows that the set of possible attractors is represented by the fixed points
discussed in Section~\ref{dynamicalsystemsanalysis} (it is easy to see that 
the equations prevent the existence of a heteroclinic cycle on the boundary 
when $0\leq n_0 \leq 3$).

In the case $n_0 > 3$ ($n_1 >3$) 
we notice the fixed point $P_3$ ($P_6$) in the interior of the plane.
To prove the fact that this is the only interior limit set as
long as $n_0\neq 5$ ($n_1\neq 5$) we resort
to the simpler intermediate dynamical system~(\ref{uqomeq}) in
the non-compactified variables $(u,q)$. Define
\begin{equation}\label{Psi}
\Psi = (u q)^{\frac{2}{n_0-1}} \:\left( \frac{n_0-5}{(n_0-1)^2} + 
\frac{ 2 q}{n_0-1} - 
\frac{q^2}{2} - \frac{q u}{n_0+1}\right)\:.
\end{equation}
Note that for the special case $n_0 =5$, apart from
the different choice of variables, $\Psi+ C = 0$ coincides with $\Phi=0$ 
(see~(\ref{n05mon})), 
i.e., for $n_0=5$, $\Psi+C=0$  describes the orbits in the polytropic subset.

\begin{figure}[htp]
	\centering
	\includegraphics[width=0.96\textwidth]{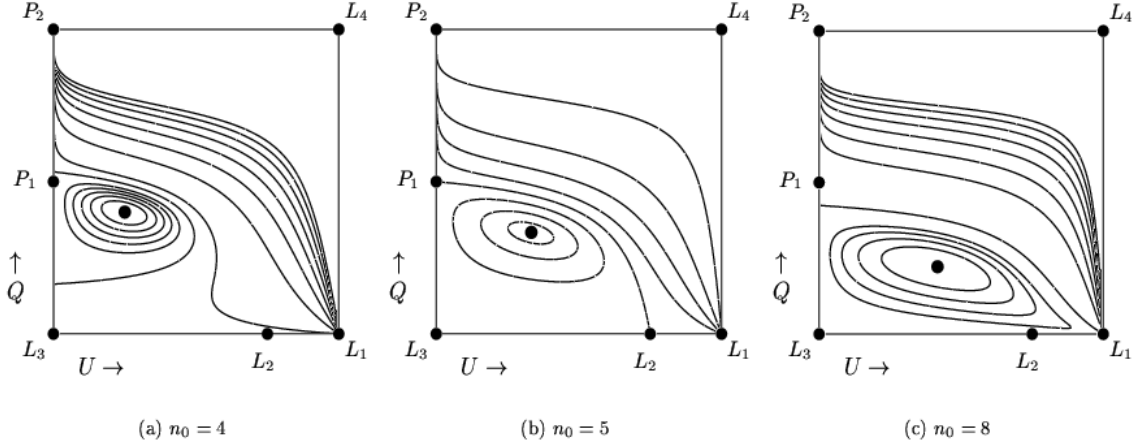}
	\caption{The curves $\Psi + C = 0$. The polytropic indices $n_0$ are:  
		(a) $n_0 = 4$, (b) $n_0=5$, (c) $n_0 = 8$.}
    	\label{closed}
\end{figure}

More generally, for all $n_0>3$ the equation $\Psi + C =0$ with 
$C_{\mathrm{min}}(n_0) < C < 0$ represents a family of closed curves with
the fixed point $P_3$ as the center, which itself is represented by $C=C_{\mathrm{min}}$. 
The shape of the $C=0$ curve depends on $n_0$. 
For $3<n_0<5$ it originates at some point below $P_1$ on the $Q$-axis,
turns around $P_3$ and ends in a point above $P_1$ on the $Q$-axis again.
Increasing $n_0$ to $n_0 =5$ this $C=0$ curve becomes the
regular orbit $L_2$-$P_1$ (plus the sections of the $Q$- and $U$-axes closing 
this curve, creating a heteroclinic cycle).
For $n_0 > 5$ the $C=0$ curve originates from $L_1$ and ends in some 
point below $P_1$ on the $Q$-axis.
The remaining curves $C>0$ always cover the second half of the $\Omega=0$ plane.
See Figure~\ref{closed}.

%\begin{figure}[htp]
%	\psfrag{L1}{{\small $L_1$}}
%	\psfrag{L2}{{\small $L_2$}}
%	\psfrag{L3}{{\small $L_3$}}
%	\psfrag{L4}{{\small $L_4$}}
%	\psfrag{L5}{{\small $L_5$}}
%	\psfrag{P1}{{\small $P_1$}}
%	\psfrag{P2}{{\small $P_2$}}
%	\psfrag{P3}{{\small $P_3$}}
%	\psfrag{P4}{{\small $P_4$}}
%	\psfrag{P5}{{\small $P_5$}}
%	\psfrag{P6}{{\small $P_6$}}
%	\psfrag{Q}[cb][Bl]{{$\begin{array}{c}\uparrow\\ Q\end{array}$}}
%	\psfrag{U}{{$U\rightarrow$}}
%	\centering
%        \subfigure[$n_0=4$]{
%		\label{monpsi4}
%		\includegraphics[width=0.3\textwidth]{monpsi4.eps}}\quad
%    	\subfigure[$n_0=5$]{
%		\label{monpsi5}
%		\includegraphics[width=0.3\textwidth]{monpsi5.eps}}\quad
%     	\subfigure[$n_0=8$]{
%		\label{monpsi8}
%		\includegraphics[width=0.3\textwidth]{monpsi8.eps}}
%	\caption{The curves $\Psi + C = 0$. The polytropic indices $n_0$ are:  
%		(a) $n_0 = 4$, (b) $n_0=5$, (c) $n_0 = 8$.}
%    	\label{closed}
%\end{figure}

The function $\Psi$ is monotonic on orbits for $n_0\neq 5$ ($n_1\neq 5$), i.e.,
\begin{equation}\label{dPsi}
\frac{d\Psi}{d\xi} = (n_0-5) \,\big( 
(n_0-1)^{-3} (u q)^{\frac{2}{n_0-1}} (-2+ (n_0-1) q)^2 \big)\:.
\end{equation}
Using the LaSalle principle, this 
implies that the limit-sets lie on $q=2/(n_0-1)$.
Taking, e.g., $dq/d\xi$ into consideration, this
guarantees that $P_3$ ($P_6$) is the only possible $\omega$-limit set 
($\alpha$-limit set)
in the interior of the $\{\Omega=0\}$ ($\{\Omega=1\}$) plane.

Together with the solution structure on the boundary, the above arguments
also imply that there are no heteroclinic cycles when $n_0 \neq 5$ $(n_1 \neq 5)$, 
and thus it follows that all orbits start and end at fixed points, with the 
exception of when $n_0=5$ $(n_1 = 5)$ when the closed orbits $C_1$ $(C_2)$ 
constitute a limit set.
\proofend

\begin{remark}
The expression~(\ref{Psi}) can be derived relatively easy
in a set of variables different from the ones previously used,
and then transformed to $(u,q)$.
Those variables are obtained from the original $\rho$, $p$, or $\eta$ by the
so-called Emden transformations (see~\cite{Chandrasekhar:1939}, p.90)  
\end{remark}

%%%%%%%%%%%%%%%%%%%%%%%%%%%%%%%%%%%%%%%%%%%%%%%%%%%
\subsection{The bifurcations at $n_0 =3$ and $n_0=5$.}
\label{appendix:thebifurcations}
%%%%%%%%%%%%%%%%%%%%%%%%%%%%%%%%%%%%%%%%%%%%%%%%%%%

Applying center manifold theory we prove that in continuation
with $n_0 < 3$, $P_1$ is excluded
as an attractor for interior orbits (and interior orbits on $\Omega=0$) 
when $n_0 = 3$. These considerations are intimately connected with 
the result that solutions have finite radii in the 
case $n_0=3$ (see Theorem~\ref{n0leq3}). 

To establish the claim it is 
preferable to investigate the dynamical system in
the uncompactified variables $(u,q,\omega)$, i.e., the system~(\ref{uqomeq}).
In this setting $P_1$ is represented by the fixed point $(0,1,0)$;
the eigenvalues of the linearization of the system at $P_1$
are $\{3-n_0, 1, -a\}$ and the corresponding eigenvectors are
$\{ (2-n_0) e_1+ e_2, e_2, e_3\}$ (where the $\{e_i\}$ form an orthonormal
basis in the $(u,q,\omega)$-space).
Introducing adapted variables $\{x\geq 0, y, z\geq 0 \}$ according to
\begin{equation}\label{oldvars}
u = x (n_0-2) \qquad q = 1-x+y \qquad \omega = z
\end{equation}
we obtain the following dynamical system:
\begin{equation}\label{sysinxyz}
\begin{array}{lcl}
\dot{x} = &(3-n_0) x & +\quad x [ 3-n(z) - x (n_0-2) - (y-x) n(z)] \\[3pt]
\dot{y} = &y & +\quad y^2 + x^2 (5 +n(z) - 2 n_0) - x y (4 +n(z) -n_0) - x (n(z)-n_0) \\[3pt]
\dot{z} = & -a z & +\quad a (x-y) z
\end{array}
\end{equation}
Setting $n_0 =3$ we observe that $\langle (1,0,0) \rangle$ is the center subspace $E^c$ of 
the system, $\langle\{(0,1,0),(0,0,1)\}\rangle$ corresponds to $E^u\oplus E^s$, so that
our system~(\ref{sysinxyz}) is of the form~(\ref{fullnonlinear}).

Solving~(\ref{hequation}), as indicated in Subsection~\ref{appendix:dynamicalsystems}, 
we see that the center manifold of the system~(\ref{sysinxyz}) can be 
represented by the following local graph $h(x)=(h_y(x), h_z(x))$:
\begin{equation}\label{hcon}
h_y(x) = -2 x^2 - 16 x^3 + O(x^4) \qquad h_z(x) = 0 \:.
\end{equation}
By Shoshitaishvili's theorem~\ref{shoshitaishvili}, in a neighborhood 
of the fixed point,
the full system~(\ref{sysinxyz}) is equivalent to the reduced system
\begin{equation}\label{redsysinxyz}
\begin{array}{lcl}
\dot{x} &= & 2 x^2 + 6 x^3 + O(x^4)  \\[3pt]
\dot{y} & = &y  \\[3pt]
\dot{z} &= & -a z  \:.
\end{array}
\end{equation}
Consider now a solution passing through a point $(x>0, y, z>0)$ (corresponding
to an orbit in the interior of the cube).
Obviously, such a solution cannot converge to $P_1 = (0,0,0)$.
Accordingly, the claim is proved.

It can also be shown that $P_4$ is the $\alpha$-limit set for a two-parameter 
family of orbits in the case $n_1=3$. The investigation of $P_4$ is almost 
identical to the above one for $P_1$.

We turn to the bifurcation at $n_0=5$ and prove that each closed orbit $C_1$ 
($C_2$) acts as an $\omega$-limit ($\alpha$-limit) set for
a one-parameter family of interior orbits of the state space. 
The method of investigation is again based on center manifold theory.

In a neighborhood of $P_3$ ($P_6$) the full linear system is equivalent
to the decoupled system
\begin{equation}\label{redsysinxyzn05}
\begin{array}{lcl}
\dot{x} &= & f_1 (x, y) \\[3pt]
\dot{y} & =& f_2 (x, y) \\[3pt]
\dot{z} &= & -a z  \:,
\end{array}
\end{equation}
where $x,y$ are variables linearly depending on $U,Q$ and $z = \Omega$.
Since we know from~(\ref{n05mon}) that for $z=0$ the orbits are periodic
we obtain the desired result. Note also that, since $dn/d\Omega=0$, when 
$\Omega \rightarrow 0$, the above type of decoupling also takes place 
asymptotically, when the orbits are not in a neighborhood of $P_3$ 
(and analogously for $P_6$).

%%%%%%%%%%%%%%%%%%%%%%%%%%%%%%%%%%%%%%%%%%%%%%%%%%%
\subsection{Proof of Theorems~\ref{spiralthm} and~\ref{MR2nd}}
\label{appendix:proofofTheorems}
%%%%%%%%%%%%%%%%%%%%%%%%%%%%%%%%%%%%%%%%%%%%%%%%%%%

{\bf Theorem~\ref{spiralthm}.}
{\it 
(Spiral structure of the $(M,R)$-diagram).
Consider an asymptotically polytropic equation of state
with asymptotic indices $0< n_0 \leq 3$ and $n_1$.
Then the mass-radius relation for high central pressures 
possesses a spiral structure, if and only if $n_1 > 5$
with the spiral given by
\begin{equation}\label{spirals2}
\left(\begin{array}{c}
R(\eta_c) \\
M(\eta_c)
\end{array}\right) \:=\:
\left(\begin{array}{c}
R_O \\
M_O
\end{array}\right) 
\:+\:
\left(\frac{1}{\eta_c}\right)^{\gamma_1} \: {\mathcal B} \: 
{\mathcal J}(\gamma_2 \log\frac{1}{\eta_c}\,)\: b 
\:+\:
o(\left(\frac{1}{\eta_c}\right)^{\gamma_1}) \:,
\end{equation}
where $R_O$ and $M_O$ are constants, ${\mathcal B}$ is a non-singular
matrix, and $b$ a non-zero vector.
The matrix ${\mathcal J}(\varphi) \in \mathrm{SO}(2)$ describes a rotation 
by an angle $\varphi$, and the constants $\gamma_1$ and $\gamma_2$ are given by
\begin{equation}\label{gamma122}
\gamma_1 = (n_1 -5) \,\frac{1}{4} \qquad 
\gamma_2 = \frac{1}{4}\, \sqrt{b}\:.
\end{equation}
Recall from Table~\ref{tab:UQcube} that $\sqrt{b} = \sqrt{-1 -22 n_1 + 7 n_1^2}$.
}

{\it Sketch of proof}.
The proof is based on the idea that the spiral structure, which is formed 
by the orbits in a neighborhood of $P_6$, is reflected
in a spiral in the $(M,R)$-diagram.

{\it Part 1}: We show that $n_1>5$ implies a spiral in the $(M,R)$-diagram
and conversely.

First assume that $n_1 > 5$, and note that in this case 
there exists one single orbit originating
from $P_6$. Like every solution for an equation of state with 
$n_0\leq 3$, it possesses finite mass, $M_O$, and finite radius, $R_O$.

Set $\Omega_\epsilon = 1 -\epsilon$, where $\epsilon$ is chosen
sufficiently small, so that $n(\Omega) > 5$ for all 
$\Omega> \Omega_\epsilon$.
Consider the $\Omega=\Omega_\epsilon$ plane, and
let $I$ be the intersection point of the $P_6$-orbit with 
this plane. Choose a small neighborhood
$U_I$ of $I$ within the plane. If $\epsilon$ and $U_I$ are
small enough then, by the approximate decoupling of the dynamical system, 
the intersection of the one-parameter set of regular solutions, 
possessing sufficiently high central pressures, forms a spiral in $U_I$.

As for equation~(\ref{universalMRcyl}),
consider a small cylinder centered around $P_2$.
The $P_6$-orbit intersects the cylinder in a point $O$,
with coordinates $(h_O, \phi_O)$, determining the mass $M_O$ and
the radius $R_O$. By the flow of the dynamical system we transport 
$U_I$ to a neighborhood $U_O$ of $O$ on the cylinder.  
Provided that $U_I$ has been chosen sufficiently small, the spiral in $U_I$
is mapped to a spiral in $U_O$, since the map is a diffeomorphism.

Inserting $h=h_O + \delta h$ and 
$\phi=\phi_O + \delta \phi$ into~(\ref{universalMRcyl}), 
leads to the following mass and radius relations in $U_O$;
\begin{equation}\label{RMO}
R = R_O + c_1 \delta h + c_2 \delta\phi \qquad
M = M_O + c_3 \delta h + c_4 \delta\phi\:,
\end{equation}
where the vectors $(c_1, c_2), (c_3, c_4)$ are linearly independent.
Accordingly, a spiral in $(h, \phi)$ results in a spiral in
the $(M,R)$-diagram.

Conversely, assume that the high pressure regular solutions
form a spiral structure in the $(M,R)$-diagram. 
We then need to show that $n_1$ must be greater than 5.

To this end, let $(M_O, R_O)$ be the center of the
spiral and $O = (h_O, \phi_O)$ be the associated point
on the cylinder near $P_2$. Choose a small neighborhood
$U_O$ of $O$ on the cylinder.
By following the orbit passing through $O$
backwards in $\lambda$ it can be seen that the 
(indirect) assumption $n_1 \leq 5$ leads to a contradiction.
Namely, there are three possible origins of the considered orbit:
If the $O$-orbit comes from $P_6$ (plus $C_2$ in the case $n_1 = 5$), 
then this must be true for 
all orbits passing through $U_O$ by continuous dependence on initial data.
Accordingly, none of these solutions is regular, which is a contradiction
to the assumption that the regular solutions form a spiral in $U_O$.
The same argument holds if the $O$-orbit originates from $L_1$.
Finally assume that the orbit corresponds to a regular solution
that comes from $\Omega_c$ on $L_2$.
However, not all regular solutions $\Omega \geq \Omega_c$ can 
pass through $U_O$, once more by continuous dependence on initial data,
which is again a contradiction to the assumed behavior of
the high pressure solutions.
Thus the first part of the theorem is established.

{\it Part 2}: We now give a sketch of the proof for 
the qualitative form of the spiral structure~(\ref{spirals}). 

Consider again the orbit that originates
from $P_6$ and passes through the points $I\in U_I\subseteq \{\Omega =\Omega_\epsilon\}$ 
and $O\in U_O\subseteq \mathrm{cylinder}$.
For initial data lying in $U_I$ close to $I$ we may linearize the full dynamical
system along the $P_6$-orbit; we get
\begin{equation}\label{alongP6orbit}
\left(\begin{array}{c}
\delta U(\lambda) \\
\delta Q(\lambda) \\
\delta \Omega(\lambda)\\
\end{array}\right) \:=\:
{\mathcal A}(\lambda) \, 
\left(\begin{array}{c}
\delta U_0 \\
\delta Q_0 \\
0 \\
\end{array}\right) \:.
\end{equation}
Below we use equation~(\ref{alongP6orbit}) to map 
the spiral of regular solutions $(\delta U_0(\Omega_c), \delta Q_0(\Omega_c))$
(parametrized by the central value $\Omega_c$) down towards the
cylinder around $P_2$, and thus obtain the $(M,R)$-spiral. The next step 
is therefore to derive an expression for the spiral of regular solutions in $U_I$.

Let $U_{\mathrm{p}}(\lambda)$, $Q_{\mathrm{p}}(\lambda)$ 
denote the regular solution of the 
polytropic system~(\ref{Ueq})+(\ref{Qeq}) with polytropic index $n_1$.
This solution gives rise to a spiral whose
asymptotic form in the limit $\lambda\rightarrow \infty$ reads
\begin{equation}\label{hatspiral}
\left(\begin{array}{c}
U_{\mathrm{p}}(\lambda)\\
Q_{\mathrm{p}}(\lambda)
\end{array}\right) \:=\:
\left(\begin{array}{c}
U_{P_6}\\
Q_{P_6}
\end{array}\right)
+ \exp(-\delta_1 \lambda) \: {\mathcal B}^\prime \:
{\mathcal J}(\delta_2 \lambda)\: b^\prime\:,
\end{equation}
where $U_{P_6}, Q_{P_6}$ are the coordinates of the
fixed point $P_6$; ${\mathcal B}^\prime$ is a non-singular
matrix, and $b^\prime$ a non-zero vector;
$\delta_1 = \frac{\beta}{4} (n_1-5)$ and $\delta_2 = \frac{\beta}{4} \sqrt{b}$.
The constants $\delta_1$ and $\delta_2$
are the real and imaginary part of the complex eigenvalue
that is associated with $P_6$,
compare with Table~\ref{tab:UQcube}.

Now consider the dynamical system~(\ref{UQOmega}) for
$\Omega\in[\Omega_\epsilon,1]$.
Since we can choose $\epsilon$ arbitrarily small (and since in addition
$\frac{d n(\Omega)}{d \Omega}|_{\Omega=1} = 0$) we are justified
to approximate $n(\Omega)$ by $n(\Omega) \equiv n_1$ in~(\ref{Ueq}), so that
the system decouples.
Accordingly,~(\ref{Omegaeq}) can be reduced to 
$\frac{d (1-\Omega)}{d\lambda} = 
a (1-\Omega) Q_{\mathrm{p}}(\lambda) (1-U_{\mathrm{p}}(\lambda))$, which
can be solved:
\begin{equation}\label{Omegaspiral} 
1-\Omega(\lambda) = (1-\Omega_c)\, \exp\left(a \int\limits_{-\infty}^{\lambda} 
Q_{\mathrm{p}}(\lambda) (1-U_{\mathrm{p}}(\lambda)) d\lambda\right)\:,
\end{equation}
where $\Omega_c$ is the initial value on $L_2$.
Inserting~(\ref{hatspiral}) into~(\ref{Omegaspiral})
we obtain the approximation
\begin{equation}\label{omegaspiralsim}
1-\Omega(\lambda) = (1-\Omega_c)\, const\, \exp\big(a Q_{P_6} (1-U_{P_6})\, 
(\lambda-\tilde{\lambda})\,\big),
\end{equation}
where $\tilde{\lambda}$ and $const$ are independent of $\Omega_c$.

We conclude that a regular solution that starts from
$\Omega_c$ near 1 reaches 
$U_I \subseteq \{\Omega=\Omega_\epsilon\}$
at $\lambda_c(\Omega_c) = \frac{1}{a \beta}\,\log\frac{1}{1-\Omega_c}+const$,
where $const$ is independent of $\Omega_c$.
Note that $\beta = Q_{P_6} (1-U_{P_6})$ (compare with Table~\ref{tab:UQcube}).
Transforming from $\Omega$ to $\eta$ yields
\begin{equation}\label{tc}
\lambda_c(\eta_c) = \frac{1}{\beta}\log\eta_c + \mathrm{const}\:.
\end{equation}

At $\lambda=\lambda_c$ the $(U,Q)$-components of the regular solution
are located at some position on the spiral~(\ref{hatspiral}),
so that we are able to explicitly give the
spiral of regular solutions in $U_I \subseteq \{\Omega =\Omega_\epsilon\}$
by inserting~(\ref{tc}) into~(\ref{hatspiral}):
\begin{equation}\label{censpiral}
\left(\begin{array}{c}
\delta U_0(\eta_c)\\
\delta Q_0(\eta_c)
\end{array}\right) \:=\:
\left(\frac{1}{\eta_c}\right)^{\gamma_1}  \: {\mathcal B}^{\prime\prime} \:
{\mathcal J}(\gamma_2 \log\eta_c\,)\: b^{\prime\prime}\:,
\end{equation}
where ${\mathcal B}^{\prime\prime}$ and $b^{\prime\prime}$ have absorbed 
the constants and a fixed rotation matrix. 

Inserting~(\ref{censpiral}) into~(\ref{alongP6orbit})
and projecting onto the small cylinder at $P_2$ affects
the spiral by an additional dis\-tortion as does 
the change of coordinates from $(h,\phi)$ to $(M,R)$.
Hence Theorem~\ref{spiralthm} is established. \proofend

\begin{remark}
Theorem~\ref{spiralthm} can easily be generalized to the case $n_0=0$;
however, as all orbits end in $L_5$ instead of $P_2$, the
$(M,R)$-relations~(\ref{universalMR}) and~(\ref{universalMRcyl}) 
must be replaced by different ones. 
Clearly these relations are not qualitatively different. 
In order to visualize things, one can again make use of the cylinder
picture: consider a cylinder lying down with small radius with $L_5$ as 
its center. Then every orbit can be uniquely characterized by its intersection
with this cylinder, i.e., by $(U_0, \phi)$.
Accordingly, all arguments used in the proof of Theorem~\ref{spiralthm} go through.
\end{remark}

{\bf Theorem~\ref{MR2nd}.}
{\it 
(Polytropic behavior of the $(M,R)$-diagram).
Consider an asymptotically polytropic equation of state with asymptotic indices
$0<n_0$, and $0<n_1<5$. To first order, 
the mass-radius relation for high central
pressures is approximated by the mass-radius relation for an exact polytrope
with polytropic index $n_1$, i.e.,
\begin{equation}
\label{highdenMR2}
R(\eta_c) = R_{\mathrm{p}:n_1}(\eta_c)\, \zeta[n(\eta)]\qquad
M(\eta_c) = M_{\mathrm{p}:n_1}(\eta_c)\, \zeta[n(\eta)]\:,
\end{equation}
where $R_{\mathrm{p}:n_1}(\eta_c)$ and
$M_{\mathrm{p}:n_1}(\eta_c)$ are the radius and 
the mass for an exact polytrope
with index $n_1$ (compare with~(\ref{MRpoly})), i.e.,
\begin{equation}\label{MRpoly22}
R_{\mathrm{p}:n_1}(\eta_c) = c_R(n_1) \, \eta_c^{1-n_1} \qquad
M_{\mathrm{p}:n_1}(\eta_c) = c_M(n_1) \, \eta_c^{3-n_1}\:,
\end{equation}
and $\zeta$ is a functional acting on the index-function $n(\eta)$, given by
\begin{equation}\label{zeta2}
\zeta[n(\eta)] = \exp\left(\frac{1}{2} \,\int\limits_0^1 \frac{n(\eta)-n_0}{\eta} \, d\eta + 
\frac{1}{2}\,
\int\limits_1^\infty \frac{n(\eta)-n_1}{\eta} \, d\eta \right)\:.
\end{equation}
}

{\it Sketch of proof.}
The proof of the claim 
is quite analogous to the proof of Theorem~\ref{spiralthm}. 
For high central pressures (i.e., close to $\{\Omega=1\}$ in the 
state space)
the dynamical system~(\ref{UQOmega}) almost decouples (recall that
$dn(\Omega)/d\Omega = 0$ at $\Omega=1$). Therefore, for $\lambda$ smaller
than some value $\tilde{\lambda}$, the $(U,Q)$ component of orbits
that originate from $\Omega_c=\frac{\eta_c^a}{1+\eta_c^a}$ close to 1,
can be approximated by the corresponding regular 
polytropic orbit $(U_{\mathrm{p}}, Q_{\mathrm{p}})$ 
in the $\{\Omega=1\}$ plane. 
Consequently the $\Omega$ component behaves like
\begin{equation}\label{omegaap}
\Omega(\lambda) = 1- (1-\Omega_c) \exp\left(
a \int_{-\infty}^\lambda Q_{\mathrm{p}}(\lambda) (1-U_{\mathrm{p}})(\lambda)) d\lambda
\right)\:.
\end{equation}
Choosing $\Omega_c$ sufficiently close to one, the corresponding orbit will
come arbitrarily close to $P_5$ at $\lambda = \tilde{\lambda}$.
Thus, linearizing the dynamical system along the $P_5$-$P_2$ orbit, we can
calculate the further evolution of the high pressure orbits.
We obtain
\begin{subequations}\label{highpU}
\begin{align}
U(\lambda) & = U_{\mathrm{p}}(\tilde{\lambda}) e^{-N(\eta(\lambda))} e^{N(\eta(\tilde{\lambda}))} \\
Q(\lambda) & = 1- \big( 1-Q_{\mathrm{p}}(\tilde{\lambda})\,\big)\, e^{-t} \\
\Omega(\lambda) &= 
\big( 1 + (1-\Omega(\tilde{\lambda})) e^{-a \tilde{\lambda}} e^{a \lambda}\big)^{-1}\:,
\end{align}
\end{subequations}
where $N(\eta)$ is the integral 
\begin{equation}\label{N}
N(\eta) := -\int n(\eta) \:\frac{1}{\eta} \:d\eta\:,
\end{equation}
and $\eta(\lambda)$ corresponds to $\Omega(\lambda)$, i.e., 
$\eta(\lambda) = \Omega(\tilde{\lambda})^{1/a} e^{\tilde{\lambda}} e^{-\lambda}$.

For large $\lambda$, when $\Omega(\lambda)$ (respectively $\eta(\lambda)$) has become small, 
we can therefore approximate the high pressure solutions by
\begin{subequations}\label{highpappr}
\begin{align}
\label{highpUappr}
U(\lambda) & = U_{\mathrm{p}}(\tilde{\lambda}) e^{n_0 \tilde{\lambda}} e^{- n_0 \lambda} 
\exp\left(-\int_0^{\eta(\tilde{\lambda})}\frac{n(\eta)-n_0}{\eta} \,d\eta\right) \\
Q(\lambda) & = 1- \big( 1-Q_{\mathrm{p}}(\tilde{\lambda})\,\big)\, e^{-t} \\
\Omega(\lambda) &= (1-\Omega(\tilde{\lambda}))^{-1}  e^{a \tilde{\lambda}} e^{-a \lambda}\:,
\end{align}
\end{subequations}
where we note that the last exponential in~(\ref{highpUappr}) can be simplified by using
\begin{eqnarray}\label{Nreg}
\nonumber
-\int_0^{\eta(\tilde{\lambda})}\frac{n(\eta)-n_0}{\eta} \,d\eta & = & 
(n_1-n_0) \big(\log\eta_c -
\int_{-\infty}^{\tilde{\lambda}} Q_{\mathrm{p}}(\lambda) 
(1-U_{\mathrm{p}})(\lambda)) d\lambda\big) \times \\[2pt]
& & \qquad \times \big(\int_0^1 \frac{n(\eta)-n_0}{\eta} \, d\eta +
\int_1^\infty \frac{n(\eta)-n_1}{\eta} \, d\eta\big) \:.
\end{eqnarray}

Equations~(\ref{highpappr}) exhibit the typical form of linearized solutions
near the fixed point $P_2$. Recall from~(\ref{universalMR}) that we can easily
read off the total mass and radius of such a solution, yielding
\begin{equation}\label{R2M2}
R^2 \propto \eta_c^{1-n_1} \qquad M^2 \propto \eta_c^{3-n_1}\:.
\end{equation}
Moreover, comparing constants appearing in the general expressions
for $M$ and $R$ and the analogue expressions for exact polytropes, we see
that they differ merely by a factor expressible as $\zeta[n(\eta)]$,
which completes the proof of the theorem. 
\proofend

%%%%%%%%%%%%%%%%%%%%%%%%%%%%%%%%%%%%%%%%%%%%%%%%%%%%%%%%%%%%%%%%%%%%%%%%
%%%%%%%%%%%%%%%%%%%%%%%%%%%%%%%%%%%%%%%%%%%%%%%%%%%%%%%%%%%%%%%%%%%%%%%%
%%%%%%%%%%%%%%%%%%%%%%%%%%%%%%%%%%%%%%%%%%%%%%%%%%%%%%%%%%%%%%%%%%%%%%%%
%%%%%%%%%%%%%%%%%%%%%%%%%%%%%%%%%%%%%%%%%%%%%%%%%%%%%%%%%%%%%%%%%%%%%%%%
\section{Miscellaneous generalizations}
\label{B}
%%%%%%%%%%%%%%%%%%%%%%%%%%%%%%%%%%%%%%%%%%%%%%%%%%%%%%%%%%%%%%%%%%%%%%%

%%%%%%%%%%%%%%%%%%%%%%%%%%%%%%%%%%%%%%%%%%%%%%%%%%%
\subsection{The negative mass cube}
\label{appendix:thenegativemasscube}
%%%%%%%%%%%%%%%%%%%%%%%%%%%%%%%%%%%%%%%%%%%%%%%%%%%

Also perfect fluid solutions with negative mass, i.e., solutions
of~(\ref{mpeq}) with negative $m(r)$ can be studied in our setting.
As before we use compactified variables $(U,Q,\Omega)$
\begin{equation}\label{boundedvarneg}
U = \frac{u}{1+|u|}\ , \quad
Q = \frac{q}{1+|q|}\ , \quad
\Omega = \frac{\omega}{1+\omega} \:,
\end{equation}
where $(u,q,\omega)$ are defined as in~(\ref{uqom}).
By our assumption $m < 0$ it follows that $\Omega$ is positive and
that $U, Q$ are negative. Including the boundaries in our discussion
we obtain the state space $[-1,0]^2\times [0,1]$.

Introducing the independent variable $\lambda$ according to 
$d\lambda/d\xi = (1 + U)^{-1}(1 + Q)^{-1}$ yields the following
equations for $U,Q,\Omega$:
\begin{subequations}\label{UQOmeganeg}
\begin{align}
\label{Unegeq}
\frac{dU}{d\lambda} &=  
U(1+U)[(1+Q)(3+2U) - n(\Omega)\,Q(1+U)] \\
\label{Qnegeq}
\frac{dQ}{d\lambda} &= 
Q(1+Q)[-(1+Q) + Q(1+U)] \\
\label{Omeganegeq}
\frac{d\Omega}{d\lambda} &= -a\Omega(1-\Omega)Q(1+U)\ . 
\end{align}
\end{subequations}

Compared to the positive mass cube the negative mass cube possesses 
a much simpler fixed point structure (see Table~\ref{tab:UQcubeneg}).

\begin{table}[htp]
  \begin{center}
    \begin{tabular}{|c|cccc|c|c|}
      \hline
      Fixed point & $U$ & $Q$ & $\Omega$ & & Eigenvalues & Restrictions\\\hline
      & & & & & & \\[-0.3cm]
      $\bar{L}_1$ & -1 & 0 & $\Omega_0$ & & $-5\ , \ -1\ , \ 0$ &\\
      $\bar{L}_3 = L_3$ & 0 & 0 & $\Omega_0$ & & $3\ , \ -1\ , \ 0$ & \\
      $\bar{L}_4$ & -1 & -1 & $\Omega_0$ & & $0 \ , \ 0 \ , \ 0$ & \\
      $\bar{L}_5$ & $U_0$ & -1 & 0 & & $0 \ , \ (1+U_0)\ , \ a (1+U_0)$ & $n_0=0$ \\
     $\bar{L}_6$ & $U_0$ & -1 & 1 & & $0 \ , \ (1+U_0)\ , \ -a (1+U_0)$ & $n_1 =0$  \\
      $\bar{P}_2$ & 0 & -1 & 0 & & $n_0\ , \ 1\ , \ a$ & \\
      $\bar{P}_5$ & 0 & -1 & 1 & & $n_1\ , \ 1\ , \ -a$ & \\  
        \hline
    \end{tabular}
  \end{center}
    \caption{Local properties of the fixed points of the negative mass cube.}
    \label{tab:UQcubeneg}
\end{table}

For orbits in the interior of the state space the flow of the 
dynamical system~(\ref{UQOmeganeg}) is characterized 
by a universal $\alpha$-limit set, namely the fixed point $\bar{P}_2$ and
a universal $\omega$-limit set, namely $\bar{L}_1$
(in the special case $n_0 =0$, $\bar{P}_2$ must be replaced by $\bar{L}_5$).
In brief: all orbits originate from $\bar{P}_2$ and end in $\bar{L}_1$.

Accordingly, using the linearization of~(\ref{UQOmeganeg}) around $\bar{P}_2$ 
and $\bar{L}_1$, we can easily derive the qualitative features
of a negative mass solution:
In a neighborhood of $\{r\leq R_-\}$ the negative mass solution is described by a 
vacuum solution with the potential $v=-M_-/r+v_-+v_S$, where $M_-$ is negative and  
$v_- = M_-/R_-$. At $r=R_-$ the quantity $\eta$ (which was defined as $\eta=-v+v_S$) 
and therefore $\Omega$ has its zero; 
here the solution enters the interior of the negative mass cube 
(at $\bar{P}_2$),
\begin{equation}\label{barP2}
v(r) = -C \delta r + v_S + O(\delta r^2),
\end{equation}
where $\delta r = \frac{r-R_-}{R_-}$ and $C=-\frac{M_-}{R_-}$. 

For $r>R_-$ the density is positive so that the solution begins to acquire
"positive mass". This happens continuously  
until the radius $r=R$ is reached where
the mass has been reduced to zero, $m(R) = 0$.
At $r=R$ the potential $v(r)$ has the (negative) minimum $v_0+v_S$.
In the state space this corresponds to the orbit reaching
$\bar{L}_1$ at a certain $\Omega_0$.

The solution can be continued into the positive mass cube by
identifying $\bar{L}_1$ with $L_1$: starting from $\Omega_0$ on $L_1$
the solution proceeds as a solution which now possesses positive mass.

Equation~(\ref{negmasssols}) gives the approximate form of the
solution near the radius $R$. Note that this expansion holds for
both $\delta r <0$ and $\delta r>0$.

%%%%%%%%%%%%%%%%%%%%%%%%%%%%%%%%%%%%%%%%%%%%%%%%%%%
\subsection{Non-increasing equations of state}
\label{appendix:non-increasingequationsofstate}
%%%%%%%%%%%%%%%%%%%%%%%%%%%%%%%%%%%%%%%%%%%%%%%%%%%

In Section~\ref{equationsofstate}, 
we require $n_0$ to be a non-negative constant
(where the special case $n_0 = 0$ corresponds
to an equation of state with $\rho\rightarrow \mbox{const} >0$ 
when $p\rightarrow 0$). (The case $n_0<0$ could be interpreted as
$\rho\rightarrow\infty$ when $p\rightarrow 0$).
However, the non-negativity requirement is not
necessary for the index-function $n(\eta)$ (or $n(\Omega)$) as a whole,
however, recall that a non-negative index-function $n(\eta)$ is 
associated with a monotonically increasing equation of state $\rho(p)$.

\begin{figure}[htp]
	\centering
	\includegraphics[width=0.4\textwidth]{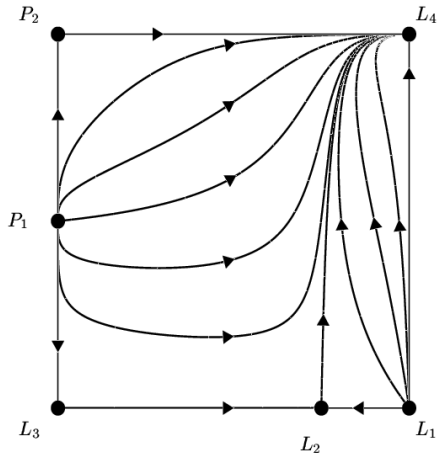}
	\caption{Orbits on the $\Omega=1$ plane in terms of the 
		variables $U$ and $Q$ for negative values of $n_1$.}
	\label{top}
\end{figure}

In particular, $n_1$ is allowed to be negative, so
that in addition to the patterns of Figure~\ref{base} the orbits
on the $\{\Omega=1\}$ plane can look like in Figure~\ref{top}.

%\begin{figure}[htp]
%	\psfrag{L1}{{\small $L_1$}}
%	\psfrag{L2}{{\small $L_2$}}
%	\psfrag{L3}{{\small $L_3$}}
%	\psfrag{L4}{{\small $L_4$}}
%	\psfrag{L5}{{\small $L_5$}}
%	\psfrag{P1}{{\small $P_1$}}
%	\psfrag{P2}{{\small $P_2$}}
%	\psfrag{P3}{{\small $P_3$}}
%	\psfrag{P4}{{\small $P_4$}}
%	\psfrag{P5}{{\small $P_5$}}
%	\psfrag{P6}{{\small $P_6$}}
%	\centering
%		\includegraphics[width=0.4\textwidth]{nminus05f.eps}
%	\caption{Orbits on the $\Omega=1$ plane in terms of the 
%		variables $U$ and $Q$ for negative values of $n_1$.}
%	\label{top}
%\end{figure}

Note that although Figure~\ref{top} qualitatively differs 
from the polytropic pictures (Figure~\ref{base}), the
global asymptotics for interior orbits does not change.

In order to prove that a closer investigation of $L_4$
is necessary, namely, we need to show that 
no interior orbits converges to $L_4$ for $\lambda\rightarrow\pm\infty$.
This can be done using polar coordinates
$(r,\phi)$ centered at $L_4$ 
and a changed independent variable.

%%%%%%%%%%%%%%%%%%%%%%%%%%%%%%%%%%%%%%%%%%%%%%%%%%%
\subsection{Superposition of solutions}
\label{appendix:superpositionofsolutions}
%%%%%%%%%%%%%%%%%%%%%%%%%%%%%%%%%%%%%%%%%%%%%%%%%%%

In this subsection we investigate superpositions of perfect fluid solutions.

Recall that $\eta(r) = v_S - v(r)$, where $v$ is the Newtonian potential. The
equations of the self-gravitating perfect fluid are the Euler-Poisson equations
leading to
\begin{equation}\label{eulerpoisson} 
\Delta \eta = - 4\pi \rho\:,
\end{equation}
where $\rho$ is to be understood as $\rho(\eta)$, the equation of state.
This formulation is equivalent to the system~(\ref{mpeq}).

If $\eta_i(r)$ ($i=1,2$) solves~(\ref{eulerpoisson}), $\Delta \eta_i = - 4\pi \rho_i$, 
then the superposition $\tilde{\eta} = \eta_1 + \eta_2 + c$ ($c$ a constant
adapted to the boundary conditions) satisfies 
$\Delta \tilde{\eta} = -4\pi (\rho_1(\eta_1) + \rho_2(\eta_2))$.
In this case there exists an equation of state $\tilde{\rho}(\tilde{\eta})$, 
such that
$\Delta \tilde{\eta} = -4 \pi \tilde{\rho}(\tilde{\eta})$. Clearly, this equation
of state $\tilde{\rho}(\tilde{\eta})$ depends on the 
explicit form of $\rho_1(r)$ and $\rho_2(r)$. 
In this (weak) sense the
superposition of two perfect fluid solutions is again a perfect fluid solution.

A particularly interesting case to study is the superposition
of a regular perfect fluid solution with a vacuum solution corresponding
to a point mass:

Consider an asymptotically polytropic equation of state $\rho(\eta)$ 
with asymptotic behavior given by the indices $n_0$ and $n_1$ and
consider an associated regular perfect fluid solution $\eta(r)$.
Let us assume that it is of finite extent with radius $R$ and total mass $M$.
A point mass solution is given by the potential $-M_0/r$, where $M_0$ is some
positive constant. It satisfies the vacuum equation on $\{r>0\}$,
i.e., $\Delta (M_0/r) = 0$.
The superposition of these two solutions is given by
$\tilde{\eta}(r) := \eta(r) + M_0/r - M_0/R$. $\tilde{\eta}(r)$ satisfies the
equation $\Delta \tilde{\eta} = -4 \pi \rho(\eta)$ and the
boundary condition $\tilde{\eta}(R) = \eta(R) = 0$.

Note that $\eta(r)$ is monotonically decreasing; hence the same is true 
for $\tilde{\eta}(r)$ and there exists the inverse function $r(\tilde{\eta})$.
Define 
\begin{equation} 
\tilde{\rho}(\tilde{\eta}) := \rho( \eta ) = 
\rho\left(\tilde{\eta} - \frac{M_0}{r(\tilde{\eta})} + \frac{M_0}{R}\right)\:,
\end{equation} 
then $\Delta \tilde{\eta} = -4 \pi \tilde{\rho}(\tilde{\eta})$ (on $\{r>0\}$), 
i.e., $\tilde{\eta}(r)$ is a perfect fluid solution corresponding to the equation
of state $\tilde{\rho}(\tilde{\eta})$.
By using the expansions (\ref{regsols}) and (\ref{finitesols}) of $\eta(r)$ 
it can easily be shown that the equation of state 
$\tilde{\rho}(\tilde{\eta})$ is asymptotically polytropic 
with asymptotic indices
$\tilde{n}_0 = n_0$ 
and $\tilde{n}_1 = 0$, namely
\begin{equation}
\tilde{\rho}(\tilde{\eta}) = 
\rho_- \left(\frac{M}{M+M_0}\right)^{n_0} 
\tilde{\eta}^{n_0} (1 + O(\tilde{\eta}^{\tilde{a}_0}))
\quad (\tilde{\eta} \rightarrow 0)
\quad,\quad
\tilde{\rho}(\tilde{\eta}) = 
\rho_c (1 + O(\tilde{\eta}^{-2}) )
\quad (\tilde{\eta} \rightarrow \infty)\:.
\end{equation}
In particular we see that $\tilde{\rho}(\tilde{\eta})$
has the same asymptotics as $\rho(\eta)$ for
low densities and
behaves like an incompressible fluid for large pressures.

In the dynamical systems formulation 
the perfect fluid solution $\tilde{\eta}(r)$ 
appears as a solution that is emitted from the fixed point
$P_4$ in the state space associated with $\tilde{\rho}(\tilde{\eta})$
(compare with (\ref{posmasssinggeneric}) for $n_1 = 0$).
We can take the viewpoint that, by superposing a 
regular solution $\eta(r)$ (associated with $\rho(\eta)$)
with a mass point solution, 
the regular solution is mapped to a solution emitted from $P_4$
(in the $\tilde{\rho}(\tilde{\eta})$-state space).

\begin{remark}
As an example we have taken regular solutions with finite extent.
However, the statement is the same for general regular solutions.
Moreover, maps as described above can be obtained for completely
different families of solutions in a similar way.
\end{remark}

%%%%%%%%%%%%%%%%%%%%%%%%%%%%%%%%%%%%%%%%%%%%%%%%%%%
\subsection{Modified formulation}
\label{appendix:modifiedformulation}
%%%%%%%%%%%%%%%%%%%%%%%%%%%%%%%%%%%%%%%%%%%%%%%%%%%

As we have already stressed earlier, 
the formulation presented in Section~\ref{dynamicalsystemsformulation},
in particular the choice of variables~(\ref{uqom}),
is adapted to asymptotically polytropic equations of state in the
proper sense, i.e., $n<\infty$.
However, in some situations one is also interested in equations
of state $\rho(p)$ that behave asymptotically like
$\rho \propto p^{\nu_0}$, $\rho \propto p^{\nu_1}$ for arbitrary 
$\nu_0,\nu_1\geq 0$. 
In particular one would like to cover equations of state that
are asymptotically linear in the low and high pressure regime
(i.e., $\nu_0=1$ or $\nu_1=1$).

In order to deal with this slightly more general class
of equations of state, we can introduce the variables
\begin{equation}\label{uvom}
u = \frac{4\pi r^3 \rho}{m}\ , \quad
v = \frac{m\rho}{r p}\ , \quad
\omega = p^a \ ,
\end{equation}
and the corresponding compactified $U, V, \Omega$.
In these variables we obtain the dynamical system
\begin{subequations}\label{UVOmega}
\begin{align}
\label{Ueqalt}
  \frac{dU}{d\lambda} &=  
  U(1-U)[(1-V)(3-4U) - \nu(\Omega)\,V(1-U)] \\
\label{Veqalt}
  \frac{dV}{d\lambda} &= 
  V(1-V)[(2U-1)(1-V) + (1-\nu(\Omega))\, V (1-U)] \\
\label{Omegaeqalt}
  \frac{d\Omega}{d\lambda} &= -a\Omega(1-\Omega)V(1-U)\, ,
\end{align}
\end{subequations}
where $\nu(\Omega)=\nu(p(\Omega))$ is the index-function defined via
$\nu(p) = \frac{d\log\rho(p)}{d\log p}$.
For polytropic equations of state 
this is a constant, $\nu(p) \equiv \nu = \frac{n}{n+1}$.
The index-function $\nu(p)$ is related to the so-called adiabatic index $\gamma_{\rm{ad}}(p)$
via $\nu(p) = 1/\gamma_{\rm{ad}}(p)$ (see, e.g., \cite{book:KippWeig1994}).
Note that for general equations of state 
$\nu(p) \neq \frac{n(p)}{n(p)+1}$ (where $n(p) = n(\eta(p))$ is
the index-function defined in~(\ref{ndef})).

\begin{figure}[htp]
	\centering
        \subfigure[$\nu=0$]{
		\label{baseg0}
		\includegraphics[height=0.2\textwidth]{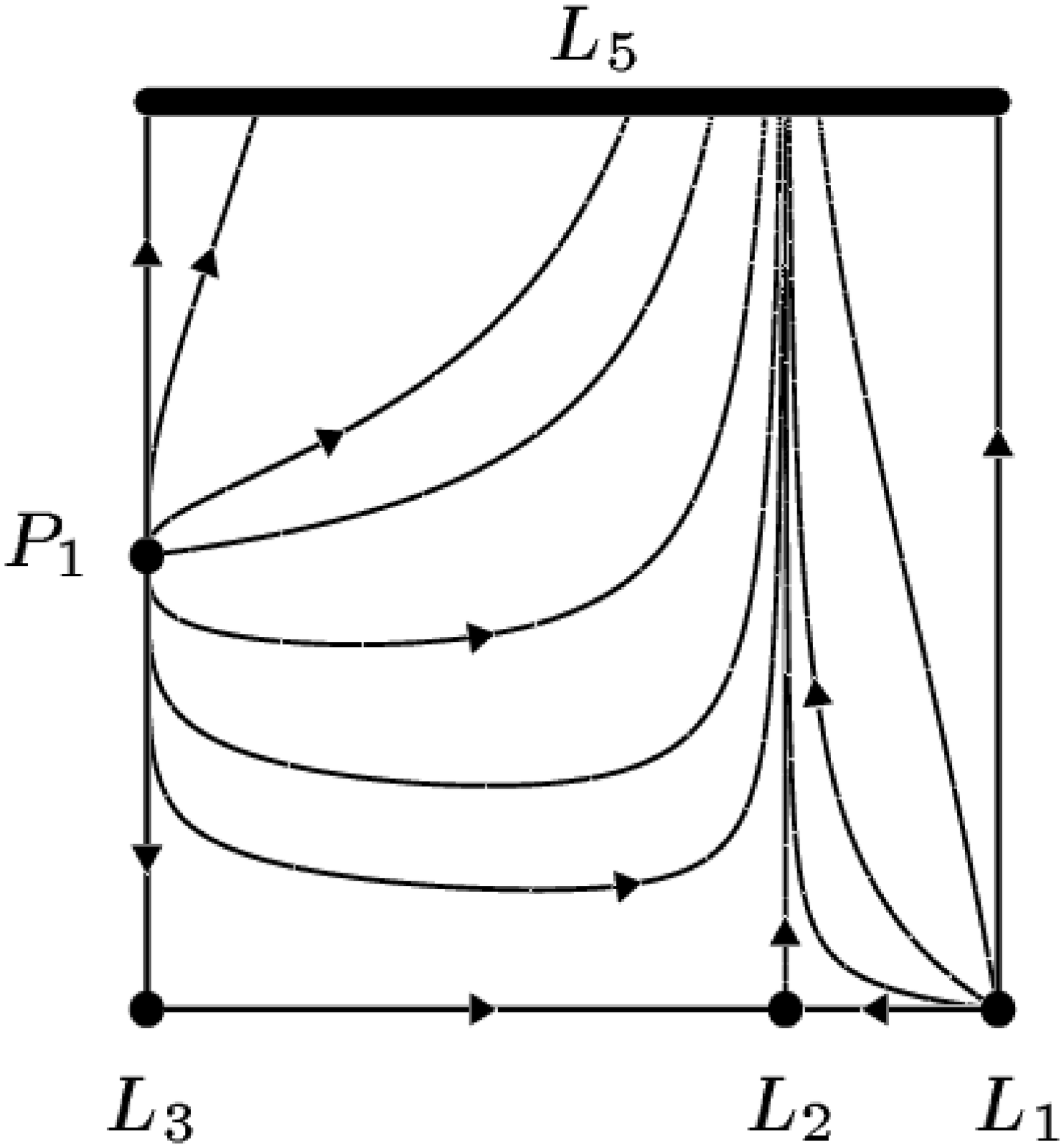}}\quad
     	\subfigure[$\nu=\frac{1}{2}$]{
		\label{baseg12}
		\includegraphics[height=0.2\textwidth]{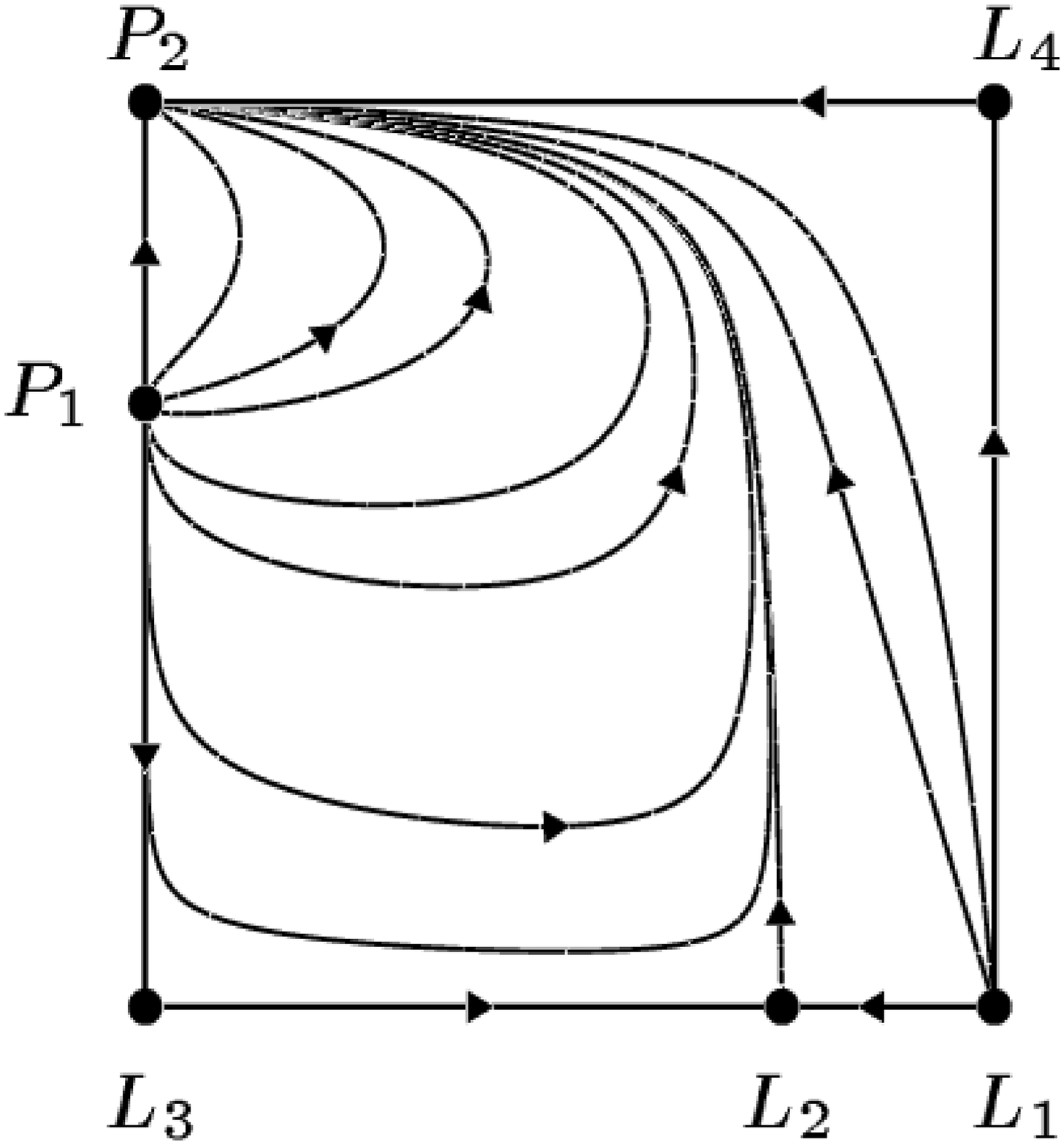}}\quad
	\subfigure[$\nu=\frac{3}{4}$]{
		\label{baseg34}
                \includegraphics[height=0.2\textwidth]{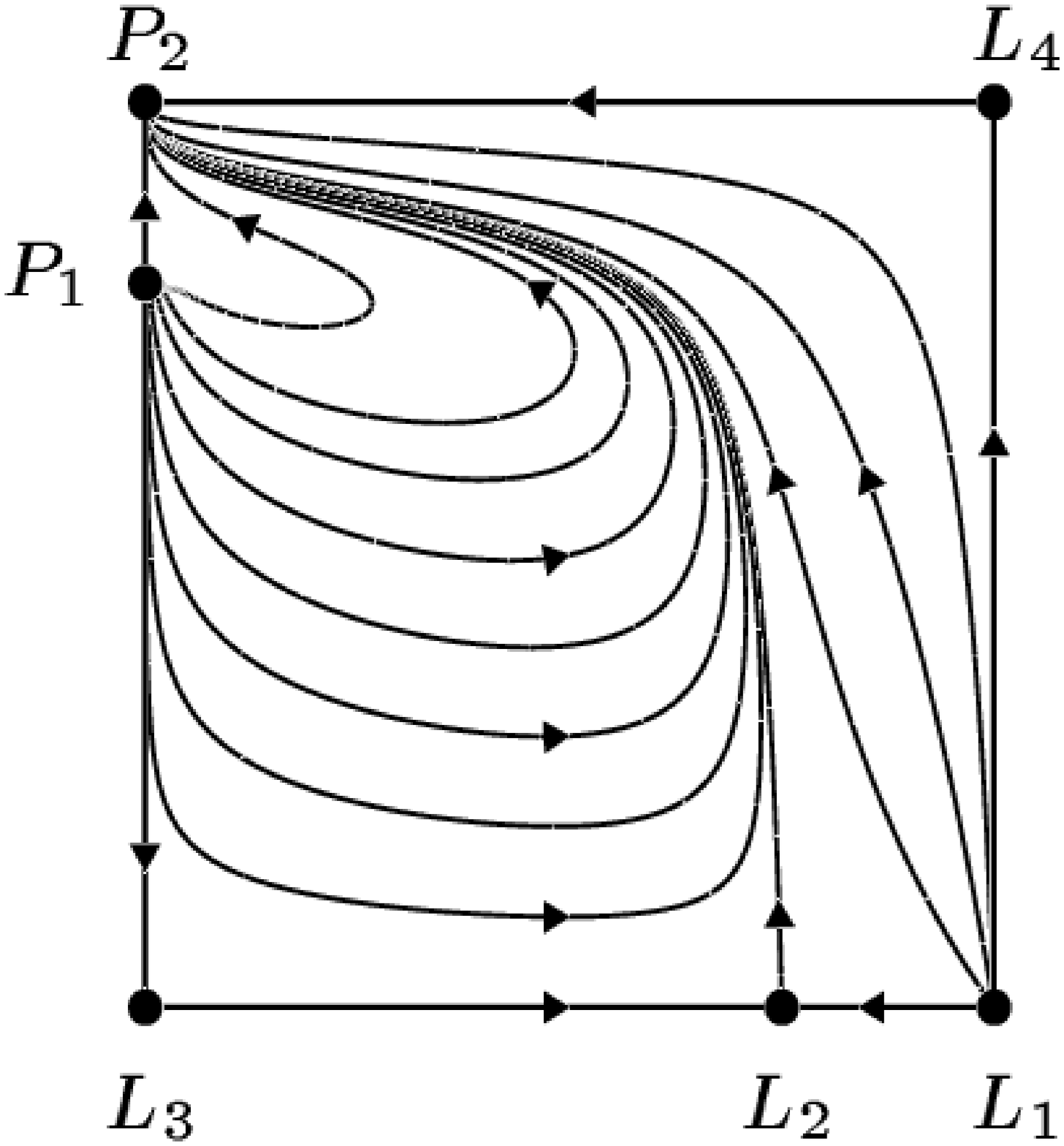}}\quad
	\subfigure[$\nu=\frac{4}{5}$]{
		\label{baseg45}
		\includegraphics[height=0.2\textwidth]{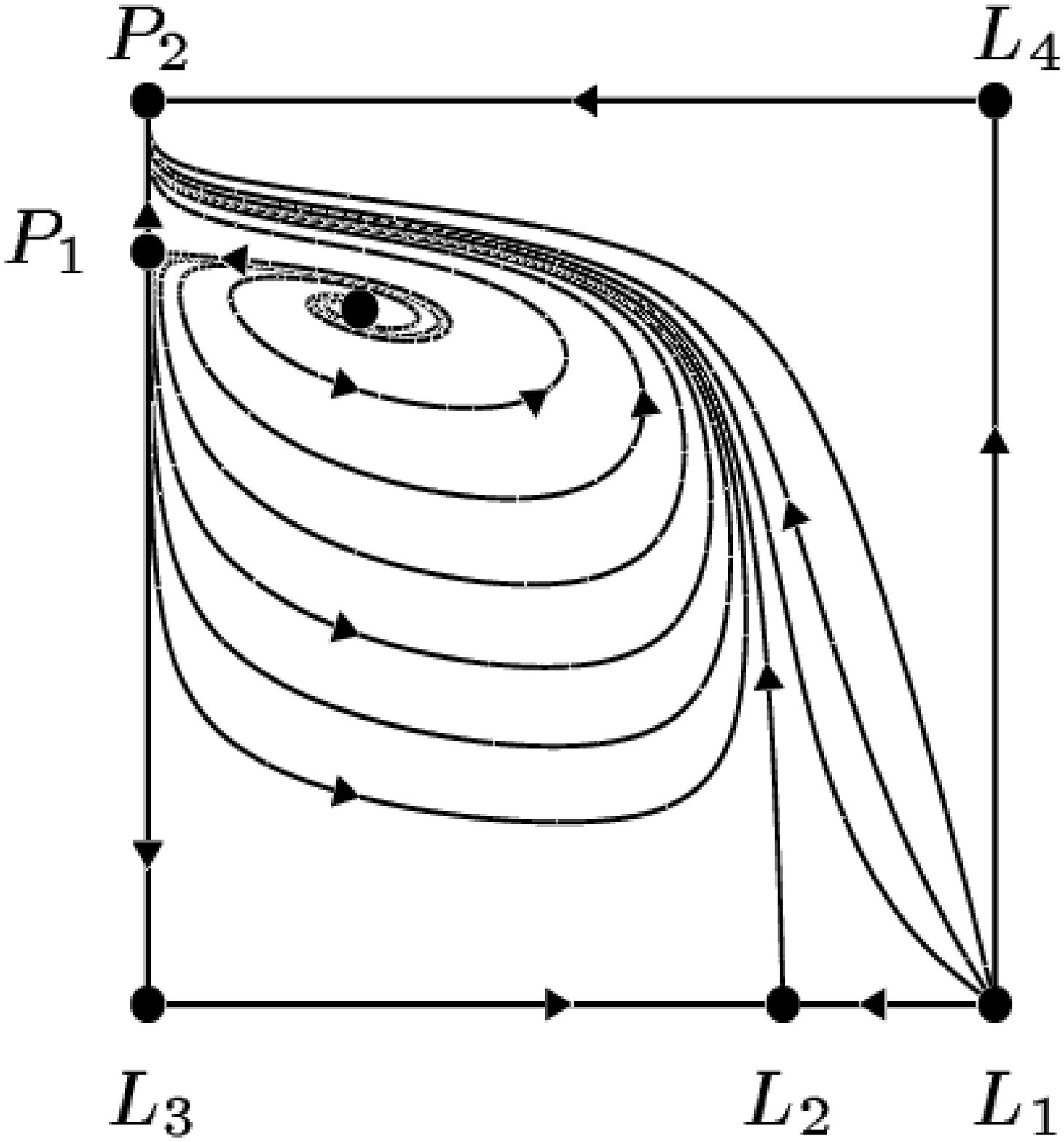}}\quad
	\subfigure[$\nu=\frac{5}{6}$]{
		\label{baseg56}
		\includegraphics[height=0.2\textwidth]{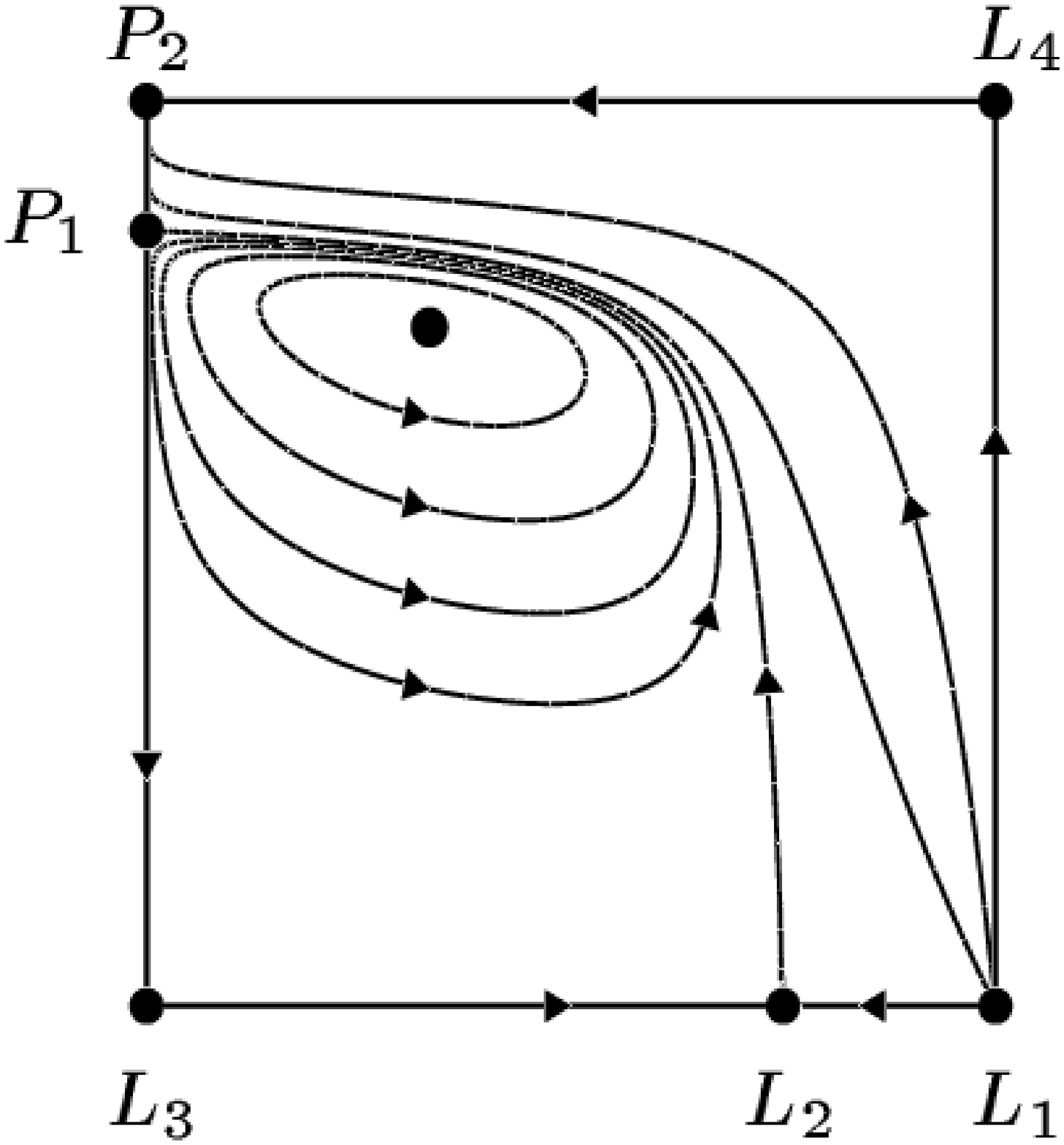}}\quad
        \subfigure[$\nu=\frac{10}{11}$]{
		\label{baseg1011}
		\includegraphics[height=0.2\textwidth]{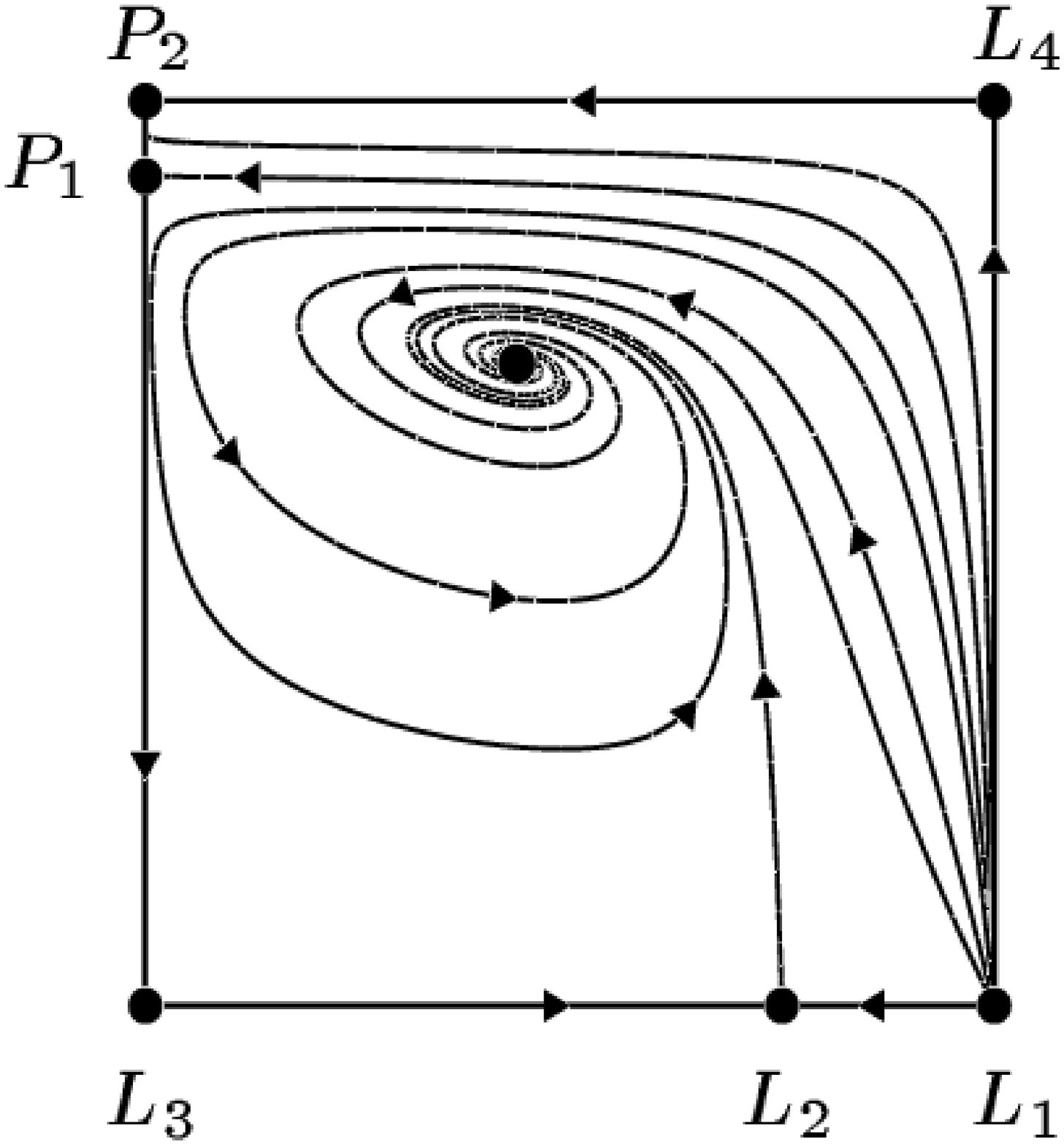}}\quad
        \subfigure[$\nu=1$]{
		\label{baseg1}
		\includegraphics[height=0.2\textwidth]{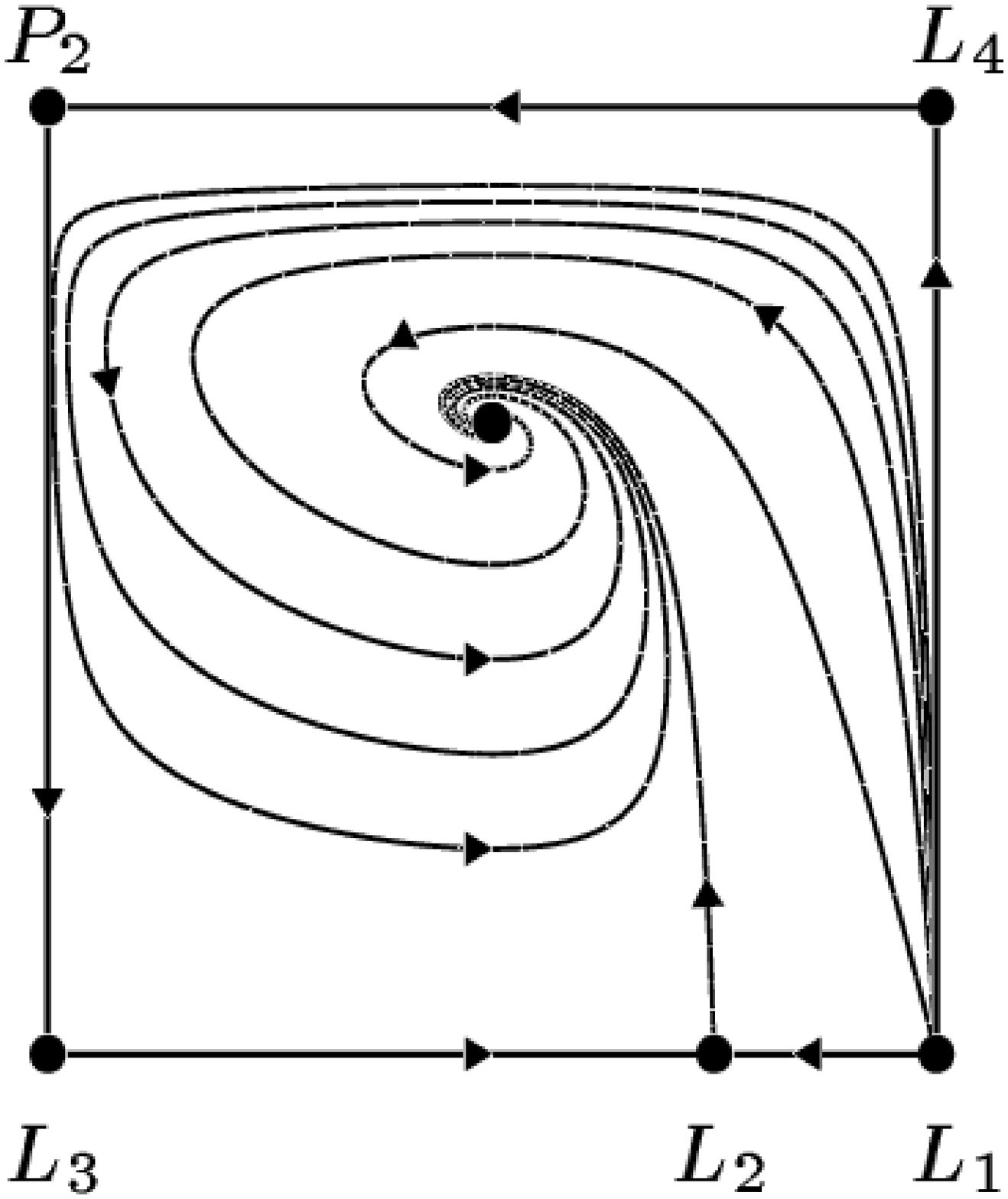}}\quad
        \subfigure[$\nu=1.5$]{
		\label{baseg15}
		\includegraphics[height=0.2\textwidth]{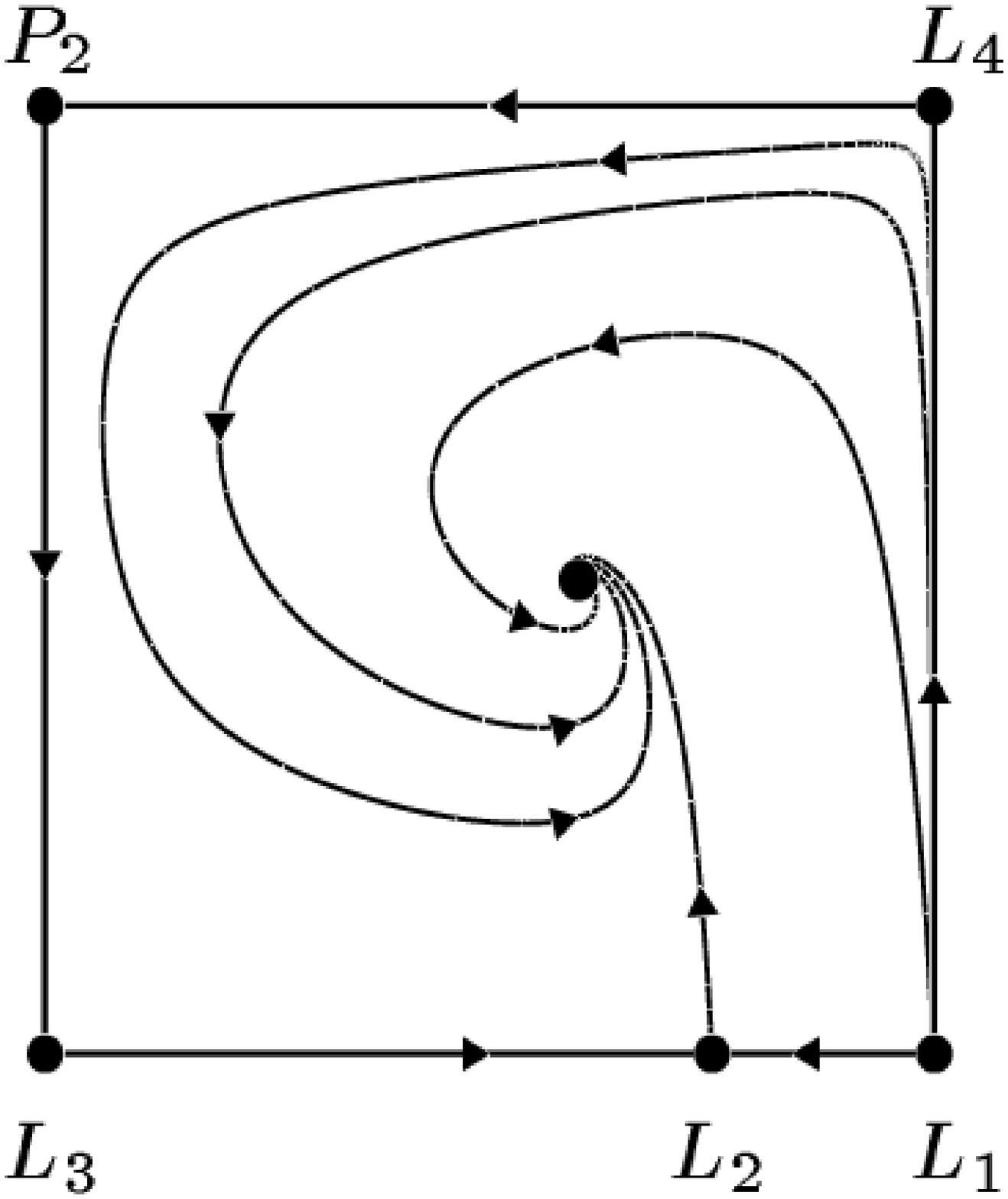}}\quad
    \caption{Orbits in the polytropic subset $\Omega=0$ in terms of the 
		alternative variables $U$ (horizontal axis) and $V$ (vertical axis).}
    \label{baseg}
\end{figure}

The flow of this dynamical system on the $\Omega=0$ plane
is depicted in Figure~\ref{baseg}; this is 
the counterpart of Figure~\ref{base} in the present formulation.

%\begin{figure}[htp]
%	\psfrag{L1}{{\scriptsize $L_1$}}
%	\psfrag{L2}{{\scriptsize $L_2$}}
%	\psfrag{L3}{{\scriptsize $L_3$}}
%	\psfrag{L4}{{\scriptsize $L_4$}}
%	\psfrag{L5}{{\scriptsize $L_5$}}
%	\psfrag{P1}{{\scriptsize $P_1$}}
%	\psfrag{P2}{{\scriptsize $P_2$}}
%	\centering
%        \subfigure[$\nu=0$]{
%		\label{baseg0}
%		\includegraphics[height=0.2\textwidth]{g0f.eps}}\quad
%     	\subfigure[$\nu=\frac{1}{2}$]{
%		\label{baseg12}
%		\includegraphics[height=0.2\textwidth]{g1over2f.eps}}\quad
%	\subfigure[$\nu=\frac{3}{4}$]{
%		\label{baseg34}
%               \includegraphics[height=0.2\textwidth]{g3over4f.eps}}\quad
%	\subfigure[$\nu=\frac{4}{5}$]{
%		\label{baseg45}
%		\includegraphics[height=0.2\textwidth]{g4over5f.eps}}\quad
%	\subfigure[$\nu=\frac{5}{6}$]{
%		\label{baseg56}
%		\includegraphics[height=0.2\textwidth]{g5over6f.eps}}\quad
%        \subfigure[$\nu=\frac{10}{11}$]{
%		\label{baseg1011}
%		\includegraphics[height=0.2\textwidth]{g10over11f.eps}}\quad
%        \subfigure[$\nu=1$]{
%		\label{baseg1}
%		\includegraphics[height=0.2\textwidth]{g1f.eps}}\quad
%        \subfigure[$\nu=1.5$]{
%		\label{baseg15}
%		\includegraphics[height=0.2\textwidth]{g1point5f.eps}}\quad
%    \caption{Orbits in the polytropic subset $\Omega=0$ in terms of the 
%		alternative variables $U$ (horizontal axis) and $V$ (vertical axis).}
%    \label{baseg}
%\end{figure}

%\begin{figure}[htp]
%	\centering
%	\includegraphics[width=0.9\textwidth]{figure14.ps}
%    \caption{Orbits in the polytropic subset $\Omega=0$ in terms of the 
%		alternative variables $U$ (horizontal axis) and $V$ (vertical axis).}
%    \label{baseg}
%\end{figure}

As indicated by Figure~\ref{baseg} 
all solutions have infinite radius
and infinite mass for $\nu_0 \geq 1$. This is in accord with
the criterion guaranteeing infinite extent if
$\eta = \int_0^p dp^\prime {\rho}^{-1}(p^\prime) \not< \infty$
(see~\cite{Rendall/Schmidt:1991} for a good presentation).
For further statements, the theorems discussed in Section~\ref{qualitativeresults}
can be reformulated and adopted to the present context.

Compared to the formulation presented in Section~\ref{dynamicalsystemsformulation}, 
the formulation~(\ref{uvom}) has certain disadvantages:
Firstly, $\nu$ appears twice in the dynamical system~(\ref{UVOmega}),
secondly, the fixed points on the boundaries of the state space
are not ``fixed'', i.e., their position depends on $\nu_i$.
Moreover, Chandrasekhar's equation of state is quite complicated
in terms of $\rho,p$ variables, while it is quite elegantly expressed
in $\rho,\eta$ variables. Thus we find that "$\eta$-based" variables,
in many contexts, is a quite natural choice.

\vspace*{3cm}

%%%%%%%%%%%%%%
\end{appendix}

\begin{thebibliography}{99}

\bibitem{art:Buchdahl1959}
H.~Buchdahl.
\newblock General Relativistic Fluid Spheres.
\newblock {\em Phys. Rev.}, 116~:~1027--1034, 1959.

\bibitem{book:Carr1981}
J.~Carr. 
\newblock {\em Applications of center manifold theory}. 
\newblock Springer Verlag, New York, 1981.

\bibitem{Chandrasekhar:1939}
S.~Chandrasekhar.
\newblock {\em An introduction to the study of stellar structure}.
\newblock University of Chicago Press, Chicago, 1939.

\bibitem{art:Crawford1991}
J.D.~Crawford.
\newblock {Introduction to bifurcation theory}.
\newblock {\em Rev. Mod. Phys.}, 63-4~:~991--1038, 1991.

\bibitem{art:Heinzle2002}
J.M.~Heinzle.
\newblock {(In)finiteness of static spherically symmetric perfect fluids}.
\newblock {\em Class. Quantum Grav.}, 19~:~2835--2853, 2002.

\bibitem{Horedt:1987}
G.P. Horedt.
\newblock {Topology of the Lane-Emden-equation}.
\newblock {\em Astron. Astrophys.}, 177~:~117--130, 1987.

\bibitem{art:Kimura1981}
H.~Kimura.
\newblock A Study of Simple Polytropes I. Fundamentals and Classification of Solutions.
\newblock {\em {P}ubl. {A}stron. {S}oc. {J}apan}, 33~:~273--298, 1981.

\bibitem{book:KippWeig1994}
R.~Kippenhahn and A.~Weigert.
\newblock {\em Stellar structure and evolution}. 
\newblock Springer, Berlin Heidelberg, 1994.

\bibitem{Lindblom:1993}
L.~Lindblom.
\newblock On the symmetries of equilibrium star models.
\newblock In S.~Chandrasekhar, editor, {\em Classical General Relativity},
  Oxford Science Publications, pages ~17--28. Oxford University Press, 1993.

\bibitem{Makino:1998}
T.~Makino.
\newblock On spherically symmetric stellar models in general relativity.
\newblock {\em J. Math. Kyoto Univ.}, 38-1~:~55--69, 1998.

\bibitem{art:Makino2000}
T.~Makino.
\newblock {On the spiral structure of the (R,M)-diagram for a 
stellar model of the Tolman-Oppenheimer-Volkoff equation}.
\newblock {\em Funkcialaj Ekvacioj}, 43~:~471--489, 2000.

\bibitem{art:Makino1984}
T.~Makino.
\newblock {Ordinary Differential Equations of Emden Type}
\newblock {\em Funkcialaj Ekvacioj}, 27~:~319--329, 1984.

\bibitem{art:grstar_leq}
U.~Nilsson and C.~Uggla.
\newblock {General relativistic stars: Linear equations of state}.
\newblock {\em Ann. Phys.}, 286~:~278-291, 2000 .

\bibitem{art:grstar_pol}
U.~Nilsson and C.~Uggla.
\newblock {General relativistic stars: Polytropic equation of state}.
\newblock {\em Ann. Phys.}, 286~:~292-319, 2000.

\bibitem{art:Rendall/Tod1999}
A.D.~Rendall and K.P.~Tod.
\newblock Dynamics of spatially homogeneous 
solutions of the Einstein-Vlasov equations which are locally rotationally symmetric.
\newblock {\em Class. Quantum Grav.}, 16~:~1705--1726, 1999.

\bibitem{Rendall/Schmidt:1991}
A.D.~Rendall and Bernd~G. Schmidt.
\newblock Existence and properties of spherically symmetric static fluid bodies
  with a given equation of state.
\newblock {\em Class. Quantum Grav.}, 8~:~985--1000, 1991.

\bibitem{Schaudt:2000}
U.M.~Schaudt.
\newblock {On Static Stars in Newtonian Gravity and Lane-Emden Type Equations}.
\newblock {\em Ann. Henri Poincar\'e}, 5~:~945--976, 2000.

\bibitem{Shapiro/Teukolsky:1983}
S.L.~Shapiro and S.A.~Teukolsky.
\newblock {\em {Black Holes, White Dwarfs and Neutron Stars}}.
\newblock John Wiley \& Sons, 1983.

\bibitem{Simon:2002}
W.~Simon.
\newblock {Criteria for (in)finite extent of static perfect fluids}.
\newblock In J.~Frauendiener and H.~Friedrich, editors, 
{\em The conformal structure of space-times: Geometry, Analysis, Numerics},
Lecture Notes in Physics {\bf 604}, Springer (2002).


\bibitem{book:WainwrightEllis1997}
J.~Wainwright and G.F.R.~Ellis.
\newblock {\em Dynamical systems in cosmology}. 
\newblock {C}ambridge {U}niversity {P}ress, Cambridge, 1997.

\end{thebibliography}
\end{document}